\documentclass[letterpaper,twocolumn,10pt]{article}
\usepackage{usenix2019,epsfig,endnotes}
\usepackage[dvipsnames]{xcolor}

\usepackage{textgreek}
\usepackage{hyperref}
\usepackage{graphicx}
\usepackage{comment}
\usepackage{authblk}
\graphicspath{ {./images/} }

\title{\LARGE \bf
SSDFS: Towards LFS Flash-Friendly File System without GC operations
}

\author[1]{Viacheslav Dubeyko}

\begin{document}

\maketitle
\pagestyle{plain}

\begin{abstract}

Solid state drives have a number of interesting characteristics. However, there are numerous file system and storage design issues for SSDs that impact the performance and device endurance. Many flash-oriented and flash-friendly file systems introduce significant write amplification issue and GC overhead that results in shorter SSD lifetime and necessity to use the NAND flash overprovisioning. SSDFS file system introduces several authentic concepts and mechanisms: logical segment, logical extent, segment's PEBs pool, Main/Diff/Journal areas in the PEB's log, Diff-On-Write approach, PEBs migration scheme, hot/warm data self-migration, segment bitmap, hybrid b-tree, shared dictionary b-tree, shared extents b-tree. Combination of all suggested concepts are able: (1) manage write amplification in smart way, (2) decrease GC overhead, (3) prolong SSD lifetime, and (4) provide predictable file system's performance.

\end{abstract}



\section{INTRODUCTION}

\textbf{Flash memory characteristics}. Flash is available in two types NOR and NAND. NOR flash is directly addressable, helps in reading but also in executing of instructions directly from the memory. NAND based SSD consists of set of blocks which are fixed in number and each block comprises of a fixed set of pages or whole pages set makes up a block. There are three types of operations in flash memory: read, write and erase. Execution of operations read and write takes place per page level. On the other hand the data is erased on block level by using erase operation. Because of the physical feature of flash memory, write operations are able to modify bits from one to zero. Hence the erase operation should be executed before rewriting as it set all bits to one. The typical latencies: (1) read operation - 20 us, (2) write operation - 200 us, (3) erase operation - 2 ms.

\textbf{Flash Translation Layer (FTL)}. FTL emulates the functionality of a block device and enables operating system to use flash memory without any modification. FTL mimics block storage interface and hides the internal complexities of flash memory to operating systems, thus enabling the operating system to read/write flash memory in the same way as reading/writing the hard disk. The basic function of FTL algorithm is to map the page number from logical to physical. However, internally FTL needs to deal with erase-before-write, which makes it critical to overall performance and lifetime of SSD.

\textbf{Garbage Collection and Wear Leveling}. The process of collecting, moving of valid data and erasing the invalid data is called as garbage collection. Through SSD firmware command TRIM the garbage collection is triggered for deleted file blocks by the file system. Commonly used erase blocks puts off quickly, slows down access times and finally burning out. Therefore the erase count of each erase block should be monitored. There are wide number of wear-leveling techniques used in FTL.

\textbf{Building blocks of SSD}. SSD includes a controller that incorporates the electronics that bridge the NAND memory components to the host computer. The controller is an embedded processor that executes firmware-level code. Some of the functions performed by the controller includes, error-correcting code (ECC), wear leveling, bad block mapping, read scrubbing and read disturb management, read and write caching, garbage collection etc. A flash-based SSD typically uses a small amount of DRAM as a cache, similar to the cache in hard disk drives. A directory of block placement and wear leveling data is also kept in the cache while the drive is operating.

\textbf{Write amplification}. For write requests that come in random order, after a period of time, the free page count in flash memory becomes low. The garbage-collection mechanism then identifies a victim block for cleaning. All valid pages in the victim block are relocated into a new block with free pages, and finally the candidate block is erased so that the pages become available for rewriting. This mechanism introduces additional read and write operations, the extent of which depends on the specific policy deployed, as well as on the system parameters. These additional writes result in the multiplication of user writes, a phenomenon referred to as “write amplification”.

\textbf{Read disturbance}. Flash data block is composed of multiple NAND units to which the memory cells are connected in series. A memory operation on a specific flash cell will influence the charge contents on a different cells. This is called disturbance, which can occur on any flash operation and predominantly this is observed during the read operation and leads to errors in undesignated memory cells. To avoid failure on reads, error-correcting codes (ECC) are widely employed. Read disturbance can occur when reading the same target cell multiple times without an erase and program operation. In general, to preserve data consistency, flash firmware reads all live data pages, erases the block, and writes down the live pages to the erased block. This process, called read block reclaiming, introduces long latencies and degrades performance.

\textbf{SSD design issues}. Solid state drives have a number of interesting characteristics that change the access patterns required to optimize metrics such as disk lifetime and read/write throughput. In particular, SSDs have approximately two orders of magnitude improvement in read and write latencies, as well as a significant increase in overall bandwidth. However, there are numerous file system and storage array design issues for SSDs that impact the performance and device endurance. SSDs suffer well-documented shortcomings: log-on-log, large tail-latencies, unpredictable I/O latency, and resource underutilization. These shortcomings are not due to hardware limitations: the non-volatile memory chips at the core of SSDs provide predictable high-performance at the cost of constrained operations and limited endurance/reliability. Providing the same block I/O interface as a magnetic disk is one of the important reason of these drawbacks.

\textbf{SSDFS features}. Many flash-oriented and flash-friendly file systems introduce significant write amplification issue and GC overhead that results in shorter SSD lifetime and necessity to use the NAND flash overprovisioning. SSDFS file system introduces several authentic concepts and mechanisms: logical segment, logical extent, segment's PEBs pool, Main/Diff/Journal areas in the PEB's log, Diff-On-Write approach, PEBs migration scheme, hot/warm data self-migration, segment bitmap, hybrid b-tree, shared dictionary b-tree, shared extents b-tree. Combination of all suggested concepts are able: (1) manage write amplification in smart way, (2) decrease GC overhead, (3) prolong SSD lifetime, and (4) provide predictable file system's performance.

The rest of this paper is organized as follows. Section II surveys the related works. Section III explains the SSDFS architecture and approaches. Section IV includes final discussion. Section V offers conclusions.

\section{RELATED WORKS}

\subsection{File Systems Statistics and Analysis}

\textbf{File size}. Agrawal, et al. \cite{c109} discovered that 1-1.5\% of files on a file system's volume have a size of zero. The arithmetic mean file size was 108 KB in 2000 year and 189 KB in 2004 year. This metric grows roughly 15\% per year. The median weighted file size increasing from 3 MB to 9 MB. Most of the bytes in large files are in video, database, and blob files, and that most of the video, database, and blob bytes are in large files. A large number of small files account for a small fraction of disk usage. Douceur, et al. \cite{c111} confirmed that 1.7\% of all files have a size of zero. The mean file size ranges from 64 kB to 128 kB across the middle two quartiles of all file systems.  The median size is 2 MB and it confirms that most files are small but most bytes are in large files. Ullah, et al. \cite{c115} have results about 1 to 10 KB the value observed is up to 32\% of the total occurrences. There are 29\% values in the range of 10 KB to 100 KB.  Gibson, et al. \cite{c116} agreed that most files are relatively small, more than half are less than 8 KB. 80\% or more of the files are smaller than 32 KB. On the other hand, while only 25\% are larger than 8 KB, this 25\% contains the majority of the bytes used on the different systems.

\textbf{File age}. Agrawal, et al. \cite{c109} stated that the median file age ranges between 80 and 160 days across datasets, with no clear trend over time. Douceur, et al. \cite{c111} has the vision of the median file age is 48 days. Studies of short-term trace data have shown that the vast majority of files are deleted within a few minutes of their creation. On 50\% of file systems, the median file age ranges by a factor of 8 from 12 to 97 days, and on 90\% of file systems, it ranges by a factor of 256 from 1.5 to 388 days. Gibson, et al. \cite{c116} showed that while only 15\% of the files are modified daily, these modifications account for over 70\% of the bytes used daily. Relatively few files are used on any one day — normally less than 5\%. 5-10\% of all files created are only used on one day, depending on the system. On the other hand, approximately 0.8\% of the files are used more than 129 times — essentially every day. These files which are used more than 129 times account for less than 1\% of all files created and approximately 10\% of all the files which were accessed or modified. 90\% of all files are not used after initial creation, those that are used are normally short-lived, and that if a file is not used in some manner the day after it is created, it will probably never be used. 1\% of all files are used daily.

\textbf{Files count per file system}. Agrawal, et al. \cite{c109} showed that the count of files per file system is going up from year to year. The arithmetic mean has grown from 30K to 90K files and the median has grown from 18K to 52K files (2000 - 2004 years). However, some percentage of file systems has achieved about 512K files already in 2004 year. Douceur, et al. \cite{c111} have vision that 31\% of all file systems contain 8k to 16k files. 30\% of file systems have fewer than 4k files. Ullah, et al. \cite{c115} found that 67\% of the occurrences are found in the range of 1-8 number of files in a directory. The percentage of the 9-16 files in a directory comprises of 15\% of the total data found. The number of file in the range 17-32 are 9\% and only 9\% occurrences are found in the data more than 32 files in a directory.

\textbf{File names}. Agrawal, et al. \cite{c109} made conclusion that dynamic link libraries (dll files) contain more bytes than any other file type. And virtual hard drives are consuming a rapidly increasing fraction of file-system space.  Ullah, et al. \cite{c115} discovered that the file name length falls in the range from 9 to 17 characters. The peak occurs for file names with length of 12 characters. The file names smaller than 8 characters are up to 11\% of the total data collected whereas the occurrences of file names larger than 16 characters and up to 32 characters is 26\% but the file names greater than 32 characters are found to be only 6\%.

\textbf{Directory size}. Agrawal, et al. \cite{c109} discovered that across all years, 23-25\% of directories contain no files. The arithmetic mean directory size has decreased slightly and steadily from 12.5 to 10.2 over the sample period, but the median directory size has remained steady at 2 files. Across all years, 65-67\% of directories contain no subdirectories. Across all years, 46-49\% of directories contain two or fewer entries. Douceur, et al. \cite{c111} shared that 18\% of all directories contain no files and the median directory size is 2 files. 69\% of all directories contain no subdirectories, 16\% contain one, and fewer than 0.5\% contain more than twenty. On 50\% of file systems, the median directory size ranges from 1 to 4 files, and on 90\% of file systems, it ranges from 0 to 7 files. On 95\% of all file systems, the median count of subdirectories per directory is zero. 15\% of all directories are at depth of 8 or greater.

\textbf{Directories count per file system}. Agrawal, et al. \cite{c109} registered that the count of directories per file system has increased steadily over five-year sample period. The arithmetic mean has grown from 2400 to 8900 directories and the median has grown from 1K to 4K directories. Douceur, et al. \cite{c111} shared that 28\% of all file systems contain 512 to 1023 directories, and 29\% of file systems have fewer than 256 directories. Ullah, et al. \cite{c115} shared that 59\% of the directories have sub-directories in the range of 1-5, 35\% occurrences are found in the range of 6-10. But the results show that only 6\% occurrences are found in the range of above 10 sub-directories in a directory.

\textbf{Namespace tree depth}. Agrawal, et al. \cite{c109} shared that there are many files deep in the namespace tree, especially at depth 7. Also, files deeper in the namespace tree tend to be orders-of-magnitude smaller than shallower files. The arithmetic mean has grown from 6.1 to 6.9, and the median directory depth has increased from 5 to 6. The count of files per directory is mostly independent of directory depth. Files deeper in the namespace tree tend to be smaller than shallower ones. The mean file size drops by two orders of magnitude between depth 1 and depth 3, and there is a drop of roughly 10\% per depth level thereafter.

\textbf{Capacity and usage}. Agrawal, et al. \cite{c109} registered that 80\% of file systems become fuller over a one-year period, and the mean increase in fullness is 14 percentage points. This increase is predominantly due to creation of new files, partly offset by deletion of old files, rather than due to extant files changing size. The space used in file systems has increased not only because mean file size has increased (from 108 KB to 189 KB), but also because the number of files has increased (from 30K to 90K). Douceur, et al. \cite{c111} discovered that file systems are on average only half full, and their fullness is largely independent of user job category. On average, half of the files in a file system have been created by copying without subsequent writes, and this is also independent of user job category. The mean space usage is 53\%.

\textbf{A File Is Not a File}. Harter, et al. \cite{c113} showed that modern applications manage large databases of information organized into complex directory trees. Even simple word-processing documents, which appear to users as a "file", are in actuality small file systems containing many sub-files.

\textbf{Auxiliary files dominate}. Tan, et al. \cite{c112} discovered that on iOS, applications access resource, temp, and plist files very often. This is especially true for Facebook which uses a large number of cache files. Also for iOS resource files such as icons and thumbnails are stored individually on the file system. Harter, et al. \cite{c113} agree with that statement. Applications help users create, modify, and organize content, but user files represent a small fraction of the files touched by modern applications. Most files are helper files that applications use to provide a rich graphical experience, support multiple languages, and record history and other metadata.

\textbf{Sequential Access Is Not Sequential}. Harter, et al. \cite{c113} stated that even for streaming media workloads, "pure" sequential access is increasingly rare. Since file formats often include metadata in headers, applications often read and re-read the first portion of a file before streaming through its contents.

\textbf{Writes are forced}. Tan, et al. \cite{c112} shared that on iOS, Facebook calls fsync even on cache files, resulting in the largest number of fsync calls out of the applications. On Android, fsync is called for each temporary write-ahead logging journal files. Harter, et al. \cite{c113} found that applications are less willing to simply write data and hope it is eventually flushed to disk. Most written data is explicitly forced to disk by the application; for example, iPhoto calls fsync thousands of times in even the simplest of tasks.

\textbf{Temporary files}. Tan, et al. \cite{c112} showed that applications create many temporary files. It might have a negative impact on the durability of the flash storage device. Also, creating many files result in storage fragmentation. The SQLite database library creates many short-lived temporary journal files and calls fsync often.

\textbf{Copied files}. Agrawal, et al. \cite{c109} shared the interesting point that over sample period (2000 - 2004), the arithmetic mean of the percentage of copied files has grown from 66\% to 76\%, and the median has grown from 70\% to 78\%. It means that more and more files are being copied across file systems rather than generated locally.  Downey \cite{c114} concluded that the vast majority of files in most file systems were created by copying, either by installing software (operating system and applications) or by downloading from the World Wide Web. Many new files are created by translating a file from one format to another, compiling, or by filtering an existing file. Using a text editor or word processor, users add or remove material from existing files, sometimes replacing the original file and sometimes creating a series of versions.

\textbf{Renaming Is Popular}. Harter, et al. \cite{c113} discovered that home-user applications commonly use atomic operations, in particular rename, to present a consistent view of files to users.

\textbf{Multiple Threads Perform I/O}. Harter, et al. \cite{c113} showed that virtually all of the applications issue I/O requests from a number of threads; a few applications launch I/Os from hundreds of threads. Part of this usage stems from the GUI-based nature of these applications; threads are required to perform long-latency operations in the background to keep the GUI responsive.

\textbf{Frameworks Influence I/O}. Harter, et al. \cite{c113} found that modern applications are often developed in sophisticated IDEs and leverage powerful libraries, such as Cocoa and Carbon. Whereas UNIX-style applications often directly invoke system calls to read and write files, modern libraries put more code between applications and the underlying file system. Default behavior of some Cocoa APIs induces extra I/O and possibly unnecessary (and costly) synchronizations to disk. In addition, use of different libraries for similar tasks within an application can lead to inconsistent behavior between those tasks.

\textbf{Applications' behavior}. Harter, et al. \cite{c113} made several conclusions about applications' nature. Applications tend to open many very small files (\textless4 KB), most of the bytes accessed are in large files (\textgreater1 MB). The vast majority of I/O is performed by reading and writing to open file descriptors. Only a few of the iBench tasks have significant pageins from memory-mapped files; most of this pagein traffic is from images. Applications perform large numbers of very small ($\leq$4 KB) reads and writes. Metadata accesses are very common, greatly outnumbering accesses to file data across all of investigated workloads. A moderate amount of reads could potentially be serviced by a cache, but most reads are to fresh data. Written data is rarely over-written, so waiting to flush buffers until data becomes irrelevant is usually not helpful. Many of the reads and writes to previously accessed data which do occur are due to I/O libraries and high-level abstractions.

\textbf{File System and Block IO Scheduler}. Hui, et al. \cite{c119} have made the estimation of interaction between file systems and block I/O scheduler. They concluded that more read or append-write may cause better performance and less energy consumption, such as in the workload of the webserver. And along with the increasing of the write operation, especially random write, the performance declines and energy consumption increases. The extent file systems express better performance and lower energy consumption. They expected that NOOP I/O scheduler is better suit for the case of SSDs because it does not sort the request, which can cost much time and decline performance. But after the test, they found that as CFQ as NOOP may be suit for the SSDs.

\textbf{NAND flash storage device}. Parthey, et al. \cite{c120} analyzed access timing of removable flash media. They found that many media access address zero especially fast. For some media, other locations such as the middle of the medium are sometimes slower than the average. Accessing very small blocks ($\leq$2 KiB) can be prohibitively slow. This is not the result of normal overhead increase for small blocks but some kind of irregularity, e.g. due to inadequate management mechanisms. The written bit pattern also influences access timing, but to a lesser degree. The overwritten value is irrelevant. The value of irregular behavior is usually 0xff. Sometimes, writing 0xff is a bit faster than all other bit patterns, in rare cases it requires considerably more access time.

Son, et al. \cite{c121} have made an empirical evaluation of NVM Express SSDs. Read performance of NVMe shows very good scalability with respect to the number of threads since the NVMe and SSD controller fully exploit the SSD channel parallelism. As the number of threads is increased, the throughput is improved almost linearly until the number of threads is 128. In case of 256 threads, the throughput is saturated at about 2.5GB/s (625K IOPS), which is a peak throughput for NVMe SSD. The latency also increases almost linearly as the throughput is increased until the number of threads is 256. However, write operations show limited scalability despite the intended scalable design of NVMe because it is limited by the flash characteristics. The performance is increased only marginally in case of more than 16 threads and saturated at 800MB/s since 256 threads. The latency reaches up to 4500us and the larger number of threads only increases contentions among threads and GC overhead. The results demonstrate that random read performance of NVMe SSD is dependent on both the number of cores and threads. For random write of NVMe SSD like random read case, overall performance increases as the number of cores increases. In both read and write cases the performance of NVMe SSD is dependent on the number of cores and reaches the peak performance when the number of cores is enough. NVMe SSD provides as many I/O queues as the number of CPU cores to improve scalability and parallelism. The impact by the number of queues is less consistent due to performance variation in write operation when compared to the random read case but it clearly shows that the maximum performance is lower when only one I/O queue is used. Fsync operation dramatically lowers the write performance of NVMe SSD. In case of sequential write, one fsync per write operation drops the write performance by 47.2\% compared to no fsync per operation. As the number of fsync decreases, the performance gradually increases. Fsync calls have more influence on random write performance than on sequential write performance. There is almost no performance improvement even if the number of fsync calls is decreased with random write operations, which means fsync call limits random write performance to the greater extent. NVMe SSD with low latency does not benefit from using large page sizes for I/O operations.

Zhou, et al. \cite{c122} concluded that when the file systems block size is the same as the SSD page size, the SSD would deliver best performance. This is because writes matching the underlying page size would generally avoid partial page updating/overwriting and correspondingly improve garbage collection efficiency. SSD would exhibit better performance when used in a log-style way. Allocation group is a unique feature to XFS file system. Intuitively, the more allocation group the file system has, the better performance it would deliver. For smaller degree of allocation group (16 allocation group), each allocation group is relatively larger and it is very likely that the workload only exercises only part of one certain allocation group which occupies only one plane, for example, and the performance suffers. For moderate degree of allocation group (32 and 64 allocation group), the working files are distributed among the allocation groups and the internal parallelism helps to improve performance. But when the allocation group is too large, then parallelism contribution disappears due to the aggressive contention for shared resources, like shared data buses, connecting circuitry. When the file system block size matches the I/O request size, the performance is consistently better than the other combinations.

\subsection{Log-Structured File System}

Rosenblum, et al. \cite{c133} introduced a new technique for disk storage management called a log-structured file system. A log-structured file system writes all modifications to disk sequentially in a log-like structure, thereby speeding up both file writing and crash recovery. The log is the only structure on disk; it contains indexing information so that files can be read back from the log efficiently. In order to maintain large free areas on disk for fast writing, they divided the log into segments and use a segment cleaner to compress the live information from heavily fragmented segments. Log-structured file systems are based on the assumption that files are cached in main memory and that increasing memory sizes will make the caches more and more effective at satisfying read requests. As a result, disk traffic will become dominated by writes. A log-structured file system writes all new information to disk in a sequential structure called the log. This approach increases write performance dramatically by eliminating almost all seeks. The sequential nature of the log also permits much faster crash recovery: current Unix file systems typically must scan the entire disk to restore consistency after a crash, but a log-structured file system need only examine the most recent portion of the log. For a log-structured file system to operate efficiently, it must ensure that there are always large extents of free space available for writing new data. This is the most difficult challenge in the design of a log-structured file system. It was presented a solution based on large extents called segments, where a segment cleaner process continually regenerates empty segments by compressing the live data from heavily fragmented segments.

\subsection{Flash-oriented File Systems}

\textbf{JFFS (The Journalling Flash File System)} \cite{c134, c135} is a purely log-structured file system [LFS]. Nodes containing data and metadata are stored on the flash chips sequentially, progressing strictly linearly through the storage space available. In JFFS v1, there is only one type of node in the log; a structure known as struct jffs\_raw\_inode. Each such node is associated with a single inode. It starts with a common header containing the inode number of the inode to which it belongs and all the current file system metadata for that inode, and may also carry a variable amount of data. There is a total ordering between the all the nodes belonging to any individual inode, which is maintained by storing a version number in each node. Each node is written with a version higher than all previous nodes belonging to the same inode. In addition to the normal inode metadata such as uid, gid, mtime, atime, mtime etc., each JFFS v1 raw node also contains the name of the inode to which it belongs and the inode number of the parent inode. Each node may also contain an amount of data, and if data are present the node will also record the offset in the file at which these data should appear. The entire medium is scanned at mount time, each node being read and interpreted. The data stored in the raw nodes provide sufficient information to rebuild the entire directory hierarchy and a complete map for each inode of the physical location on the medium of each range of data. Metadata changes such as ownership or permissions changes are performed by simply writing a new node to the end of the log recording the appropriate new metadata. File writes are similar; differing only in that the node written will have data associated with it.

The oldest node in the log is known as the head, and new nodes are added to the tail of the log. In a clean filesystem which on which garbage collection has never been triggered, the head of the log will be at the very beginning of the flash. As the tail approaches the end of the flash, garbage collection will be triggered to make space. Garbage collection will happen either in the context of a kernel thread which attempts to make space before it is actually required, or in the context of a user process which finds insufficient free space on the medium to perform a requested write. In either case, garbage collection will only continue if there is dirty space which can be reclaimed. If there is not enough dirty space to ensure that garbage collection will improve the situation, the kernel thread will sleep, and writes will fail with −ENOSPC errors. The goal of the garbage collection code is to erase the first flash block in the log. At each pass, the node at the head of the log is examined. If the node is obsolete, it is skipped and the head moves on to the next node. If the node is still valid, it must be rendered obsolete. The garbage collection code does  so by writing out a new data or metadata node to the tail of the log.

While the original JFFS had only one type of node on the medium, JFFS2 is more flexible, allowing new types of node to be defined while retaining backward compatibility through use of a scheme inspired by the compatibility bitmasks of the ext2 file system. Every type of node starts with a common header containing the full node length, node type and a cyclic redundancy checksum (CRC). Aside from the differences in the individual nodes, the high-level layout of JFFS2 also changed from a single circular log format, because of the problem caused by strictly garbage collecting in order. In JFFS2, each erase block is treated individually, and nodes may not overlap erase block boundaries as they did in the original JFFS. This means that the garbage collection code can work with increased efficiency by collecting from one block at a time and making intelligent decisions about which block to garbage collect from next.

In traditional file systems the index is usually kept and maintained on the media, but unfortunately, this is not the case for JFFS2. In JFFS2, the index is maintained in RAM, not on the flash media. And this is the root of all the JFFS2 scalability problems. Of course, having the index in RAM JFFS2 achieves extremely high file system throughput, just because it does not need to update the index on flash after something has been changed in the file system. And this works very well for relatively small flashes, for which JFFS2 was originally designed. But as soon as one tries to use JFFS2 on large flashes (starting from about 128MB), many problems come up. JFFS2 needs to build the index in RAM when it mounts the file system. For this reason, it needs to scan the whole partition in order to locate all the nodes which are present there. So, the larger is JFFS2 partition, the more nodes it has, the longer it takes to mount it. The second, it is evidently that the index consumes some RAM. And the larger is the JFFS2 file system, the more nodes it has, the more memory is consumed.

\textbf{UBIFS (Unsorted Block Image File System)} \cite{c136, c137} follows a node-structured design, that enables their garbage collectors to read eraseblocks directly and determine what data needs to be moved and what can be discarded, and to update their indexes accordingly. The combination of data and metadata is called a node. Each node records which file (more specifically inode number) that the node belongs to and what data (for example file offset and data length) is contained in the node. The big difference between JFFS2 and UBIFS is that UBIFS stores the index on flash whereas JFFS2 stores the index only in main memory, rebuilding it when the file system is mounted. Potentially that places a limit on the maximum size of a JFFS2 file system, because the mount time and memory usage grow linearly with the size of the flash. UBIFS was designed specifically to overcome that limitation.

The master node stores the position of all on-flash structures that are not at fixed logical positions. The master node itself is written repeatedly to logical eraseblocks (LEBs) one and two. LEBs are an abstraction created by UBI. UBI maps physical eraseblocks (PEBs) to LEBs, so LEB one and two can be anywhere on the flash media (strictly speaking, the UBI device), however UBI always records where they are. Two eraseblocks are used in order to keep two copies of the master node. This is done for the purpose of recovery, because there are two situations that can cause a corrupt or missing master node. LEB zero stores the superblock node.

The superblock node contains file system parameters that change rarely if at all. For example, the flash geometry (eraseblock size, number of eraseblocks etc) is stored in the superblock node. The other UBIFS areas are: the log area (or simply the log), the LEB properties tree (LPT) area, the orphan area and the main area. The log is a part of UBIFS's journal.

The purpose of the UBIFS journal is to reduce the frequency of updates to the on-flash index. The index consists of the top part of the wandering tree that is made up of only index nodes, and that to update the file system a leaf node must be added or replaced in the wandering tree and all the ancestral index nodes updated accordingly. It would be very inefficient if the on-flash index were updated every time a leaf node was written, because many of the same index nodes would be written repeatedly, particularly towards the top of the tree. Instead, UBIFS defines a journal where leaf nodes are written but not immediately added to the on-flash index. Note that the index in memory (see TNC) is updated. Periodically, when the journal is considered reasonably full, it is committed. The commit process consists of writing the new version of the index and the corresponding master node.

After the log area, comes the LPT area. The size of the log area is defined when the file system is created and consequently so is the start of the LPT area. At present, the size of the LPT area is automatically calculated based on the LEB size and maximum LEB count specified when the file system is created. Like the log area, the LPT area must never run out of space. Unlike the log area, updates to the LPT area are not sequential in nature - they are random. In addition, the amount of LEB properties data is potentially quite large and access to it must be scalable. The solution is to store LEB properties in a wandering tree. In fact the LPT area is much like a miniature file system in its own right. It has its own LEB properties - that is, the LEB properties of the LEB properties area (called ltab). It has its own form of garbage collection. It has its own node structure that packs the nodes as tightly as possible into bit-fields. However, like the index, the LPT area is updated only during commit. Thus the on-flash index and the on-flash LPT represent what the file system looked like as at the last commit. The difference between that and the actual state of the file system, is represented by the nodes in the journal.

The next UBIFS area to describe is the orphan area. An orphan is an inode number whose inode node has been committed to the index with a link count of zero. That happens when an open file is deleted (unlinked) and then a commit is run. In the normal course of events the inode would be deleted when the file is closed. However in the case of an unclean unmount, orphans need to be accounted for. After an unclean unmount, the orphans' inodes must be deleted which means either scanning the entire index looking for them, or keeping a list on flash somewhere. UBIFS implements the latter approach.

The final UBIFS area is the main area. The main area contains the nodes that make up the file system data and the index. A main area LEB may be an index eraseblock or a non-index eraseblock. A non-index eraseblock may be a bud (part of the journal) or have been committed. A bud may be currently one of the journal heads. A LEB that contains committed nodes can still become a bud if it has free space. Thus a bud LEB has an offset from which journal nodes begin, although that offset is usually zero.

There are three important differences between UBIFS and JFFS2. The first has already been mentioned: UBIFS has an on-flash index, JFFS2 does not - thus UBIFS is potentially scalable. The second difference is implied: UBIFS runs on top of the UBI layer which runs on top of the MTD subsystem, whereas JFFS2 runs directly over MTD. UBIFS benefits from the wear-leveling and error handling of UBI at the cost of the flash space, memory and other resources taken by UBI. The third important difference is that UBIFS allows writeback.

\textbf{Yaffs (Yet Another Flash File System)} \cite{c138} contains objects. The object is anything that is stored in the file system. These are: (1) Regular data files, (2) Directories, (3) Hard-links, (4) Symbolic links, (5) Special objects (pipes, devices etc). All objects are identified by a unique integer object Id. In Yaffs, the unit of allocation is the chunk. Typically a chunk will be the same as a NAND page, but there is flexibility to use chunks which map to multiple pages.

Many, typically 32 to 128 but as many as a few hundred, chunks form a block. A block is the unit of erasure. NAND flash may be shipped with bad blocks and further blocks may go bad during the operation of the device. Thus, Yaffs is aware of bad blocks and needs to be able to detect and mark bad blocks. NAND flash also typically requires the use of some sort of error detection and correction code (ECC). Yaffs can either use existing ECC logic or provide its own.

Yaffs2 has a true log structure. A true log structured file system only ever writes sequentially. Instead of writing data in locations specific to the files, the file system data is written in the form of a sequential log. The entries in the log are all one chunk in size and can hold one of two types of chunk: (1) Data chunk - a chunk holding regular data file contents, (2) Object Header - a descriptor for an object (directory, regular data file, hard link, soft link, special descriptor,...). This holds details such as the identifier for the parent directory, object name, etc. Each chunk has tags associated with it. The tags comprise the following important fields: (1) ObjectId - identifies which object the chunk belongs to, (2) ChunkId - identifies where in the file this chunk belongs, (3) Deletion Marker - (Yaffs1 only) shows that this chunk is no longer in use, (4) Byte Count - number of bytes of data if this is a data chunk, (5) Serial Number - (Yaffs1 only) serial number used to differentiate chunks with the same objectId and chunkId.

When a block is made up only of deleted chunks, that block can be erased and reused. However, it needs to copy the valid data chunks off a block, deleting the originals and allowing the block to be erased and reused. This process is referred to as garbage collection. If garbage collection is aggressive, the whole block is collected in a single garbage collection cycle. If the collection is passive then the number of copies is reduced thus spreading the effort over many garbage collection cycles. This is done to reduce garbage collection load and improve responsiveness. The rationale behind the above heuristics is to delay garbage collection when possible to reduce the amount of collection that needs to be performed, thus increasing average system performance. Yet there is a conflicting goal of trying to spread the garbage collection so that it does not all happen at the same causing fluctuations in file system throughput. These conflicting goals make garbage tuning quite challenging.

Mount scanning takes quite a lot of time and slows mounting. Checkpointing is a mechanism to speed the mounting by taking a snapshot of the Yaffs runtime state at unmount or sync() and then reconstituting the runtime state on re-mounting. The actual checkpoint mechanism is quite simple. A “stream” of data is written to a set of blocks which are marked as holding checkpoint data and the important runtime state is written to the stream.

\textbf{NAFS (NAND flash memory Array File System)} \cite{c145} consists of a Conventional File System and the NAND Flash Memory Array Interface; the former provides the users with basic file operations while the latter allows concurrent accesses to multiple NAND flash memories through a striping technique in order to increase I/O performance. Also, parity bits are distributed across all flash memories in the array to provide fault tolerance like RAID5.

The NAND flash memory is partitioned into two areas: one for the superblock addresses and the other for the superblock itself, inodes, and data. In order to provide uniform wear-leveling, the superblock is stored at the random location in the Data/Superblock/Inode-block Partition while its address is stored in the Superblock Address Partition. NAFS attempts to write file data consecutively into each block of NAND flash memory for better read and write performance.

In addition, NAFS adopts a new double list cache scheme that takes into account the characteristics of both large-capacity storage and NAND flash memory in order to increase I/O performance. The double list cache makes it possible to defer write operations and increase the cache hit ratio by prefetching relevant pages through data striping of NAND Flash Memory Array Interface. The double list cache consists of the clean list for the actual caching and the dirty list for monitoring and analyzing page reference patterns. The dirty list maintains dirty pages in order to reduce their search times, and the clean list maintains clean pages. All the pages that are brought into memory by read operations are inserted into the clean list. If clean pages in the clean list are modified by write operations, they are removed from the clean list and inserted into the head of the dirty list. Also, if a new file is created, its new pages are inserted into the head of the dirty list. If a page fault occurs, clean pages are removed  from the tail of the clean list.

When NAFS performs delayed write operations using the cache, since two blocks are assigned to each NAND flash memory, it is always guaranteed that a file data can be written contiguously within at least one block. In addition, NAFS performs delayed write operations in the dirty list, resulting in reduction of the number of write operations and consecutive write operations of file data in each block of NAND flash memory.

\textbf{CFFS (Core Flash File System)} \cite{c148}, which is another file system based on YAFFS, stores index entries and metadata into index blocks which are distinct from data blocks. Since CFFS just reads in the index blocks during mount, its mount time is faster than YAFFS2's. Furthermore, since frequently modified metadata are collected and stored into index blocks, garbage collection performance of CFFS is better than YAFFS2's. However, since CFFS stores the physical addresses of index blocks into the first block of NAND flash memory in order to reduce mount time, wear-leveling performance of CFFS is worse than others due to frequent erasure of the first block.

\textbf{NAMU (NAnd flash Multimedia file system)} \cite{c147} takes into consideration the characteristics of both NAND flash memory and multimedia files. NAMU utilizes an index structure that is suitable for large-capacity files to shorten the mount time by scanning only index blocks located in the index area during mount. In addition, since NAMU manages data in the segment unit rather than in the page unit, NAMU's memory usage efficiency is better than JFFS2's, YAFFS2's.

\textbf{MNFS (novel mobile multimedia file system)} \cite{c149} introduces (1) hybrid mapping, (2) block-based file allocation, (3) an in-core only Block Allocation Table (iBAT), and (4) upward directory representation. Using these methods, MNFS achieves uniform write-responses, quick mounting, and a small memory footprint.

The hybrid mapping scheme means that MNFS uses a page mapping scheme (log-structured method) for the metadata by virtue of the frequent updates. On the other hand, a block mapping scheme is used for user data, because it is rarely updated in mobile multimedia devices. The entire flash memory space is logically divided into two variable-sized areas: the Metadata area and the User data area.

MNFS uses a log structure to manage the file system metadata. The metadata area is a collection of log blocks that contain file system metadata; the page mapping scheme is used for this area. The user data area is a collection of data blocks that contains multimedia file data; the block mapping scheme is used for this area. A multimedia file, e.g. a music or video clip, is an order of magnitude larger than the text-based file. Therefore, MNFS uses a larger allocation unit than the block size (usually 4 Kbytes) typically found in a legacy general purpose file system. MNFS defines the allocation unit of the file system as a block of NAND flash memory. The block size of NAND flash memory ranges from 16 Kbyte to 128 Kbyte, and this size is device specific.

MNFS uses the iBAT, which is similar to the File Allocation Table in the FAT file system, for both uniform write-responses and for robustness of the file system. There are two important differences between the FAT and the iBAT. First, the iBAT is not stored in the flash memory. Like the in-memory tree structure in YAFFS, the iBAT is dynamically constructed, during the mount time, in the main memory (RAM) through scanning the spare area of all the blocks. Secondly, the iBAT uses block-based allocation whereas the FAT uses cluster-based allocation. In the FAT file system, as the file size grows, a new cluster is allocated, requiring modification of the file allocation table in the storage device. Access to the storage device for the metadata update not only affects the response time of the write request, but it can also invoke file system inconsistency when the system crashes during the update. In MNFS, the iBAT is not stored separately in the flash memory, and the block allocation information is stored in the spare area of the block itself while the block is allocated to a file. These two differences make MNFS more robust than the FAT file system.

MNFS  uses upward directory representation method. In this method, each directory entry in the log block has its parent directory entry ID. That is, the child entry points to its parent entry. The directory structure of the file system can be represented using this parent directory entry ID. For the upward directory representation method, it is necessary to read all of the directory entries in order to  construct the directory structure of the file system in the memory.

\subsection{Flash-friendly File Systems}

\textbf{NILFS (New Implementation of a Log-structured File System)} \cite{c139, c140} has the on-disk layout is divided into several parts: (1) superblock, (2) full segment, (3) partial segment, (4) logical segment, (5) segment management block.

Superblock has the parameters of the file system, the disk block address of the latest segment being written, etc. Each full segment consists of a fixed length of disk blocks. This is a basic management unit of the garbage collector. Partial segment is write units. Dirty buffers are written out as partial segments. The partial segment does not exceed the full segment boundaries. The partial segment sequence includes inseparable directory operations. For example, a logical segment could consist of two partial segments. In the recovery operations, the two partial segments are treated as one inseparable segment. There are two flag bits, Logical Begin and Logical End, at the segment summary of the partial segment.

The NILFS adopts the B-tree structure for both file block mapping and inode block mapping. The two mappings are implemented in the common B-tree operation routine. The B-tree intermediate node is used to construct the B-tree. It has 64-bit-wide key and 64-bit-wide pointer pairs. The file block B-tree uses a file block address as its key, whereas the inode block B-tree uses an inode number as its key. The root block number of the file block B-tree is stored to the corresponding inode block. The root block number of the inode block B-tree is stored to the superblock of the file system. So, there is only one inode block B-tree in the file system. File blocks, B-tree blocks for file block management, inode blocks, and B-tree blocks for inode management are written to the disk as logs. A newly created file first exists only in the memory page cache. Because the file must be accessible before being written to the disk, the B-tree structure exists even in memory. The B-tree intermediate node in memory is on the memory page cache, the data structures are the same as those of the disk blocks. The pointer of the B-tree node stored in memory holds the disk block number or the memory address of the page cache that reads the block. When looking up a block in the B-tree, if the pointer of the B-tree node is a disk block number, the disk block is read into a newly allocated page cache before the pointer is rewritten. The original disk block number remains in the buffer-head structure on the page cache.

The partial segment consists of three parts: (1) The segment summary keeps the block usage information of the partial segment. The main contents are checksums of the data area, the segment summary, the length of the partial segment, and partial segment creation time. (2) Data area contains file data blocks, file data B-tree node blocks, inode blocks, and inode block B-tree node blocks in order. (3) A checkpoint is placed on the last tail of the partial segment. The checkpoint includes a checksum of the checkpoint itself. The checkpoint accuracy means successfully writing the partial segment to the disk. The most important information in the checkpoint is the root block number of the inode block B-tree. The block number is written out last, and the whole file system state is updated.

The data write process started by the sync system call and NILFS kernel thread, advances in the following order: (1) Lock the directory operations, (2) The dirty pages of the file data are gathered from its radix-tree, (3) The dirty B-tree intermediate node pages of both file block management and inode management are gathered, (4) The dirty inode block pages are gathered, (5) The B-tree intermediate node pages which will be dirty for registered block address being renew are gathered, (6) New disk block addresses are assigned to those blocks in order of file data blocks, B-tree node blocks for file data, inode blocks, B-tree node blocks for inodes, (7) Rewrite the disk block addresses to new ones in the radix-tree and B-tree nodes, (8) Call block device input/output routine to writing out the blocks, (9) Unlock the directory operations. The NILFS snapshot is a whole consistent file system at some time instant.

In LFS, all blocks remain as is (until they are collected by garbage collection), therefore, no new information is needed to make a snapshot. In NILFS, the B-tree structure manages the file and inode blocks, and B-tree nodes are written out as a log too. So, the root block number of the inode management B-tree is the snapshot of the NILFS file system. The root block number is stored in the checkpoint position of a partial segment. The NILFS checkpoint is the snapshot of the file system itself. Actually, user can specify the disk block address of the NILFS checkpoint to Linux using the "mount" command, and the captured file system is mounted as a read-only file system. However, when the user use all checkpoints as the snapshot, there is no disk space for garbage collection. The user can select any checkpoint as a snapshot, and the garbage collector collects other checkpoint blocks.

\textbf{F2FS (Flash-Friendly File System)} \cite{c142} employs three configurable units: segment, section and zone. It allocates storage blocks in the unit of segments from a number of individual zones. It performs "cleaning" in the unit of section. These units are introduced to align with the underlying FTL's operational units to avoid unnecessary (yet costly) data copying.

F2FS introduced a cost-effective index structure in the form of  node address table with the goal to attack  the "wandering tree" problem. Also multi-head logging was suggested. F2FS uses an effective hot/cold data separation scheme applied during logging time (i.e., block allocation time). It runs multiple active log segments concurrently and appends data and metadata to separate log segments based on their anticipated update frequency. Since the flash storage devices exploit media parallelism, multiple active segments can run simultaneously without frequent management operations. F2FS builds basically on append-only logging to turn random writes into sequential ones. At high storage utilization, however, it changes the logging strategy to threaded logging to avoid long write latency. In essence, threaded logging writes new data to free space in a dirty segment without cleaning it in the foreground. F2FS optimizes small synchronous writes to reduce the latency of fsync requests, by minimizing required metadata writes and recovering synchronized data with an efficient roll-forward mechanism.

F2FS divides the whole volume into fixed-size segments. The segment is a basic unit of management in F2FS and is used to determine the initial file system metadata layout. A section is comprised of consecutive segments, and a zone consists of a series of sections. F2FS splits the entire volume into six areas: (1) Superblock (SB), (2) Checkpoint (CP), (3) Segment Information Table (SIT), (4) Node Address Table (NAT), (5) Segment Summary Area (SSA), (6) Main Area.

Superblock (SB) has the basic partition information and default parameters of F2FS, which are given at the format time and not changeable. Checkpoint (CP) keeps the file system status, bitmaps for valid NAT/SIT sets, orphan inode lists and summary entries of currently active segments. Segment Information Table (SIT) contains per-segment information such as the number of valid blocks and the bitmap for the validity of all blocks in the "Main" area. The SIT information is retrieved to select victim segments and identify valid blocks in them during the cleaning process. Node Address Table (NAT) is a block address table to locate all the "node blocks" stored in the Main area. Segment Summary Area (SSA) stores summary entries representing the owner information of all blocks in the Main area, such as parent inode number and its node/data offsets. The SSA entries identify parent node blocks before migrating valid blocks during cleaning. Main Area is filled with 4KB blocks. Each block is allocated and typed to be node or data. A node block contains inode or indices of data blocks, while a data block contains either directory or user file data. Note that a section does not store data and node blocks simultaneously.

F2FS utilizes the "node" structure that extends the inode map to locate more indexing blocks. Each node block has a unique identification number, "node ID". By using node ID as an index, NAT serves the physical locations of all node blocks. A node block represents one of three types: inode, direct and indirect node. An inode block contains a file's metadata, such as file name, inode number, file size, atime and dtime. A direct node block contains block addresses of data and an indirect node block has node IDs locating another node blocks. In F2FS, a 4KB directory entry ("dentry") block is composed of a bitmap and two arrays of slots and names in pairs. The bitmap tells whether each slot is valid or not. A slot carries a hash value, inode number, length of a file name and file type (e.g., normal file, directory and symbolic link). A directory file constructs multi-level hash tables to manage a large number of dentries efficiently.

F2FS maintains six major log areas to maximize the effect of hot and cold data separation. F2FS statically defines three levels of temperature—hot, warm and cold—for node and data blocks. Direct node blocks are considered hotter than indirect node blocks since they are updated much more frequently. Indirect node blocks contain node IDs and are written only when a dedicated node block is added or removed. Direct node blocks and data blocks for directories are considered hot, since they have obviously different write patterns compared to blocks for regular files. Data blocks satisfying one of the following three conditions are considered cold: (1) Data blocks moved by cleaning, (2) Data blocks labeled "cold" by the user, (3) Multimedia file data.

F2FS performs cleaning in two distinct manners, foreground and background. Foreground cleaning is triggered only when there are not enough free sections, while a kernel thread wakes up periodically to conduct cleaning in background. A cleaning process takes three steps: (1) Victim selection, (2)  Valid block identification and migration, (3) Post-cleaning process.

The cleaning process starts first to identify a victim section among non-empty sections. There are two well-known policies for victim selection during LFS cleaning—greedy and cost-benefit. The greedy policy selects a section with the smallest number of valid blocks. Intuitively, this policy controls overheads of migrating valid blocks. F2FS adopts the greedy policy for its foreground cleaning to minimize the latency visible to applications. Moreover, F2FS reserves a small unused capacity (5\% of the storage space by default) so that the cleaning process has room for adequate operation at high storage utilization levels. On the other hand, the cost-benefit policy is practiced in the background cleaning process of F2FS. This policy selects a victim section not only based on its utilization but also its "age". F2FS infers the age of a section by averaging the age of segments in the section, which, in turn, can be obtained from their last modification time recorded in SIT. With the cost-benefit policy, F2FS gets another chance to separate hot and cold data.

After selecting a victim section, F2FS must identify valid blocks in the section quickly. To this end, F2FS maintains a validity bitmap per segment in SIT. Once having identified all valid blocks by scanning the bitmaps, F2FS retrieves parent node blocks containing their indices from the SSA information. If the blocks are valid, F2FS migrates them to other free logs. For background cleaning, F2FS does not issue actual I/Os to migrate valid blocks. Instead, F2FS loads the blocks into page cache and marks them as dirty. Then, F2FS just leaves them in the page cache for the kernel worker thread to flush them to the storage later. This lazy migration not only alleviates the performance impact on foreground I/O activities, but also allows small writes to be combined. Background cleaning does not kick in when normal I/O or foreground cleaning is in progress.

After all valid blocks are migrated, a victim section is registered as a candidate to become a new free section (called a "pre-free" section in F2FS). After a checkpoint is made, the section finally becomes a free section, to be reallocated. We do this because if a pre-free section is reused before checkpointing, the file system may lose the data referenced by a previous checkpoint when unexpected power outage occurs.

\textbf{SFS (SSD File System)} \cite{c161} is based on three design principles. A file system should exploit the file block semantics directly. It needs to take a log-structured approach based on the observation that the random write bandwidth is much slower than the sequential one. The existing lazy data grouping in LFS during segment cleaning fails to fully utilize the skewness in write patterns and argue that an eager data grouping is necessary to achieve sharper bimodality in segment utilization. SFS takes a log-structured approach that turns random writes at the file level into sequential writes at the LBA level. Moreover, in order to utilize nearly 100\% of the raw SSD bandwidth, the segment size is set to a multiple of the clustered block size. The result is that the performance of SFS will be limited by the maximum sequential write performance regardless of random write performance.

It shows that, if hot data and cold data are grouped into separate segments, the segment utilization distribution becomes bimodal: most of the segments are almost either full or empty of live blocks. Therefore, because the segment cleaner can almost always work with nearly empty segments, the cleaning overhead will be drastically reduced. To form a bimodal distribution, LFS uses a cost-benefit policy for segment cleaning that prefers cold segments to hot segments. However, previous studies show that even the cost-benefit policy performs poorly under the large segment size (e.g., 8 MB), because the increased segment size makes it harder to find nearly empty segments. With SSD, the cost-benefit policy encounters a dilemma: small segment size enables LFS to form a bimodal distribution, but small random writes caused by the small segment severely degrades write performance of SSD. Instead of separating the data lazily on segment cleaning after writing them regardless of their hotness, SFS classifies data proactively on writing using file block level statistics, as well as on segment cleaning. In such eager data grouping, since segments are already composed of homogeneous data with similar update likelihood, the segment cleaning overhead will be significantly reduced. In particular, the I/O skewness commonly found in many real workloads will make this more attractive.

SFS has four core operations: segment writing, segment cleaning, reading, and crash recovery. The first step of segment writing in SFS is to determine the hotness criteria for block grouping. This is, in turn, determined by segment quantization that quantizes a range of hotness values into a single hotness value for a group. It is assumed that there are four segment groups: hot, warm, cold, and read-only groups. The second step is to calculate the block hotness for each dirty block and assign them to the nearest quantized group by comparing the block hotness and the group hotness. At this point, those blocks with similar hotness levels should belong to the same group. The third step is to fill a segment with blocks belonging to the same group. If the number of blocks in a group is not enough to completely fill a segment, the segment writing of the group is deferred until the group grows to completely fill a segment. This eager grouping of file blocks according to the hotness serves to colocate blocks with similar update likelihoods in the same segment.

Segment cleaning in SFS consists of three steps: select victim segments, read the live blocks from the victim segments into the page cache and mark the live blocks as dirty, and trigger the writing process. The writing process treats the live blocks from victim segments the same as normal blocks; each live block is classified into a specific quantized group according to its hotness. After all the live blocks are read into the page cache, the victim segments are then marked as free so that they can be reused for writing. For better victim segment selection, cost-hotness policy is introduced, which takes into account both the number of live blocks in the segment (i.e., cost) and the segment hotness.

\subsection{GC Overhead}

Wu, et al. \cite{c231} proposed the greedy algorithm (GR) for garbage collection The greedy algorithm selects the block with the fewest valid pages as the victim block for garbage collection. This approach can reduce the overhead required for copying valid pages within the victim block to free space during garbage collection. However, the GR algorithm does not take into account wear leveling in flash-based consumer electronic devices. It has been shown that the GR algorithm performs well in terms of wear leveling for random memory accesses but does not perform well for memory accesses with a high spatial locality of reference.

Kawaguchi, et al. \cite{c232} proposed the cost-benefit (CB) algorithm for flash memory. CB calculates a cost-benefit value for each block and selects the block with the highest value as a victim. The cost-benefit value for a block is calculated as (age * (1−u))/2u, where age is the elapsed time since the last modification of a page within the block and u is the percentage of valid pages within the block. Because the CB algorithm takes into account both the age of invalid pages and the percentage of valid pages in a block, it could provide improved wear leveling in flash-based consumer electronic devices. However, because the CB algorithm does not take into account the erase count for each block, its wear leveling performance is not sufficient.

Chiang, et al. \cite{c233} proposed the cost-age-time (CAT) algorithm, which extends the CB algorithm by considering the erase count for each block when selecting a victim block. The CAT algorithm attempts to maintain a balance between reducing the garbage collection overhead and improving wear leveling in flash memory.

Syu, et al. \cite{c186} developed a mechanism that takes advantage of FTL inactive time between requests and actively launching tasks to reclaim invalidated space. The underlying concept is that the timing impact could be reduced through distributing the one-time cost of space recycling.

The active space recycling mechanism is developed based on a task model composed of four tasks, namely the manager task, the collector task, the eraser task, and the read/write handler task. Jobs of these four tasks are released in a fixed time interval. The jobs of the first three tasks follow a precedence constraint that the manager task precedes the collector task while the collector task precedes the eraser task. The read/write handler task is assigned the lowest priority among the four tasks.

At the start of each time interval, a job of the manager task is released. It is responsible for determining the amount of invalidated space to be reclaimed. It gathers necessary statistic information, and computes the number of pages holding invalidated data to be reclaimed. The mission of the collector task is to collect dirty blocks holding invalid pages and maintains a garbage queue holding these blocks. The collector task first selects "right" dirty blocks to form a candidate list. It then moves, from the candidate list, the block with the maximum invalidated space into the garbage queue. The valid data, if any, in the dirty block is copied to other free space before it is put into the garbage queue. The action carried out by the eraser task is quite straightforward. It removes and erases the dirty blocks in the garbage queue maintained by the collector task. When a block is erased, it is marked as a free block and added into the free block list. Its associated erase count is also updated.

The read/write handler task is responsible for carrying out the requests of read or write operations. It is active only when the other three jobs have finished and there is pending or unfinished read/write request. When executing, the read/write handler task calculates how much time can be used for handling the read/write request before the end of current time period. If the time left is not enough to complete reading or writing a page, the remaining read/write operations of this request will be postponed.

Yan, et al. \cite{c183} proposed an efficient file-aware garbage collection algorithm, called FaGC. The FaGC algorithm copies valid pages in a victim block to clusters in free blocks according to the calculated update frequency of the associated chunk. The FaGC algorithm adopts a hybrid wear-leveling policy to improve the lifespan of NAND flash memory. To avoid unnecessary garbage collection, a scattering factor is defined and calculated to determine when to trigger the garbage collection policy.

A file consists of a series of chunks mapped to physical pages in the flash memory. Each file is assigned a unique number, called File ID, and each chunk in a file is assigned a unique number, called Chunk ID. In general, different files have different update frequencies, and different chunks in the same file have different update frequencies. In a file-aware system structure, an update frequency table (UFT) is built into random access memory (RAM) to record the update frequency for each chunk in a file. Each UFT entry contains four values: File ID, Chunk ID, Time, and Freq. Time records the most recent time that a chunk in a file has been updated, and Freq records the frequency with which a chunk has been updated.

Simultaneously, the physical-to-logical translation table (PLT) maintains the File ID and Chunk ID for each block and physical page in the flash memory. When a chunk in a file is modified or updated, it is rewritten to another physical page in flash memory according to the out-of-place update scheme. At that time, File ID and Chunk ID in the PLT are updated. Additionally, when a chunk in a file is modified or updated, the current time is recorded in the UFT, and Freq is calculated and recorded as follows.

In general, the block with the fewest valid pages is selected as the victim block to minimize the overhead for the copy operation, as in the GR algorithm. After the victim block is selected, the valid pages in the victim block are copied to free space, and the victim block is then erased and reclaimed. Before the valid pages are copied, the update frequency of the chunk associated with each valid page in the victim block will be checked in the PLT and UFT. Therefore, wear leveling is improved using this clustering procedure based on the update frequency. Additionally, the decision of when to trigger garbage collection affects the performance of NAND flash-based consumer electronic devices.

\subsection{Write Amplification Management}

Kuo, et al. \cite{c167} suggested an efficient on-line hot-data identification and a multi-hash-function framework with the goal to manage the write amplification issue. The proposed framework adopts K independent hash functions to hash a given LBA into multiple entries of a M-entry hash table to track the write number of the LBA, where each entry is associated with a counter of C bits. Whenever a write is issued to the FTL, the corresponding LBA is hashed simultaneously by K given hash functions. Each counter corresponding to the K hashed values (in the hash table) is incremented by one to reflect the fact that the LBA is written again. Whenever an LBA needs to be verified to see if it is associated with hot data, the LBA is hashed simultaneously and in the same way by the K hash functions. The data addressed by the given LBA is considered as hot data if the H most significant bits of every counter of the K hashed values contain a non-zero bit value.

Jagmohan, et al. \cite{c165} proposed a NAND Flash system which uses multi-write coding to reduce write amplification. Multi-write coding allows a NAND Flash page to be written more than once without requiring an intervening block erase. They presented a novel two-write coding technique based on enumerative coding, which achieves linear coding rates with low computational complexity. The proposed technique also seeks to minimize memory wear by reducing the number of programmed cells per page write.

\subsection{SSD Lifetime Management}

Chen, et al. \cite{c199} implemented CAFTL (A Content-Aware Flash Translation Layer). CAFTL eliminates duplicate writes and redundant data  through a combination of both in-line and out-of-line deduplication. Inline deduplication refers to the case where CAFTL proactively examines the incoming data and cancels duplicate writes before committing a write request to flash. As a 'best-effort' solution, CAFTL does not guarantee that all duplicate writes can be examined and removed immediately. Thus CAFTL also periodically scans the flash memory and coalesces redundant data out of line. When a write request is received at the SSD, (1) the incoming data is first temporarily maintained in the on-device buffer; (2) each updated page in the buffer is later computed a hash value, also called fingerprint, by a hash engine, which can be a dedicated processor or simply a part of the controller logic; (3) each fingerprint is looked up against a fingerprint store, which maintains the fingerprints of data already stored in the flash memory; (4) if a match is found, which means that a residing data unit holds the same content, the mapping tables, which translate the host-viewable logical addresses to the physical flash addresses, are updated by mapping it to the physical location of the residing data, and correspondingly the write to flash is canceled; (5) if no match is found, the write is performed to the flash memory as a regular write.

Wang, et al. \cite{c225} proposed to develop a real-time, per-process per-stream based pattern detection scheme that identifies various write patterns. These patterns are then used to guide the write buffer to improve the write performance of SSDs that employ a log-structured block-based FTL. They classify fine-grained write patterns into the following three categories: (a) sequential, (b) clustered (page or block), and (c) random. Each of the above patterns is defined as follows: 1) A sequential pattern is defined as a series of requests with consecutive logical addresses in an ascending order; 2) A page clustered pattern is defined as a process repeatedly updates a specific page; 3) A block clustered pattern is defined as a process repeatedly updates a specific block; 4) A pattern which falls in none of above is classified as random. In order to filter out transition "noise", each pattern is allocated with a bit map. The number of bits n will determine how many times in a row a pattern has to be detected, before the algorithm decides that the I/O stream has entered into a new pattern. For each pattern, they devised an adaptive dirty flush policy. Each dirty page in the buffer cache is associated with a pattern type. The dirty pages with the same pattern will be linked together in a linked list, so that dirty pages in the buffer cache are virtually partitioned according to their associated patterns. When a page is written by a process, it will be moved to the head of the pattern list. As a result, in each list, the head will be the most recently written page while the tail will be the least recently written one. When the system needs to flush dirty pages, it will scan the lists according to the following priorities. The sequential pattern will be given the highest priority to be flushed since its pages are most likely written only once, therefore there is no point to keep them in the cache. The random pattern will be given the next priority. The page clustered and block clustered dirty pages will have the lowest priority, since they may be overwritten in the future hence we want to keep them in the cache. The suggested schemes reduce SSD erase cycles which is directly translated to a major improvement on the life-span of SSDs.

Huang, et al. \cite{c154} presented a Content and Semantics Aware File System (CSA-FS) which is able to reduce write traffic to SSDs. It employs deduplication and delta-encoding techniques to file system data blocks and semantic blocks, respectively. It is motivated by two important observations: (1) there exists a huge amount of content redundancy within primary storage systems, and (2) semantic blocks are visited much more frequently than data blocks, with each update bringing very minimal changes. By separately deduplicating redundant data blocks and delta-encoding similar semantic blocks, CSA-FS can significantly reduce the total write traffic to SSDs and greatly improve their lifetime correspondingly. CSA-FS applies deduplication to data blocks and delta-encoding to semantic blocks, respectively. Semantic blocks are extracted from the file system and exported for lookups. Semantic blocks mainly include super-blocks, group descriptors, data block bitmap, inode bitmap and inode tables. For every block write request, CSA-FS checks whether it accesses semantic block or data block by consulting the exported semantic blocks. For data block write, it computes its MD5 digest and looks up the hash value in a hash table to determine whether it is a duplicate block write. If it is a duplicate write, CSA-FS simply returns the block number in the found hash entry, and then uses that block number to update the block pointer table of the file's inode. If it is a new write request, it first goes through the normal procedure, i.e., allocating a free block, updating the corresponding bitmap block and performing necessary accounting statistics, and finally inserts a new entry containing the block number, its MD5 value and some housekeeping information to the hash table. For metadata  block write, CSA-FS calculates the content delta relative to its original content, and then appends the delta to a delta-logging region.

\subsection{Deduplication}

Fu, et al. \cite{c197} made research of cloud backup services in the personal computing environment. They concluded that the majority of storage space is occupied by a small number of compressed files with low sub-file redundancy. About 61\% of all files are smaller than 10KB, accounting for only 1.2\% of the total storage capacity, and only 1.4\% files are larger than 1MB but occupy 75\% of the storage capacity. This suggests that tiny files can be ignored during the deduplication process as so to improve the deduplication efficiency, since it is the large files in the tiny minority that dominate the deduplication efficiency. Static chunking (SC) method can outperform content defined chunking (CDC) in deduplication effectiveness for static application data and virtual machine images. The computational overhead for deduplication is dominated by data capacity. The amount of data shared among different types of applications is negligible. They suggested AA-Dedupe (An Application-Aware Source Deduplication Approach) where tiny files are first filtered out by file size filter for efficiency reasons, and backup data streams are broken into chunks by an intelligent chunker using an application-aware chunking strategy. Data chunks from the same type of files are then deduplicated in the application-aware deduplicator by looking up their hash values in an application-aware index that is stored in the local disk. If a match is found, the metadata for the file containing that chunk is updated to point to the location of the existing chunk. If there is no match, the new chunk is stored based on the container management in the cloud, the metadata for the associated file is updated to point to it and a new entry is added into the application-aware index to index the new chunk.

Meister, et al. \cite{c201} showed that most files are very small, but the minority of very large files occupies most of the storage capacity: 90\% of the files occupy less than 10\% of the storage space, in some cases even less than 1\%. In most data sets, between 15\% and 30\% of the data is stored redundantly and can be removed by deduplication techniques. Often small files have high deduplication rates, but they contribute little to the overall savings. Middle-sized files usually have a high deduplication ratio. Full file duplication typically reduces the data capacity by 5\% - 10\%. In most data sets, most of the deduplication potential is lost if only full file elimination is used. The deduplication slowly decreases slowly with increasing chunk sizes. Fixed size chunking detects around 6-8\% less redundancies than content-defined chunking. Between 3.1\% and 9.4\% of the data are zeros. Of all chunks, 90\% were only referenced once. This means that the chunks are unique and do not contribute to deduplication. The most referenced chunk is the zero chunk. The mean number of references is 1.2, the median is 1 reference. The most referenced 5\% of all chunks account for 35\% and the first 24\% account for 50\% of all references. Of all multi-referenced chunks, about 72\% were only referenced twice. A small fraction of the chunks causes most of the deduplication. The evaluation shows that typically 20\% to 30\% of online data can be removed by applying data deduplication techniques, peaking up to 70\% for some data sets. This reduction can only be achieved by a subfile deduplication approach, while approaches based on whole-file comparisons only lead to small capacity savings.

Xia, et al. \cite{c200} suggested DARE (a Deduplication-Aware Resemblance detection and Elimination scheme) for compressing backup datasets. DARE is designed to improve resemblance detection for additional data reduction in deduplication-based backup/archiving storage systems. For an incoming backup stream, DARE goes through the following four key steps: (1) Duplicate Detection - the data stream is first chunked by the CDC approach, fingerprinted by SHA-1, duplicate-detected, and then grouped into container of sequential chunks to preserve the backup-stream locality. (2) Resemblance Detection - the DupAdj resemblance detection module in DARE first detects duplicate-adjacent chunks in the containers formed in Step 1. After that, DARE’s improved super-feature module further detects similar chunks in the remaining non-duplicate and non-similar chunks that may have been missed by the DupAdj detection module when the duplicate-adjacency information is lacking or weak. (3) Delta Compression - for each of the resembling chunks detected in Step 2, DARE reads its base-chunk, then delta encodes their differences by the Xdelta algorithm. In order to reduce the number of disk reads, an LRU and locality-preserved cache is implemented to prefetch the base-chunks in the form of locality-preserved containers. (4) Storage Management - the data NOT reduced, i.e., non-similar or delta chunks, will be stored as containers into the disk. The file mapping information among the duplicate chunks, resembling chunks, and non-similar chunks will also be recorded as the file recipes to facilitate future data restore operations in DARE. They concluded that supplementing delta compression to deduplication can effectively enlarge the logical space of the restoration cache, but the data fragmentation in data reduction systems remains a serious problem.

Kim, et al. \cite{c209} designed a deduplication layer on FTL. It consists of three components, namely, fingerprint generator, fingerprint manager, and mapping manager. The fingerprint generator creates a hash value, called fingerprint, which summarizes the content of written data. The fingerprint manager manipulates generated fingerprints and conducts fingerprint lookups for detecting deduplication. Finally, the mapping manager deals with the physical locations of duplicate data. They proposed two acceleration techniques: sampling-based filtering and recency-based fingerprint management. The former selectively applies deduplication based upon sampling and the latter effectively exploits limited controller memory while maximizing the deduplication ratio. Experimental results have shown that they achieve the duplication rate ranging from 4\% to 51\%, with an average of 17\%, for the nine considered workloads. The response time of a write request can be improved by up to 48\% with an average of 15\%, while the lifespan of SSDs is expected to increase up to 4.1 times with an average of 2.4 times.

Ha, et al. \cite{c202} proposed a new deduplication scheme called block-level content-aware chunking to extend the lifetime of SSDs. The proposed scheme divides the data within a fixed-size block into a set of variable-sized chunks based on its contents and avoids storing duplicate copies of the same chunk. Evaluations on a real SSD platform showed that the proposed scheme improves the average deduplication rate by 77\% compared to the previous block-level fixed-size chunking scheme. Additional optimizations reduce the average memory consumption by 39\% with a 1.4\% gain in the  average deduplication rate.

Li \cite{c194} presented Flash Saver, which is coupled with the ext2/3 file system and aims to significantly reduce the write traffic to SSDs. Flash Saver deploys deduplication and delta-encoding to reduce the write traffic. Specifically, Flash Saver applies deduplication to file system data blocks and delta-encoding to file system meta-data blocks, based on two important observations which are: (1) there exist large amounts of duplicate data blocks (2) metadata blocks are accessed/modified much more frequently than data blocks, but with very minimal changes for each update. Specifically, Flash Saver semantically identifies the fixed-size (e.g. 4KB) content blocks of the file system into data blocks and meta blocks. For data blocks, it computes the SHA-1 hash value of the block and uses the hash value to examine whether the same block has already been stored in order to avoid storing multiple copies of the block having the same content. For metablocks, it logs the incremental changes relative to the corresponding meta block to save I/Os and storage space. Obviously, under most cases, file system meta blocks are frequently modified with minor changes. The experimental results have shown that Flash Saver can save up to 63\% of the total write traffic, which implies reasonably prolonged lifetime, larger effective flash space and higher reliability than that of the original counterpart within their allowable lifespan.

Rozier, et al. \cite{c211} have modeled the fault tolerance consequences of deduplication. They concluded that deduplication has a net negative impact on reliability, both due to its impact on unrecoverable data loss, and the impact of silent data corruptions, though the former is easily countered by using higher level RAID configurations. In both cases, system reliability can be increased by maintaining additional copies of deduplicated instances typically by keeping multiple copies for a very small percentage of the deduplicated instances in a given category.

Nam, et al. \cite{c206} introduced two reliability parameters for deduplication storage: chunk reliability and chunk loss severity. To provide a demanded reliability for an incoming data stream, most deduplication storage systems first carry out deduplication process by eliminating duplicates from the data stream and then apply erasure coding for the remaining (unique) chunks. A unique chunk may be shared (i.e., duplicated) at many places of the data stream and shared by other data streams. That is why deduplication can reduce the required storage capacity. However, this occasionally becomes problematic to assure certain reliability levels required from different data streams. The chunk reliability means each chunk's tolerance level in the face of any failures. The chunk loss severity represents an expected damage level in the event of a chunk loss, formally defined as the multiplication of actual damage by the probability of a chunk loss. They proposed a reliability-aware deduplication solution that not only assures all demanded chunk reliability levels by making already existing chunks sharable only if its reliability is high enough, but also mitigates the chunk loss severity by adaptively reducing the probability of having a chunk loss.

\subsection{Compression}

A distinction is made between lossless and lossy algorithms, i.e. those algorithms that will preserve the original data exactly, and those that will discard parts of the data, reducing the quality. The latter type is typically domain specific, i.e. knowledge about what type of data is being compressed is needed to determine what to discard. For general compression three of the most often used algorithms are: (1) Run length coding — a simple and fast scheme that replaces repeating patterns with the patterns and number of repetitions; (2) Huffman \cite{c234} coding — Huffman coding analyzes the frequency of different fixed length symbols in a data set, and to the symbols assigns codes whose lengths correspond to the frequency of the respective symbol in the data set, i.e. frequent symbols get short codes, infrequent get long codes; (3) Lempel-Ziv \cite{c235, c236} — these algorithms basically replace strings (variable length symbols) found in a dictionary with codes representing those strings.

The efficiency of these algorithms is determined by the size of the dictionary and how much effort is spent searching in the dictionaries. Gzip \cite{c237} and others use general compression algorithms based on the two Lempel-Ziv algorithms LZ77 \cite{c235} and LZ78 \cite{c236} (or Lempel-Ziv-Welch \cite{c238}, which is based on LZ78). These algorithms are sometimes augmented with Huffman \cite{c234} coding. Huffman coding on its own is typically faster than Lempel-Ziv based algorithms, although it will typically yield less compression. Run length encoding is extremely fast, but the gain is often small compared to Lempel-Ziv or Huffman coding.

Mannan, et al. \cite{c239} introduced the concept of block Huffman coding. Their main idea is to break the input stream into blocks and compress each block separately. They choose block size in such a way that it can store one full single block in main memory. They use a block size as moderate as 5 KiB, 10 KiB or 12 KiB. Finally, they observed that to obtain better efficiency from block Huffman coding, a moderate sized block is better and the block size does not depend on file types.

Chang, et al. \cite{c219} have made the performance evaluation of block LZSS compression algorithm. They studied the block LZSS algorithm and investigated the relationship between the compression ratio of block LZSS and the value of index or length. They found that as the block size increases, the compression ratio becomes better. To obtain better efficiency from block LZSS, a moderate sized block which is greater than 32KiB, may be optimal, and the optimal block size does not depend on file types. They have found that, in some cases, the block Huffman coding has a better compression ratio than no blocking Huffman coding, and with the increasing block size, the compression ratio deteriorates. The optimal block size in which it obtains the best compression ratio is about 16KiB. The reason for the better efficiency may be attributed to the principle of locality of data.

\subsection{Delta-encoding}

Douglis, et al. \cite{c208} have found that the benefits of application-specific deltas vary depending on the mix of content types. For example, HTML and email messages display a great deal of redundancy across large datasets, resulting in deltas that are significantly smaller than simply compressing the data, while mail attachments are often dominated by non-textual data that do not lend themselves to the technique. A few large files can contribute much of the total savings if they are particularly amenable to delta-encoding. Application-specific techniques, such as delta-encoding an unzipped version of a zip or gzip file and then zipping the result, can significantly improve results for a particular file, but unless an entire dataset consists of such files, overall results improve by just a couple of percent.

For web content, it have been found substantial overlap among pages on a single site. For the five web datasets they considered, deltas reduced the total size of the dataset to 8-19\% of the original data, compared to 29-36\% using compression. For files and email, there was much more variability, and the overall benefits are not as dramatic, but they are significant: two of the largest datasets reduced the overall storage needs by 10-20\% beyond compression. There was significant skew in at least one dataset, with a small fraction of files accounting for a large portion of the savings. Factors such as shingle size and the number of features compared do not dramatically affect these results. Given a particular number of maximal matching features, there is not a wide variation across base files in the size of the resulting deltas. A new file will often be created by making a small number of changes to an older file; the new file may even have the same name as the old file. In these cases, the new file can often be delta-encoded from the old file with minimal overhead.

\subsection{Refactored Design of I/O Architecture for Flash Storage}

\textbf{REDO (REfactored Design of I/O architecture)} \cite{c158} refactors two main components of the I/O subsystem—the file system and the storage device. REDO removes logical-to-physical mapping and garbage collection from the storage device. Instead, a refactored file system (RFS) directly manages the storage address space, including the garbage collection. Unlike host-based FTL, all those functions are conducted by RFS without any helps from an intermediate host layer like a device driver. This eliminates the need for maintaining a large logical-to-physical page-map table, allowing us to perform garbage collection more efficiently at the file system level. A refactored storage device controller (RSD) becomes simpler because it runs a small number of essential flash management functions. RSD maintains a much smaller logical-to-physical segment-map table to manage wear-leveling and bad blocks. Unlike FFS, REDO provides interoperability with block I/O subsystems, allowing SSD vendors to hide all the details of their devices and NAND characteristics. 

RFS is designed differently from the conventional LFS in two ways; it only issues out-place update commands and informs a storage device about which blocks have become erasable via TRIM commands. This frees the flash controller from the task of garbage collection all together. RFS writes file data, inodes, and the pieces of the inode map in an out-place update manner. Unlike LFS, the incoming data are written to a physical segment corresponding to a logical segment, and their relative offsets in the logical segment are preserved in the physical one. For check-pointing, RFS reserves two fixed logical segments, called check-point segments. RFS then appends new check-points with different version numbers, so that the overwrites never happen. RFS manages all the obsolete data at the level of a file system and triggers garbage collection when free space is exhausted. RFS chooses the logical segment 2 as a victim and copies live data to free space. The victim segment becomes free for future use. To inform that the physical segment for the victim has obsolete data, RFS delivers a TRIM command to RSD. Finally, RSD marks the physical segment out-of-date and erases flash blocks. 

RSD maintains the segment-map table, and each entry of the table points to physical blocks that are mapped to a logical segment. When write requests come, RSD calculates a logical segment number (i.e., 100) using the logical file-system page number (i.e., 1,600). Then, it looks up the remapping table to find the physical blocks mapped to the logical segment. If physical blocks are not mapped yet, RSD builds the physical segment by allocating new flash blocks. RSD picks up free blocks with the smallest P/E cycles in the corresponding channel/way. A bad block is ignored. If there are flash blocks already mapped, RSD writes the data to the fixed location in the physical segment. Block erasure commands are not explicitly issued from RFS. But, RSD easily figures out which blocks are out-of-date and are ready for erasure because RFS informs RSD of physical segments only with obsolete data via a TRIM command. RSD handles overwrites like block-level FTL.

\textbf{LightNVM (The Linux Open-Channel SSD Subsystem)} \cite{c226, c227, c228, c229, c230} proposes that SSD management trade-offs should be handled through Open-Channel SSDs, a new class of SSDs, that give hosts control over their internals. It introduces a new Physical Page Address I/O interface that exposes SSD parallelism and storage media characteristics. LightNVM integrates into traditional storage stacks, while also enabling storage engines to take advantage of the new I/O interface.

The Physical Page Address (PPA) I/O interface is based on a hierarchical address space. It defines administration commands to expose the device geometry and let the host take control of SSD management, and data commands to efficiently store and retrieve data. The interface is independent of the type of non-volatile media chip embedded on the open-channel SSD. Open-channel SSDs expose to the host a collection of channels, each containing a set of Parallel Units (PUs), also known as LUNs. A PU may cover one or more physical die, and a die may only be a member of one PU. Each PU processes a single I/O request at a time. Regardless of the media, storage space is quantized on each PU. NAND flash chips are decomposed into blocks, pages (the minimum unit of transfer), and sectors (the minimum unit of ECC). Byte-addressable memories may be organized as a flat space of sectors. PPAs are organized as a decomposition hierarchy that reflects the SSD and media architecture.

The PPA address space can be organized logically to act as a traditional logical block address (LBA), e.g., by arranging NAND flash using "block, page, plane, and sector". This enables the PPA address space to be exposed through traditional read/write/trim commands. In contrast to traditional block I/O, the I/Os must follow certain rules. Writes must be issued sequentially within a block. Trim may be issued for a whole block, so that the device interprets the command as an erase.

LightNVM is organized in three layers, each providing a level of abstraction for open-channel SSDs: (1) NVMe Device Driver. A LightNVM-enabled NVMe device driver gives kernel modules access to open-channel SSDs through the PPA I/O interface. (2) LightNVM Subsystem. An instance of the subsystem is initialized on top of the PPA I/O-supported block device. The instance enables the kernel to expose the geometry of the device through both an internal nvm\_dev data structure and sysfs. (3) High-level I/O Interface. A target gives kernel-space modules or user-space applications access to open-channel SSDs through a high-level I/O interface, either a standard interface like the block I/O interface provided by pblk, or an application-specific interface provided by a custom target.

\section{SSDFS ARCHITECTURE}

\subsection{Segment Concept}

\begin{figure}[h]
\centering
 
 \includegraphics[width=0.80\columnwidth,keepaspectratio]{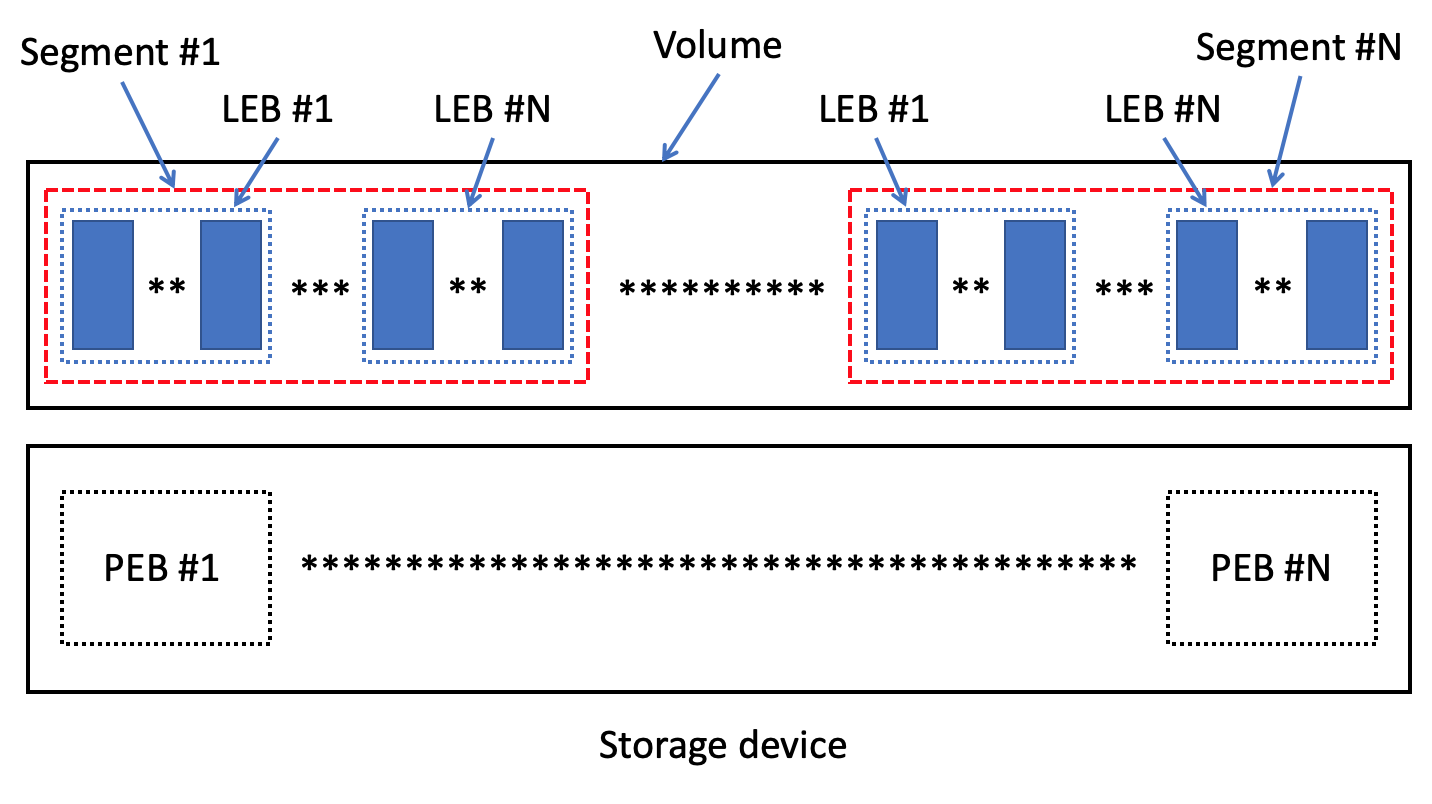}
\caption{Logical segment concept.}
\label{fig:fig001}
\end{figure}

\textbf{Segment} is the cornerstone concept of any Log-structured File System (LFS). This notion (segment concept) points out the reality of presence of Physical Erase Blocks (PEB) on the storage device (SSD) side. Generally speaking, segment can be imagined like a portion of storage device that includes one or several erase blocks. The erase block is very important item of any NAND-based storage device because it is the unit of the erase operation. Finally, any SSDFS file system’s volume can be imagined like a sequence of segments (Fig. \ref{fig:fig001}).

\textbf{Logical segment}. Generally speaking, segment would represent the real physical unit(s) (for example, one or several PEBs are identified by LBAs on the storage device). However, SSDFS operates by logical segments. The logical segment is the unit that is always located on some offset from the volume's beginning for the whole lifetime of file system volume (Fig. \ref{fig:fig001}). Segment is capable to include a variable number of PEBs. However, SSDFS file system's volume includes a fixed number of segments with identical size after the definition of segment size (during a file system volume creation). Very important goal of the segment concept is the capability to execute the erase operation for the whole segment. However, segment is the aggregation of several PEBs in the case of SSDFS file system. It means that such segment construction provides the opportunity to execute the erase operation on the basis of particular PEB(s) inside of the same segment. Finally, the aggregation of several PEBs inside of one segment has several goals: (1) exploitation of operation parallelism for different PEBs inside of the segment, (2) capability to execute the partial erase operation on the PEB basis instead of the whole segment, (3) capability to use a RAID-like or erasure coding scheme inside of the segment, (4) capability to select a proper segment size for a particular workload.

\begin{figure}[h]
\centering
 
 \includegraphics[width=0.80\columnwidth,keepaspectratio]{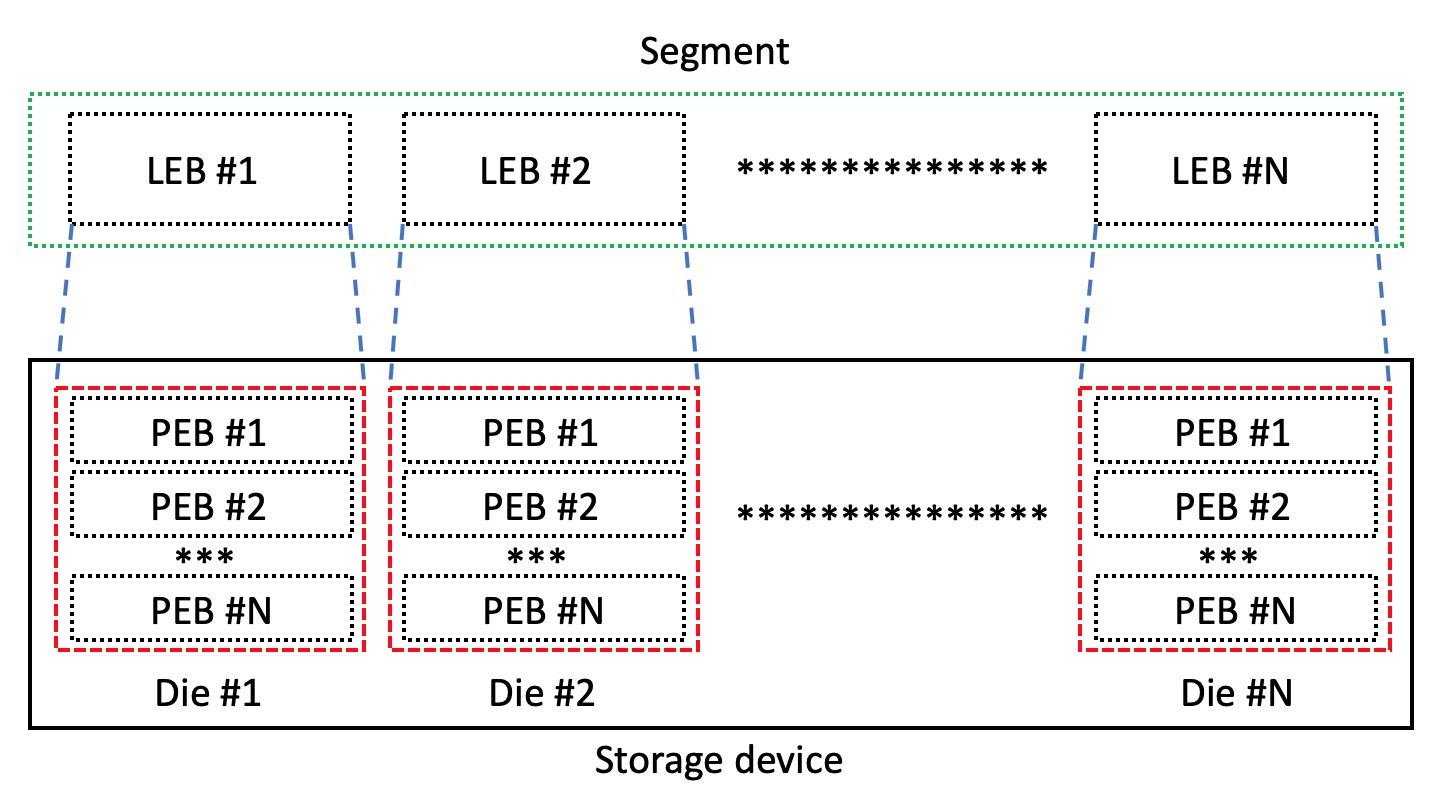}
\caption{Logical extent concept.}
\label{fig:fig002}
\end{figure}

\textbf{Logical extent}. Usually, segment is associated with a PEB (flash-oriented file system) or with a LBA (flash-friendly file system). However, segment is the pure logical entity without the strict relation with PEB or LBA in the case of SSDFS file system. Generally speaking, the segment is simply some portion of the file system volume is always located on some offset from the volume’s beginning (Fig. \ref{fig:fig001}).

The nature of SSDFS file system's segment has the goal to implement a logical extent concept. This concept implies that logical extent (segment ID + logical block + length) is always located in the same logical position inside of the same segment (Fig. \ref{fig:fig002}). Generally speaking, the goal of logical extent is to exclude the necessity to update the metadata about a logical block's position in the case of data migration from the initial PEB into another one (in the case of update or GC operation, for example).

The logical extent concept is the technique of resolving the write amplification issue for the case of LFS file system. It means that any metadata structure keeping a logical extent doesn't need in updating the logical extent value in the case of data migration between the PEBs because the logical extent remains the same until the data is living in the same segment (Fig. \ref{fig:fig002}). The implementation of logical extent concept needs in introduction of Logical Erase Block (LEB) concept (Fig. \ref{fig:fig001} - \ref{fig:fig002}). LEB represents the logical analogue of PEB on storage device side.

Generally speaking, segment is a sequence of LEBs and every LEB is equal to the size of one or multiple PEBs. As a result, the LEB can be imagined like a container that could be associated with any PEB on storage device side. Finally, every LEB always has the same index in the particular segment. And the problem of association the particular LEB with a PEB is resolving by means of a special mapping table. The PEB mapping table has the several important goals in the SSDFS file system.

LEB and PEB represent different notions. PEB represents the physical erase block on a storage device side. Generally speaking, PEB is really allocated portion of storage device is able to receive read and write I/O requests. Also it is possible to apply the erase operation for this portion of storage device. Oppositely, LEB represents a fixed logical portion of file system's volume is identified by the segment ID and the index in a segment. It is possible to say that LEB is pre-allocated portion of the file system's volume space.

However, the real association of LEB with PEB takes place only if some LEB's logical block/extent was allocated and filled by data. Otherwise, empty LEB doesn't need in association with any PEB. It means that if LEB hasn't any data then no PEB is linked with such LEB. Generally speaking, SSDFS file system's volume represents a sequence of logical segments. Every logical segment contains a set of LEBs that provide opportunity to allocate some number of logical blocks (Fig. \ref{fig:fig001} - \ref{fig:fig002}). As a result, internal metadata structures of SSDFS file system operates by logical extents that need to be updated only in the case of segment ID change. 

\begin{figure}[h]
\centering
 
 \includegraphics[width=0.80\columnwidth,keepaspectratio]{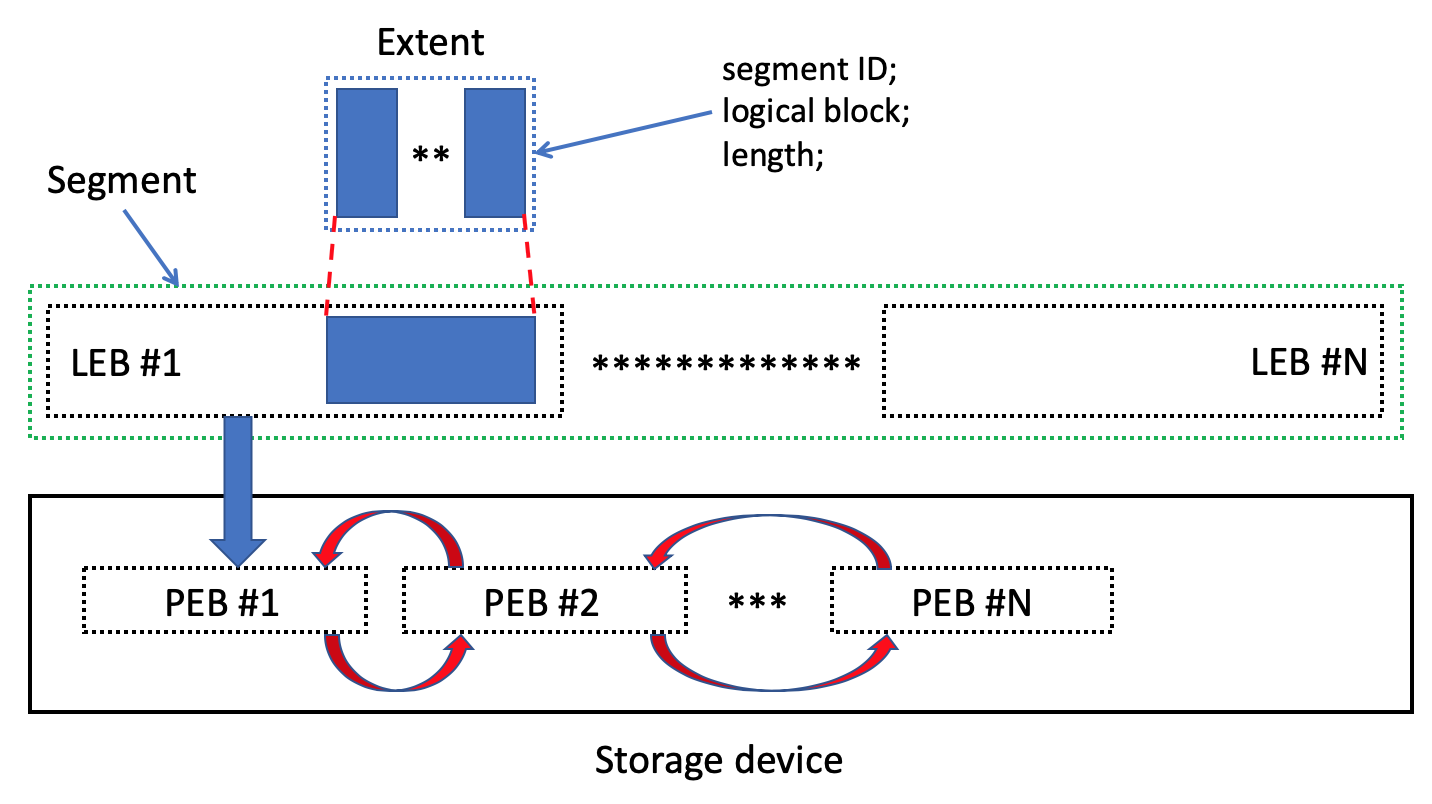}
\caption{Segment parallelism.}
\label{fig:fig003}
\end{figure}

\textbf{Segment parallelism}. One of the important goal to have several LEBs/PEBs in one segment is the trying to employ the parallelism of operation with PEBs are located on different dies. Usually, any SSD contains a set of dies are able to execute various operations independently and concurrently (for example, erase operation). Moreover, multi-channel SSD architecture is capable to deliver commands and data to different dies by means of independent channels. Generally speaking, LEBs of the same segment are able to be associated with PEBs are located on different dies (Fig. \ref{fig:fig003}). As a result, the operation parallelism in one segment is able to improve the file system performance at whole. From another point of view, the opportunity to associate any LEB with any PEB creates the flexibility in policy of distribution of LEBs of the same segment in the different areas of the storage device.

The critical point is the capability to have the knowledge about a distribution of PEBs' ranges amongst the different dies. Generally speaking, such distribution can be implemented on the basis of static or dynamic policies. The static policy means that the whole address space of storage device is distributed among the different dies in static manner. Otherwise, the storage device itself should be able to inform the host about such distribution by means of special protocol. For example, Open-channel SSD could be able to provide such data on the host side.

\subsection{LEB/PEB Architecture}

\begin{figure}[h]
\centering
 
 \includegraphics[width=0.80\columnwidth,keepaspectratio]{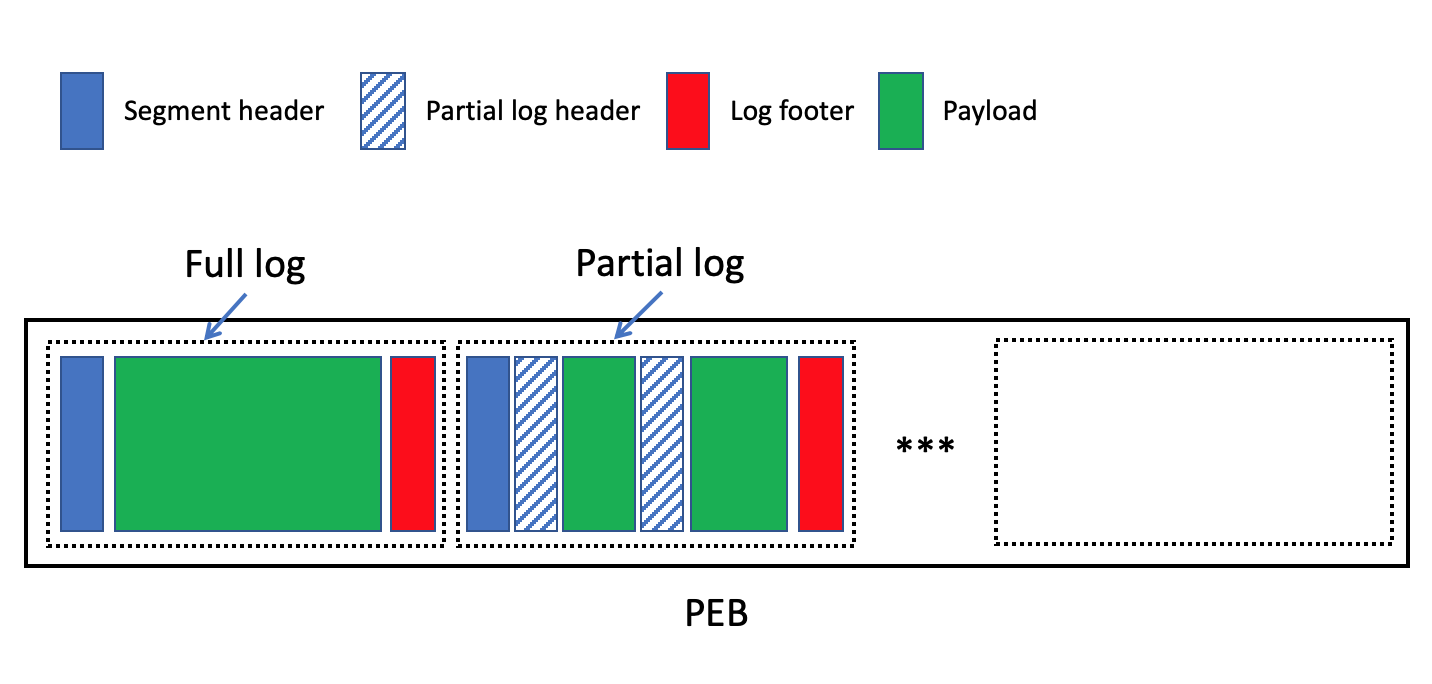}
\caption{Log concept.}
\label{fig:fig004}
\end{figure}

\textbf{Log concept}. Log is the fundamental basic structure of SSDFS file system (Fig. \ref{fig:fig004}). Any user data or metadata are stored in the log of SSDFS file system's volume. Generally speaking, the log concept tries to achieve the several very important goals: (1) replication of critical metadata structures that characterize a file system volume, (2) creation the opportunity to recover the log's payload (user data or metadata) on the basis of log's metadata even if all other logs are corrupted, (3) localization of block bitmap by scope of one PEB, (4) implementation the concept of an offsets translation table, (5) implementation the concept of main, diff updates, and journal areas.

\begin{figure}[h]
\centering
 
 \includegraphics[width=0.80\columnwidth,keepaspectratio]{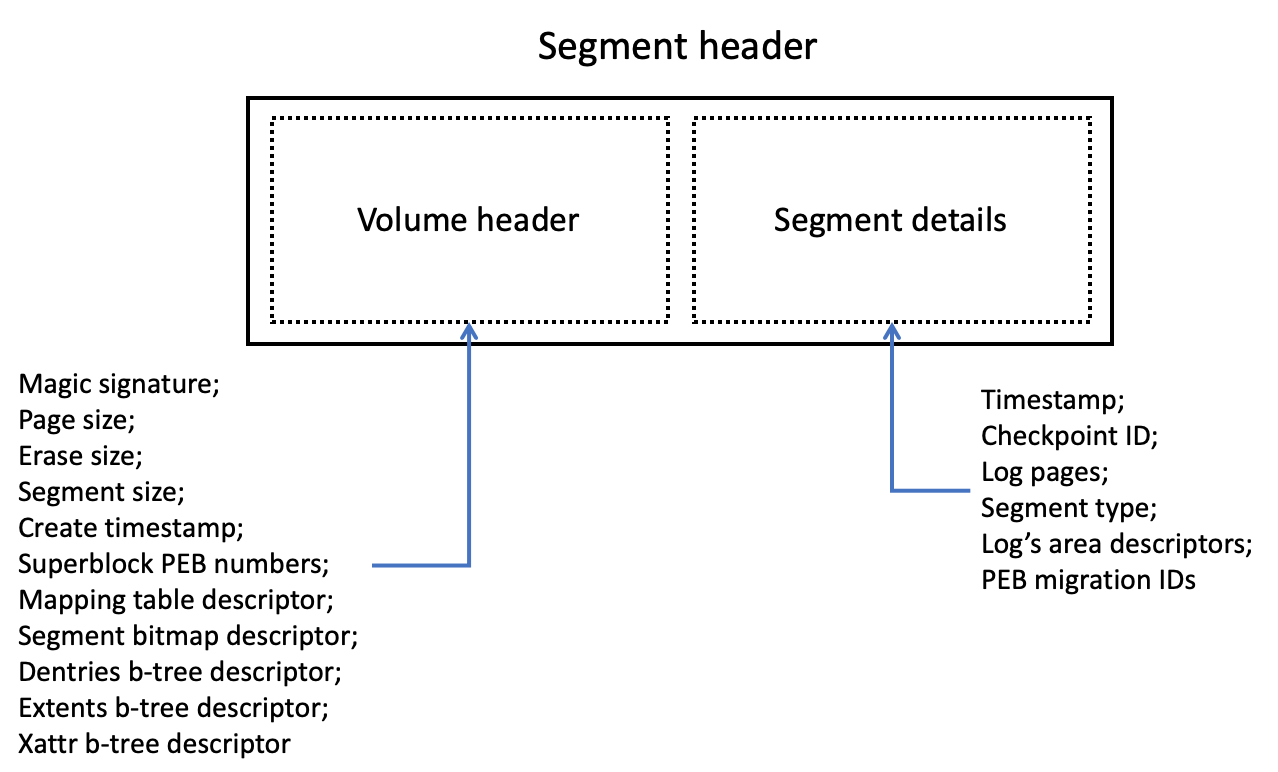}
\caption{Log header.}
\label{fig:fig005}
\end{figure}

It is possible to imagine the log like a container that includes a header, a payload, and a footer (Fig. \ref{fig:fig004}). The responsibility of header (Fig. \ref{fig:fig005}) is the identification of file system type and the log's beginning because the header is capable to play the role of file system's superblock. Any PEB could contain one or several logs and end-user is able to define the size of the log. Moreover, various segment types are able to have the different log's size. However, mount type, unmount operation, segment type, or workload type could result in the necessity to commit the log without enough data in the log's payload. As a result, it needs to distinguish full and partial logs. The full log has to contain such number of logical blocks (or NAND flash pages) that were defined by end-user for this particular segment type during the file system volume creation. Every full log contains the segment header, payload, and footer (Fig. \ref{fig:fig004}).

\begin{figure}[h]
\centering
 
 \includegraphics[width=0.80\columnwidth,keepaspectratio]{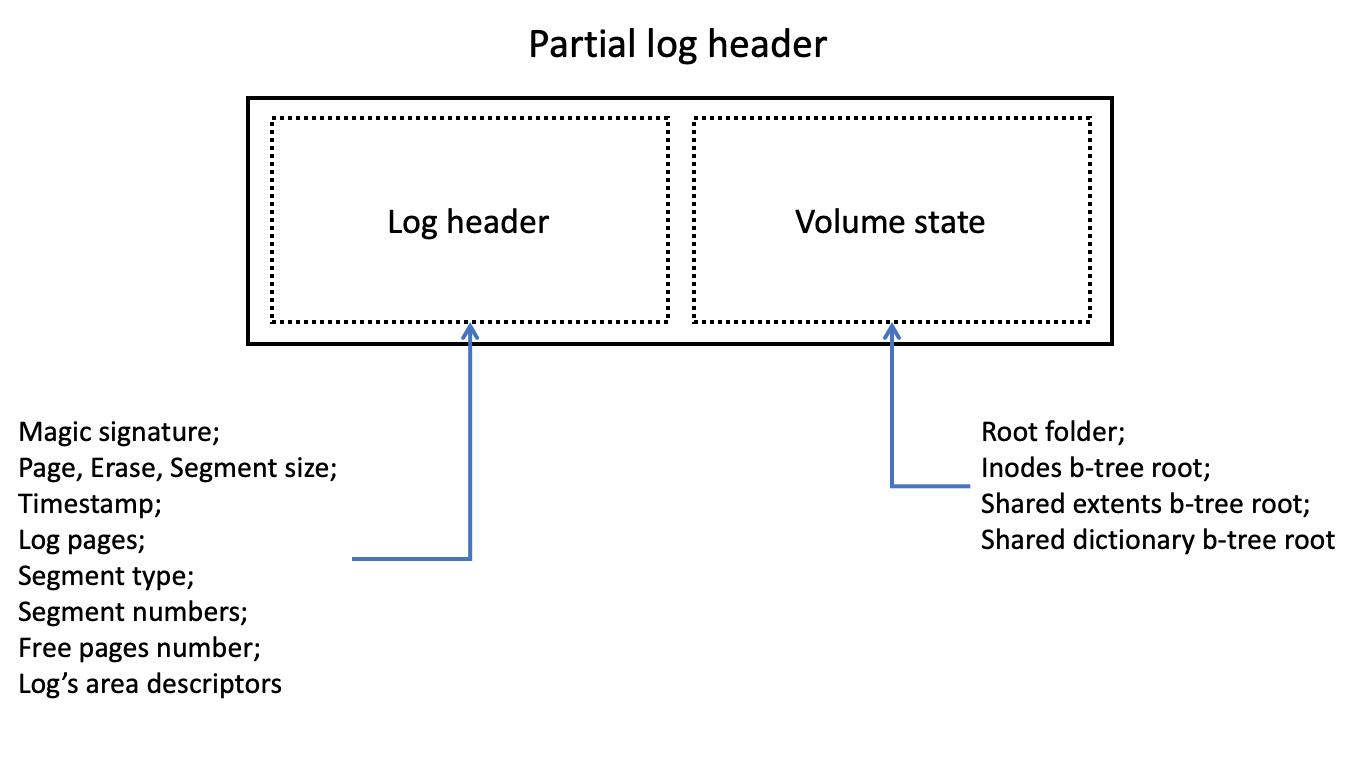}
\caption{Partial log header.}
\label{fig:fig006}
\end{figure}

\begin{figure}[h]
\centering
 
 \includegraphics[width=0.80\columnwidth,keepaspectratio]{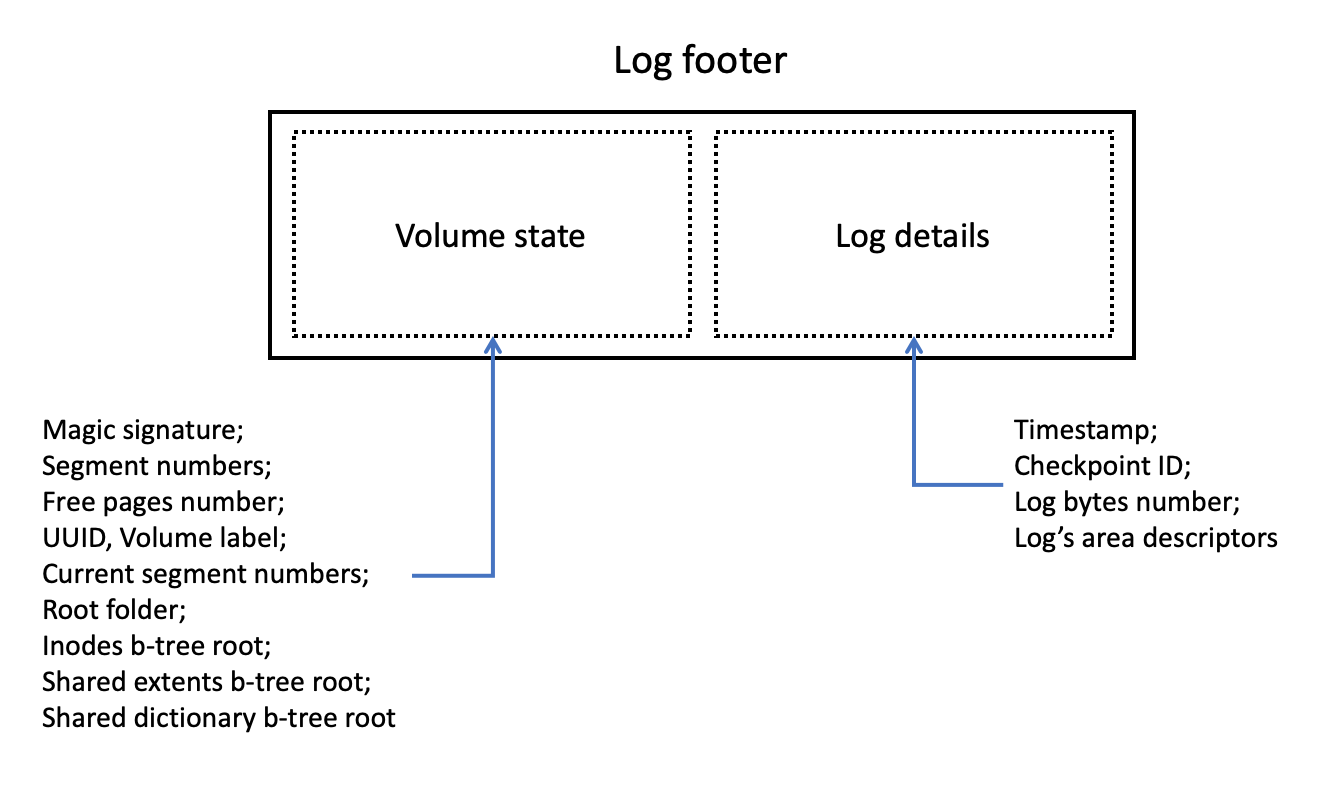}
\caption{Log footer.}
\label{fig:fig007}
\end{figure}

If it exists the necessity to commit a log without the presence of enough data in the payload then it needs to create a chain of partial logs in a PEB (Fig. \ref{fig:fig004}). The first partial log contains the segment header (Fig. \ref{fig:fig005}), the partial log header (Fig. \ref{fig:fig006}), and the payload. Every next partial log includes only the partial log header and the payload. Finally, the last partial log ends with the log footer (Fig. \ref{fig:fig004}, Fig. \ref{fig:fig007}). Generally speaking, to select an optimal value of log's size could be not easy task because different workloads is able to need in specialized log's size. It means that collecting statistics about partial logs' size could be the basis for searching and gradual correction of the full log's size with the goal to achieve the local or global optimum value.

\textbf{Superblock}. Usually, any file system starts from a superblock that is located in one or several fixed position(s) on the file system's volume. The responsibility of the superblock is to identify the file system's type and to provide the description of the key file system's metadata structures. SSDFS represents the LFS file system type that is using the Copy-On-Write (COW) policy for updating the state of any data or metadata. Generally speaking, it means that the superblock cannot be located in the fixed position of SSDFS file system's volume. Oppositely, every log of SSDFS file system contains the “superblock” copy. It means that extracting the header (Fig. \ref{fig:fig005} - \ref{fig:fig006}) and footer (Fig. \ref{fig:fig007}) of any log provides the state of file system's superblock is actual for some timestamp. SSDFS file system is using the special algorithm of fast searching the last actual superblock's state.

SSDFS file system splits the superblock's state on: (1) static data, (2) dynamic data. The static part of superblock is characterized the key parameters/features of a file system's volume (for example, logical block size, erase block size, segment size, creation timestamp and so on) that are defined during the volume creation. This part of superblock is kept in the volume header of log's header (Fig. \ref{fig:fig005}). Oppositely, the dynamic part of superblock is represented by mutable parameters/features of file system's volume (for example, segment numbers, free logical blocks number, volume UUID, volume label and so on). The volume state of the log footer (Fig. \ref{fig:fig007}) keeps the dynamic part of superblock. And, finally, the partial log header (Fig. \ref{fig:fig006}) represents the restricted combination of static and dynamic parts of the superblock. Every metadata structure of the log (segment header, partial log header, log footer) starts from a magic signature (Fig. \ref{fig:fig005} - \ref{fig:fig007}). Generally speaking, the responsibility of magic signature is to identify the file system type and the type of metadata structure. Another very important field is the log's area descriptors (Fig. \ref{fig:fig005} - \ref{fig:fig007}). These descriptors describe the position and the size of every existing area (user data or metadata) in a log.

\textbf{Block bitmap}. One of the very important log's metadata structure is a block bitmap. Usually, a file system uses the block bitmap as a single metadata structure for the whole volume. The responsibility of block bitmap is to track the state of logical blocks (free or used). As a result, the block bitmap is frequently accessed and modified metadata structure. However, this compact and efficient metadata structure cannot be used in traditional way for the case of LFS file system by virtue of: (1) frequent updates of the block bitmap is able to increase the write amplification, (2) logical block of LFS file system needs in more states (free, used, invalid), (3) the volume capacity could change because of necessity to track the presence of bad erase blocks.

\begin{figure}[h]
\centering
 
 \includegraphics[width=1.00\columnwidth,keepaspectratio]{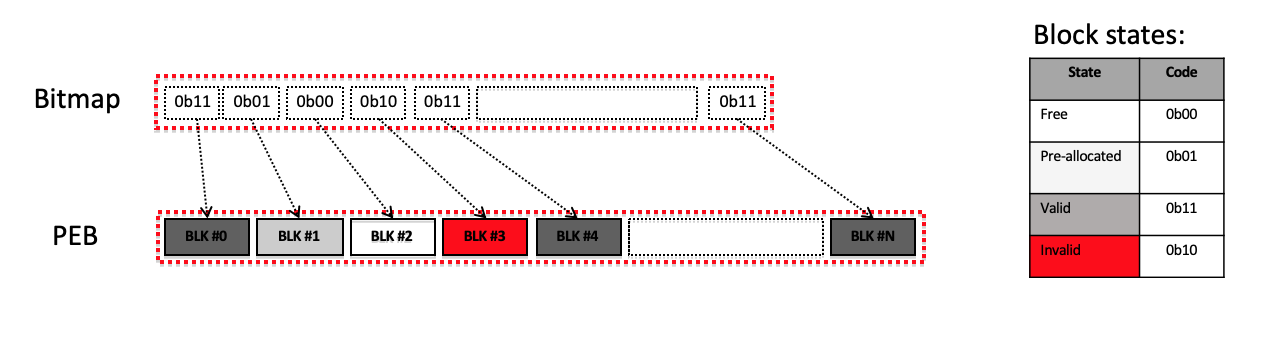}
\caption{Block bitmap concept.}
\label{fig:fig008}
\end{figure}

SSDFS file system introduces the PEB-based block bitmap because of proven efficiency and compactness of this metadata structure. First of all, the block bitmap (Fig. \ref{fig:fig008}) tracks such states of logical block: (1) free, (2) used, (3) pre-allocated, (4) invalid. The free state means that logical block is ready for allocation and write operation. Oppositely, the used state means that the logical block was allocated and the write operation has taken place for this logical block. The invalid state represents the case when the update or GC operation invalidates (makes not actual) the state of a logical block in one PEB and to store the actual state into another one. And, finally, the pre-allocated state can be used for representing the case when several fragments of different logical blocks can be stored into one NAND flash page (for example, in the case of compression or delta-encoding).

\begin{figure}[h]
\centering
 
 \includegraphics[width=0.80\columnwidth,keepaspectratio]{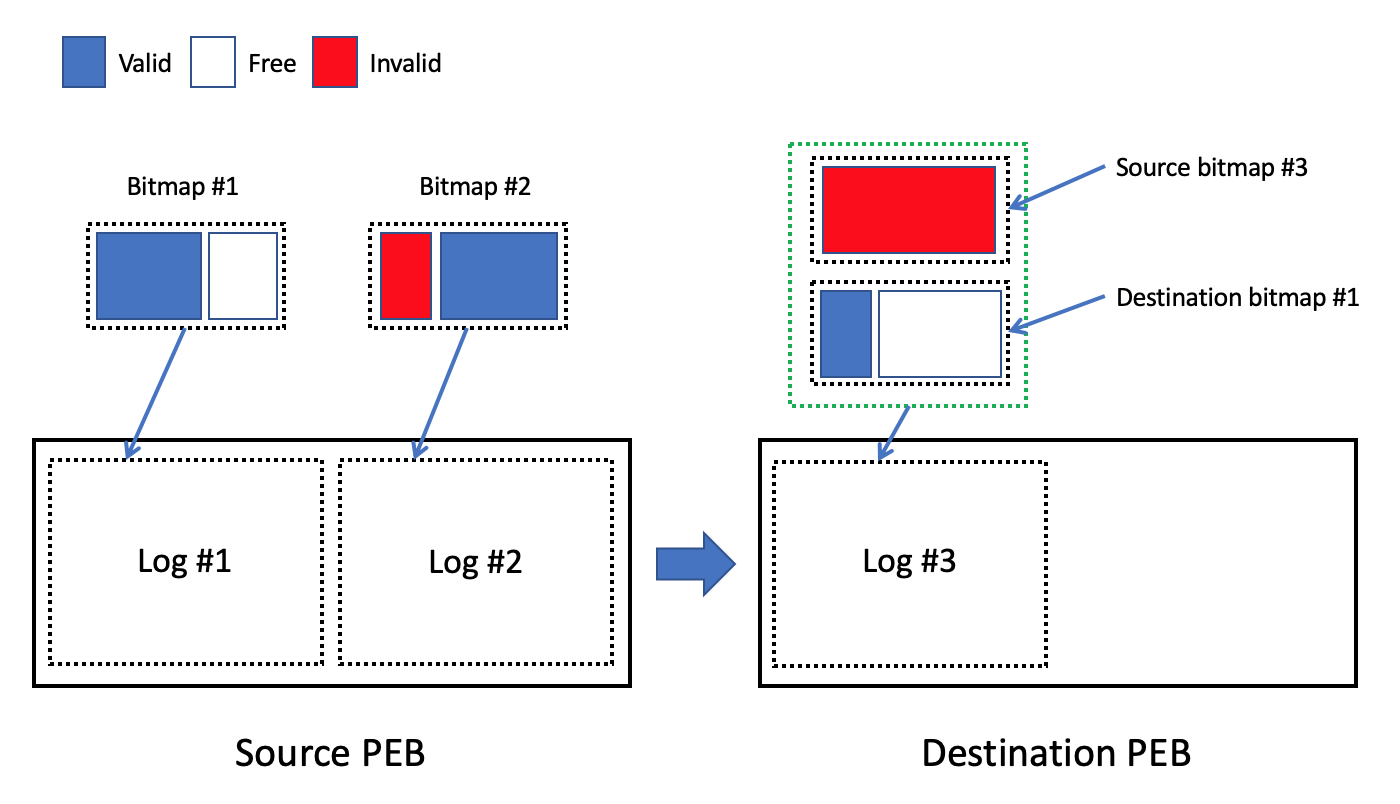}
\caption{Technique of using the block bitmap.}
\label{fig:fig009}
\end{figure}

Block bitmap is the PEB-based metadata structure in the case of SSDFS file system (Fig. \ref{fig:fig008} - \ref{fig:fig009}). The goals of such approach are: (1) opportunity to access/modify block bitmaps of different PEBs without the necessity to use any synchronization primitives, (2) capability to lose the bad erase blocks without the necessity to rebuild the block bitmap, (3) capability to allocate the logical blocks and to execute GC operations concurrently for different PEBs in the same segment or for the file system's volume at whole, (4) opportunity to track the state of logical blocks only inside the log's payload. Every log keeps the actual state of the block bitmap for the case of some timestamp. It means that previous PEB's logs play the role of block bitmap's checkpoints or snapshots (Fig. \ref{fig:fig009}). As a result, it is possible to use the block bitmaps of previous logs in the case of corruption of particular PEB's log. If some LEB is under active migration then every log of the destination PEB has to store block bitmap as source PEB as destination PEB (Fig. \ref{fig:fig009}) because migration could be executed in several phases.

\begin{figure}[h]
\centering
 
 \includegraphics[width=0.80\columnwidth,keepaspectratio]{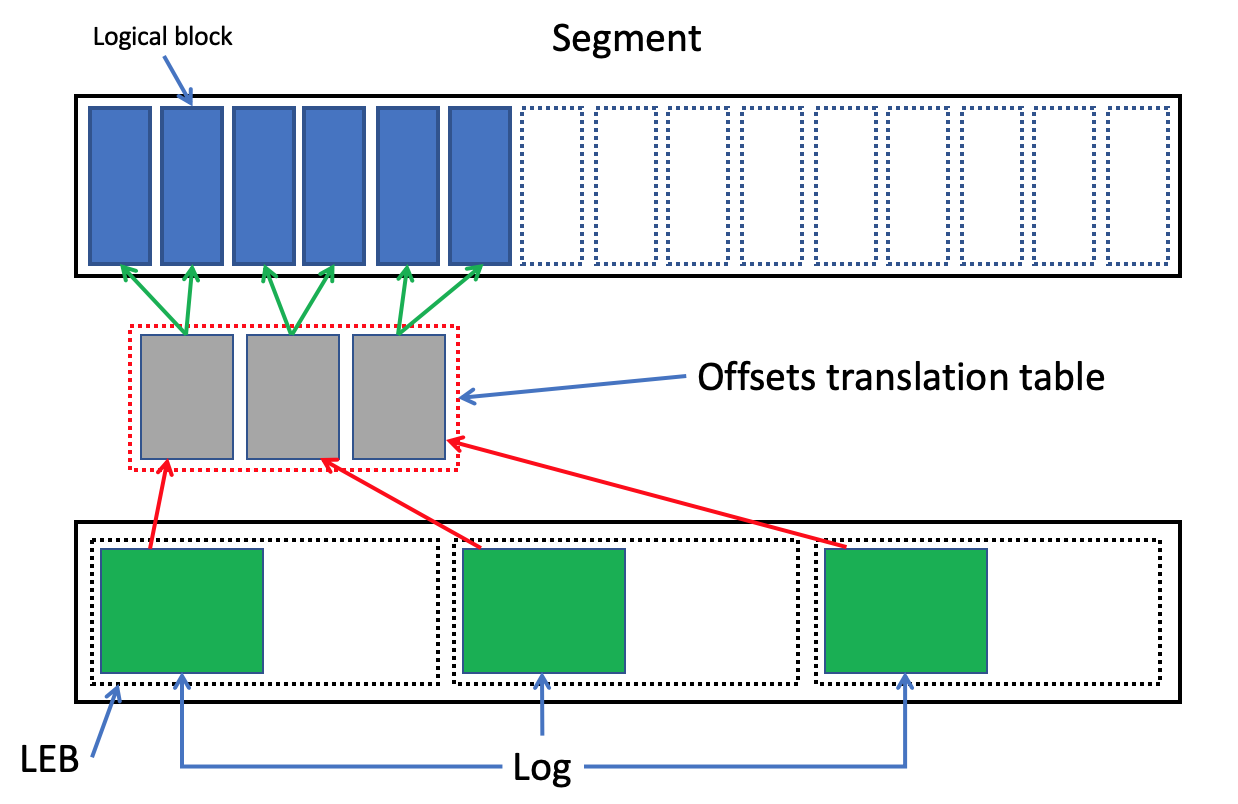}
\caption{Offsets translation table concept.}
\label{fig:fig010}
\end{figure}

\textbf{Offsets translation table}. Any subsystem of SSDFS file system's driver that needs to store user data or metadata treats the segment like a sequence of logical blocks. Generally speaking, the goal of such approach is to provide the opportunity to access the stored data by means of segment ID and logical block number without the knowledge what PEB keeps the actual state of data for the requested logical block. The SSDFS file system uses an offsets translation table (Fig. \ref{fig:fig010}) for implementation this approach. Generally speaking, offsets translation table looks like a sequence of fragments and every fragment is stored in the particular log (Fig. \ref{fig:fig010}). The fragment keeps a portion of the table that associates a logical block number with a descriptor is keeping the offset to the data in the log's payload. As a result, if someone would like to retrieve the actual state of data then it needs to find the latest record in the sequence of fragments of offsets translation table for the requested logical block number. The found record will identify the PEB, log index, and byte offset to the actual state of data in the log's payload.

\begin{figure}[h]
\centering
 
 \includegraphics[width=0.80\columnwidth,keepaspectratio]{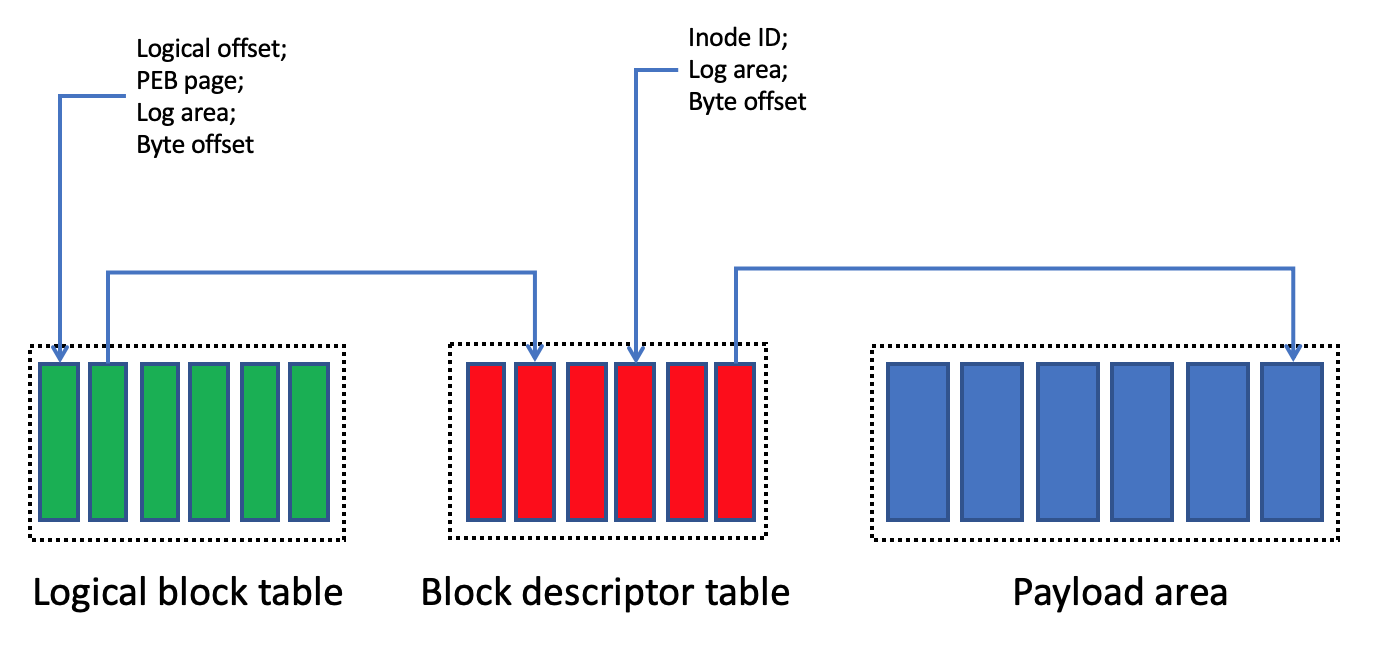}
\caption{Offsets translation table architecture.}
\label{fig:fig011}
\end{figure}

Generally speaking, the offsets translation table includes several metadata structures inside of the log (Fig. \ref{fig:fig011}): (1) logical block table, (2) block descriptor table, (3) payload area. The logical block table represents an array of descriptors where the logical block number can be used as the index. Every descriptor of logical block table keeps: (1) logical offset from the beginning of file or metadata structure, (2) PEB page number that identifies an index of logical block in the block bitmap, (3) log's area that identifies metadata area or payload is keeping the content of logical block, (4) byte offset from the area's beginning till the data portion. Finally, the descriptor of logical block table could point out directly in the payload (for example, in the case of full plain logical block) or into the block descriptor table (Fig. \ref{fig:fig011}). Every record of block descriptor table keeps the inode ID and several descriptors on logical block's states in the payload area(s). The goal of keeping the several descriptors on logical block's states in one record of block descriptor table is to provide the capability to represent the several sequential modifications of a logical block or the various delta-encoded fragments of the same logical block. Finally, the payload could keep the plain full logical block or compressed (delta-encoded) fragment with associated checkpoint and parent snapshot IDs. It needs to point out that the checkpoint and parent snapshot IDs can be extracted from the segment header for the case of plain full logical block. Generally speaking, the knowledge of logical offset from file's beginning, inode ID, checkpoint ID, and parent snapshot ID provides the capability to recover the stored data from the log's payload on the basis of log's metadata only.

\begin{figure}[h]
\centering
 
 \includegraphics[width=0.80\columnwidth,keepaspectratio]{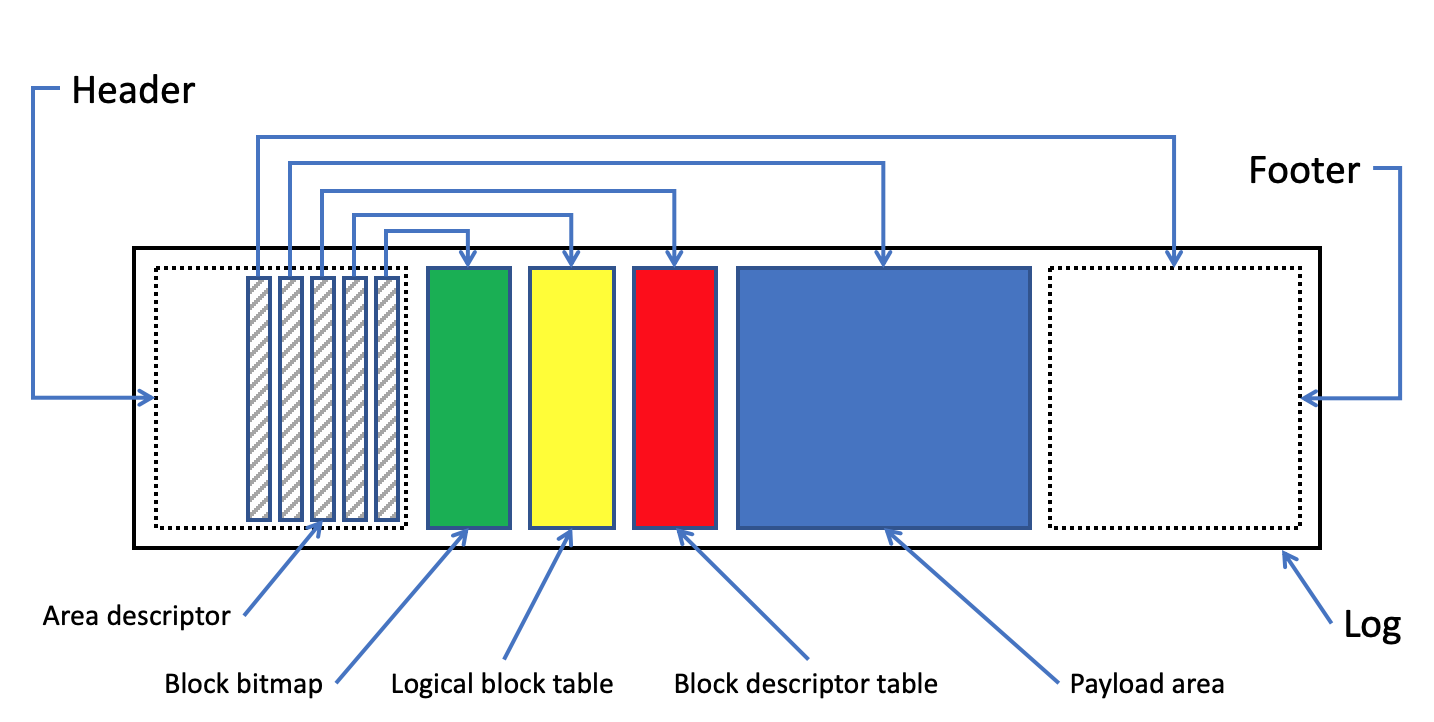}
\caption{Log structure.}
\label{fig:fig012}
\end{figure}

\textbf{Log structure}. As a result, log's structure (Fig. \ref{fig:fig012}) begins with the header that identifies the file system's type and the log's beginning by means of magic signature. Moreover, the header contains an array of area descriptors that describes the existing areas in the log: (1) block bitmap, (2) logical block table, (3) block descriptor table, (4) payload, (5) footer. The footer is also able to include the array of area descriptors with the goal to replicate the critical metadata structures of the log (for example, block bitmap and logical block table).

\begin{figure}[h]
\centering
 
 \includegraphics[width=0.90\columnwidth,keepaspectratio]{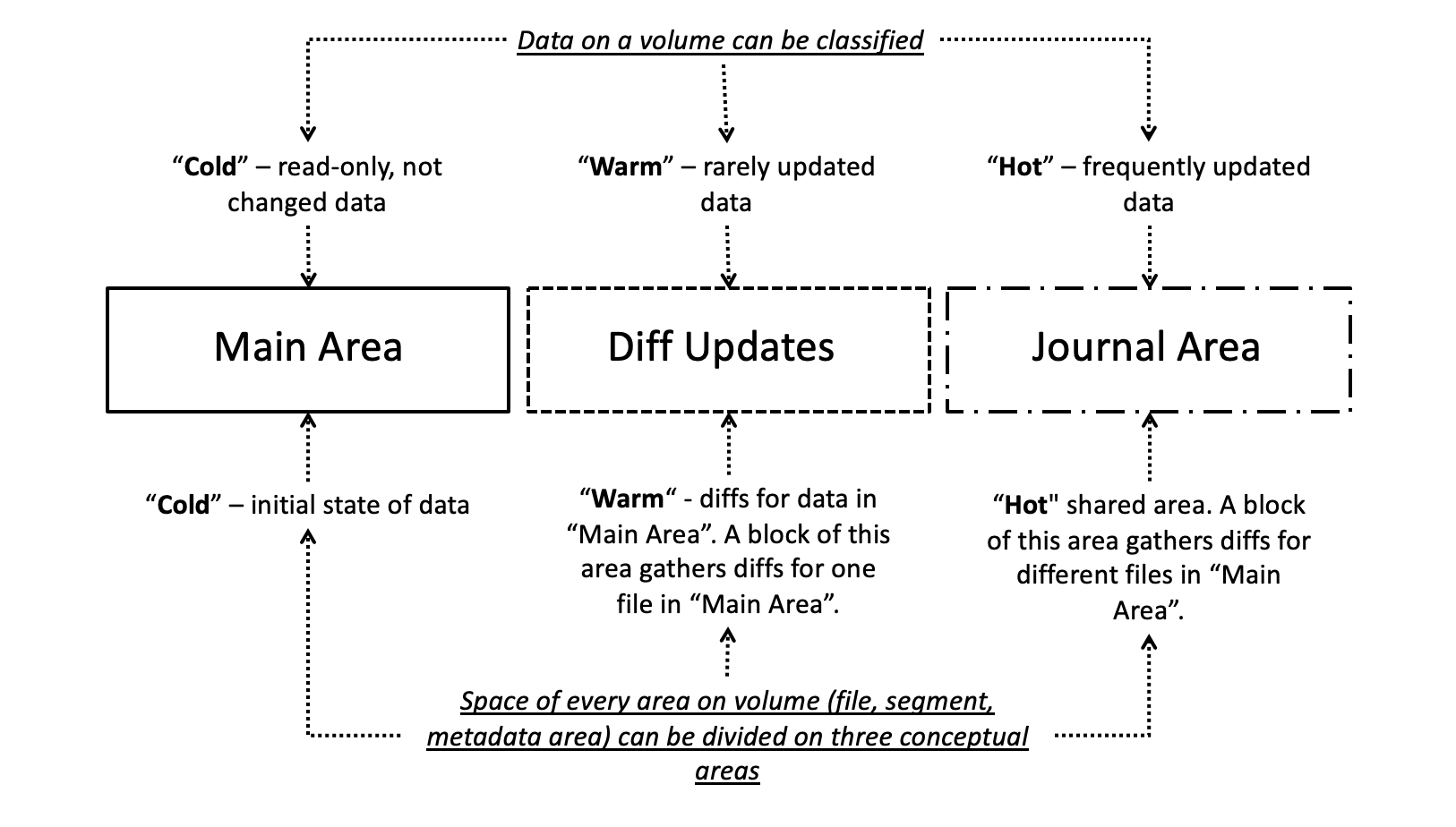}
\caption{Main, diff and journal payload areas.}
\label{fig:fig013}
\end{figure}

\textbf{Main/Diff/Journal areas}. It is very important to distinguish "cold" and "hot" data for the case of LFS file system. Because the identification of "cold" and "hot" types of data provides the opportunity to implement an efficient data management scheme, especially, for the case of GC operations. As a result, SSDFS file system introduces (Fig. \ref{fig:fig013}) the three types of payload areas: (1) main area, (2) diff updates area, (3) journal area. The main area is used for storing the plain full blocks. Generally speaking, the write operation for any logical block takes place in the main area only once and the following updates are stored into the diff updates or journal area. As a result, such write/update policy creates area (main area) with "cold" data because all following updates of any logical block in the main area will be stored into another area(s) (diff updates or journal area). If the diff updates or the journal area gathers significant amount of updates for some logical block in the main area of some log then this logical block could be stored in the main area of another log with applying of all existing updates.

The diff updates area (Fig. \ref{fig:fig013}) could play the role of area with the "warm" data. The responsibility of diff updates area is to gather into one NAND flash page the compressed blocks or delta-encoded fragments of the same file. It means that this area is able to store the significant amount of updates for logical blocks in the main area. However, the updates of the data in the diff updates area could be not so significant like for the main or journal areas. Finally, the diff updates area will be hotter than main area but it could be colder that the journal area. The goal of journal area (Fig. \ref{fig:fig013}) is to represent the area with the "hot" data. One NAND flash page of journal area is used for compaction of several small files or updates of logical blocks of different files. Finally, it means that the NAND flash page with fragments of various files is able to receive more updates than main or diff updates areas.

From one point of view, the compaction of several fragments of different logical blocks into one NAND flash page creates the capability to move more data for one GC operation. From another viewpoint, warm/hot areas introduce the areas with high frequency of update operations. Generally speaking, it is possible to expect that high frequency of update operations (in diff updates and journal areas) creates the natural migration of data between PEBs without the necessity to use the extensive GC operations.

\subsection{Segment Types}

Usually, user data and metadata are based on different granularity of items and very different frequency of updates. Moreover, various metadata structures have different architectures and live under different workloads. SSDFS file system distinguishes various type of segments with the goal to guarantee a predictable and deterministic nature of data management. As a result, there are several type of segments on any SSDFS file system's volume: (1) superblock segment, (2) snapshot segment, (3) PEB mapping table segment, (4) segment bitmap, (5) b-tree segment, (6) user data segment. Generally speaking, the goal to distinguish the different type of segments is to localize the peculiarities of different types of data (user data and metadata, for example) inside of specialized segments. Another important responsibility of the segments' specialization is to provide a reliable basis for data and metadata recovering in the case of file system's volume corruption. It means that if a PEB keeps a specialized type of metadata or user data then it simplifies the task of data/metadata recovering in the case of file system's volume corruption.

\begin{figure}[h]
\centering
 
 \includegraphics[width=0.90\columnwidth,keepaspectratio]{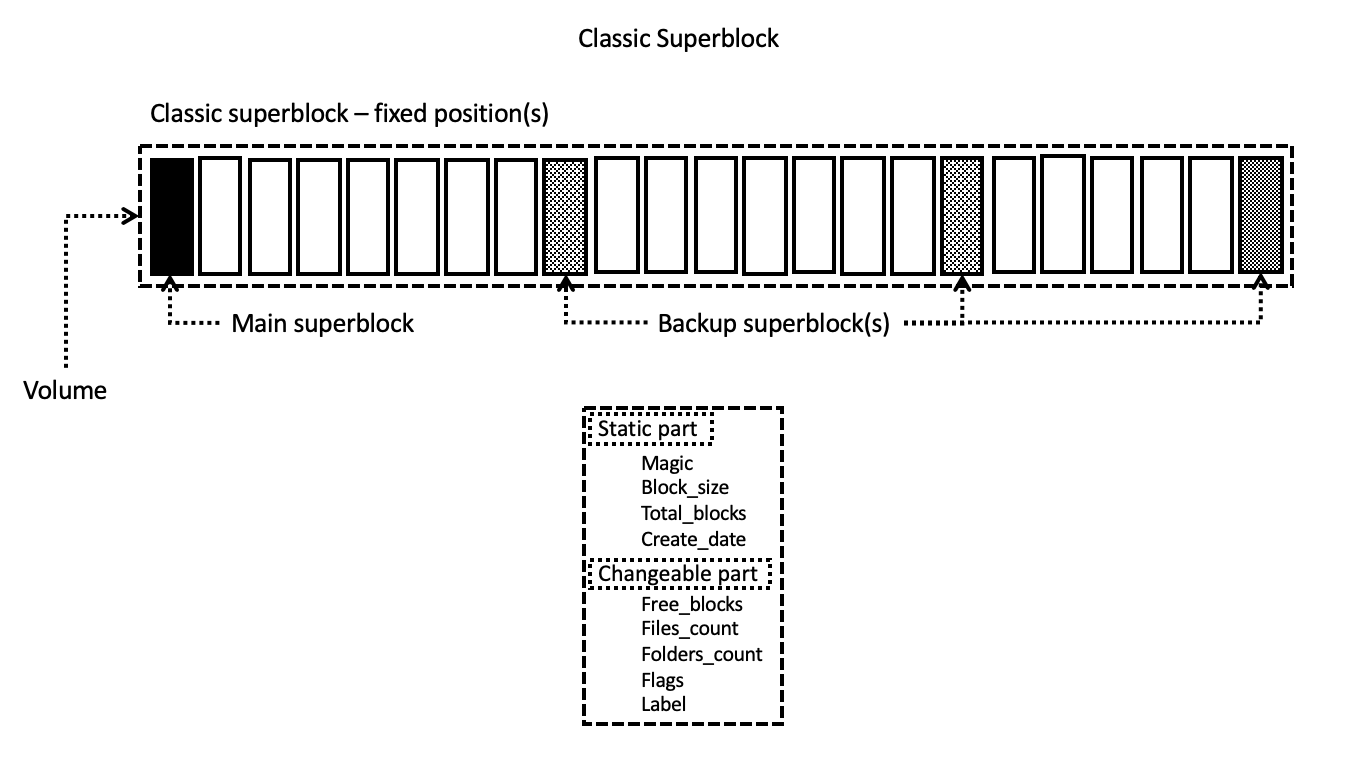}
\caption{Classic superblock approach.}
\label{fig:fig014}
\end{figure}

\textbf{Superblock segment}. Superblock is one of the critical metadata structure of any file system (Fig. \ref{fig:fig014}). First of all, the superblock identifies a type of file system (ext4, xfs, btrfs, for example). From another viewpoint, the superblock's responsibility is to describe the crucial features of a file system's volume (logical block size, number of free blocks, number of folders, for example). And, finally, file system driver extracts from superblock the knowledge about position of the key metadata structures (block bitmap, inodes array, for example) on the volume. Usually, superblock is stored into one or several fixed position(s) on the file system's volume (Fig. \ref{fig:fig014}). Generally speaking, the fixed position of the superblock provides the opportunity to find the superblock easily and to identify a file system's type on the volume. However, SSDFS is Log-structured (LFS) and flash-friendly file system. It means that the fixed position of the superblock is not suitable solution for the case of SSDFS file system. If anybody considers the superblock metadata structure (Fig. 14) then it is possible to distinguish the two principal types of fields: (1) static metadata - describe basic and unchangeable features of file system's volume (logical block size, for example), (2) mutable metadata - describe the volume's features that are modified during the mounted state of file system's volume (number of free blocks, for example).

\begin{figure}[h]
\centering
 
 \includegraphics[width=0.90\columnwidth,keepaspectratio]{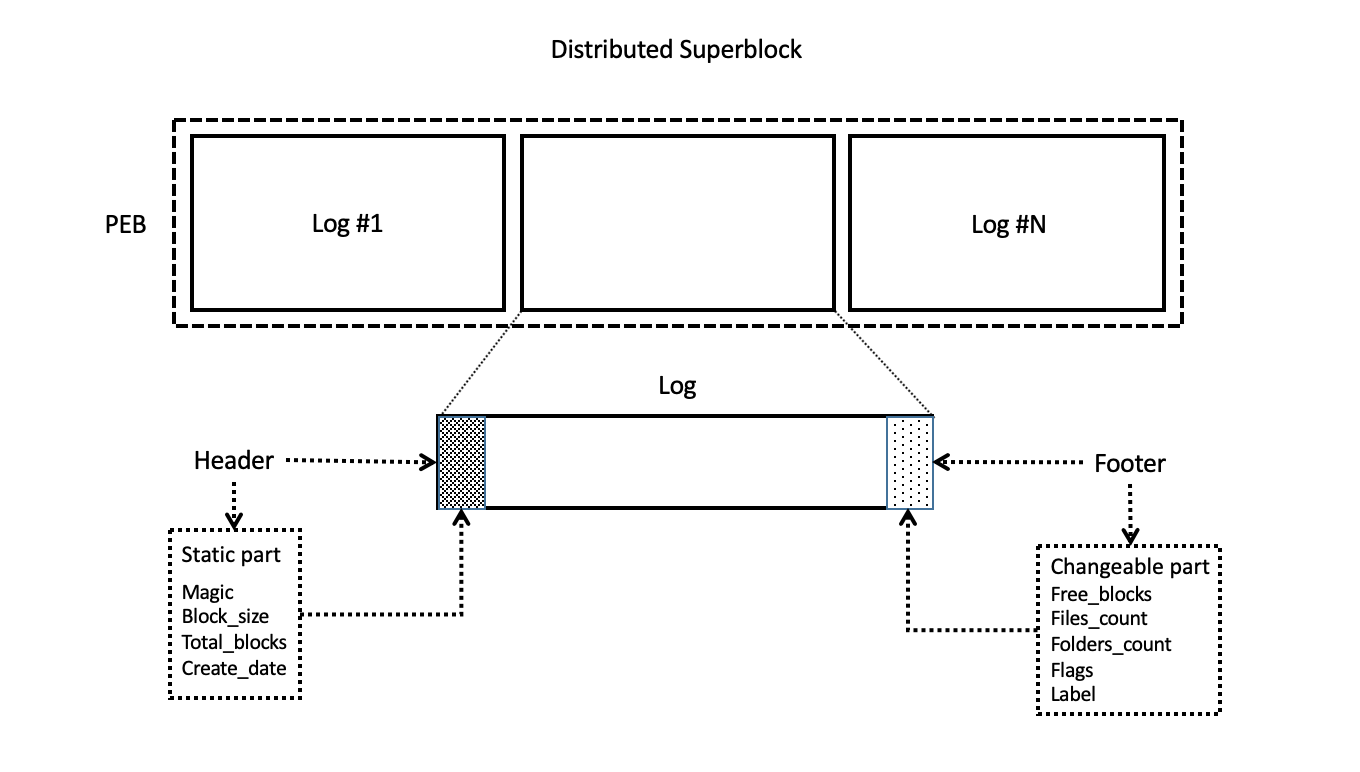}
\caption{Distributed superblock approach.}
\label{fig:fig015}
\end{figure}

Any SSDFS file system's volume represents a sequence of logical segments. Every segment contains some number of LEBs. Finally, it needs to associate a LEB with a PEB in the case of necessity to store any data in the segment. As a result, any associated PEB stores a sequence of logs. Moreover, every full log is started from the fixed position in the PEB and the full log contains the segment header and log footer (Fig. \ref{fig:fig015}). Generally speaking, segment header and log footer are located in fixed positions if anybody knows the size of full log. SSDFS file system keeps the static part of superblock's metadata in the segment header but the mutable part in the log footer of every full log (Fig. \ref{fig:fig015}). The massive replication of superblock's metadata has the goal to increase the reliability of storing the superblock inside of SSDFS file system's volume. From another point of view, keeping the superblock's metadata in every full log creates the opportunity to start the recovery of the corrupted file system's volume from any particular PEB on the volume. Moreover, even if it survives only one log from the whole SSDFS file system's volume then it will be possible to extract and to recover the data or metadata from the survived log on the basis of metadata in the header and footer.

\begin{figure}[h]
\centering
 
 \includegraphics[width=0.90\columnwidth,keepaspectratio]{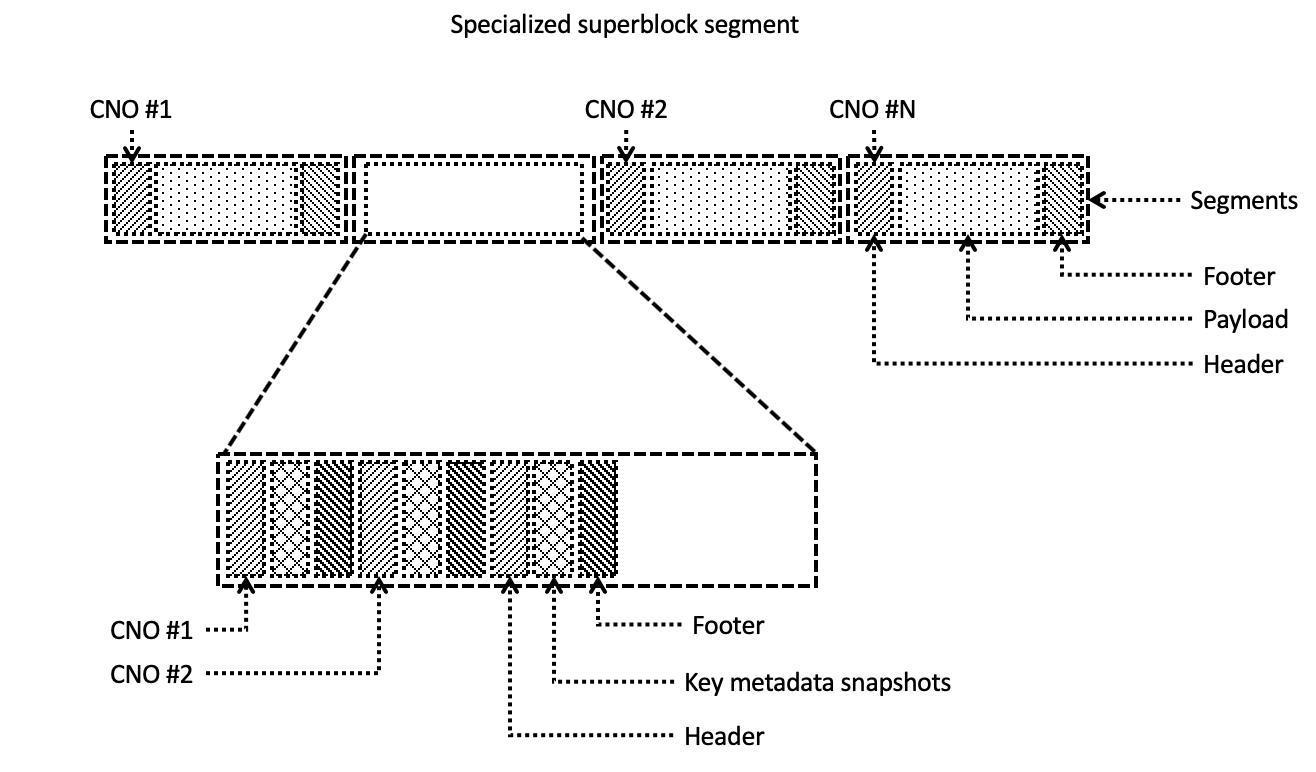}
\caption{Specialized superblock concept.}
\label{fig:fig016}
\end{figure}

However, the massive replication of superblock's metadata creates the problem to find the last actual state of mutable part of superblock's metadata. To resolve this problem the SSDFS file system introduces a special type of segment - the superblock segment (Fig. \ref{fig:fig016}). Generally speaking, the goal of the superblock segment is to keep a sequence of superblock's states are stored for every mount and unmount operations. As a result, the superblock segment contains a sequence of logs that keep the state of superblock in header and footer (Fig. \ref{fig:fig016}). Moreover, every log of superblock segment is able to store in the payload a snapshot of some critical metadata structures. The key goals of superblock segment are: (1) to store the superblock's state for every mount and unmount operations, (2) to provide the mechanism of fast search the last actual state of the superblock.

\begin{figure}[h]
\centering
 
 \includegraphics[width=0.90\columnwidth,keepaspectratio]{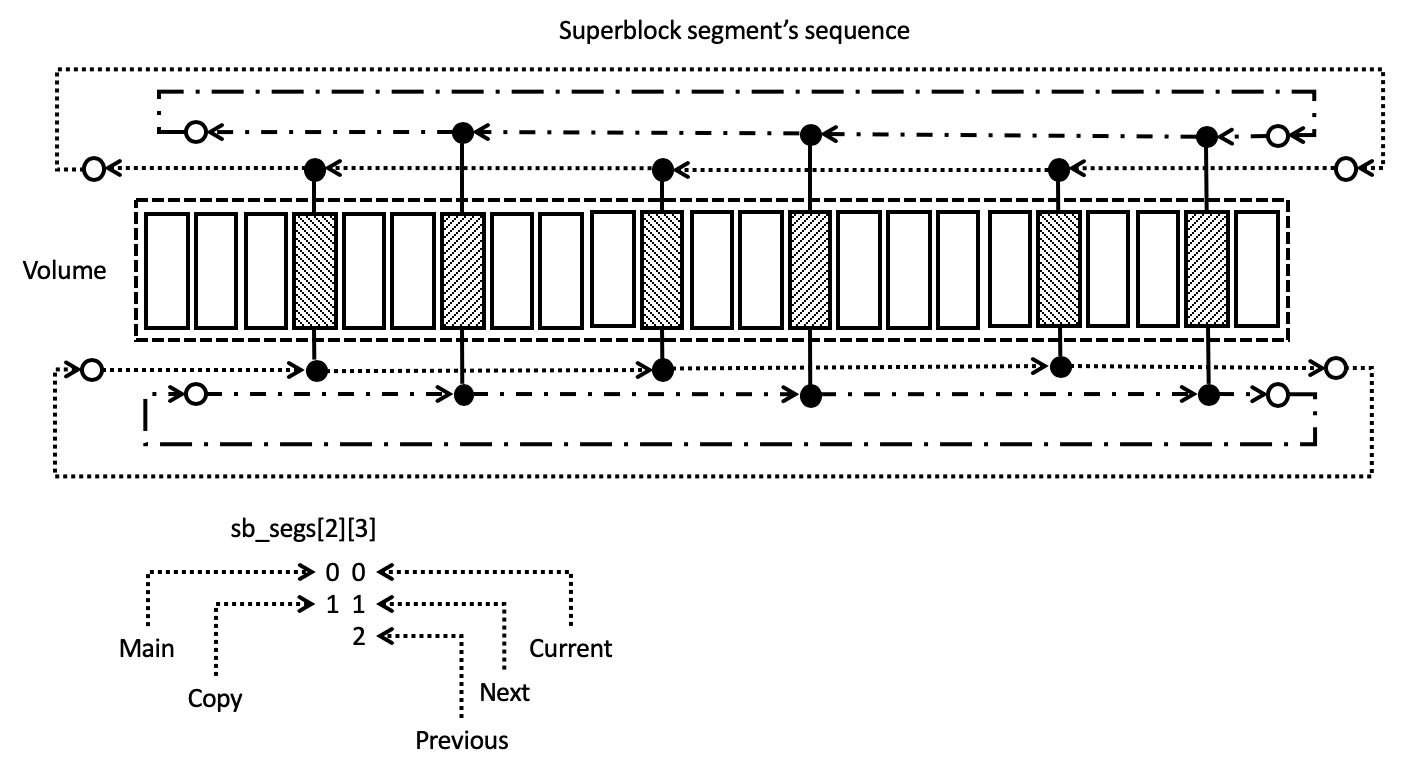}
\caption{Superblock segments' migration scheme.}
\label{fig:fig017}
\end{figure}

The fast lookup method is based on the knowledge of numbers of current and next superblock segments that are stored in every segment header (Fig. \ref{fig:fig017}). It means that the segment header of any valid PEB with logs is able to provide the number of current and next superblock segments that were actual for some timestamp. The operation of checking the available numbers of segments is able to discover the actual superblock segments or more actual numbers of superblock segments. Finally, it is possible to find the actual superblock segment by means of passing through the chain of segment numbers. As a result, it needs to find the latest log in the found actual superblock segment with the goal to retrieve the actual superblock's state. Moreover, SSDFS file system keeps two copies of the superblock segment with the goal to improve the reliability and to increase the performance of the lookup operation.

The segment header of any full log keeps the previous, current, and next numbers of superblock segment (Fig. \ref{fig:fig017}). These numbers creates the basis for migration technique of superblock segment. Initially, file system driver is using the number of current superblock segment for storing the logs with superblocks' state. The next superblock segment's number plays the role of reserve for the case of exhaustion of the current superblock segment. Finally, the file system driver moves the number of superblock segment from the current to the previous state in the case of exhaustion. If the previous state contained the number of some superblock segment then this superblock segment saved as snapshot or the erase operation is applied for the old superblock segment. As a result, the next superblock segment's number becomes the current superblock segment. It needs to allocate some clean segment for the new reservation of space for superblock segment. Otherwise, it needs to use the number of segment that was previous superblock segment in the case of inability to allocate a clean segment. Generally speaking, distributed superblock approach and specialized superblock segment concept provide the reliable way of superblock storing and efficient technique of searching the last actual superblock's state for the case of mount operation.

\begin{figure}[h]
\centering
 
 \includegraphics[width=0.80\columnwidth,keepaspectratio]{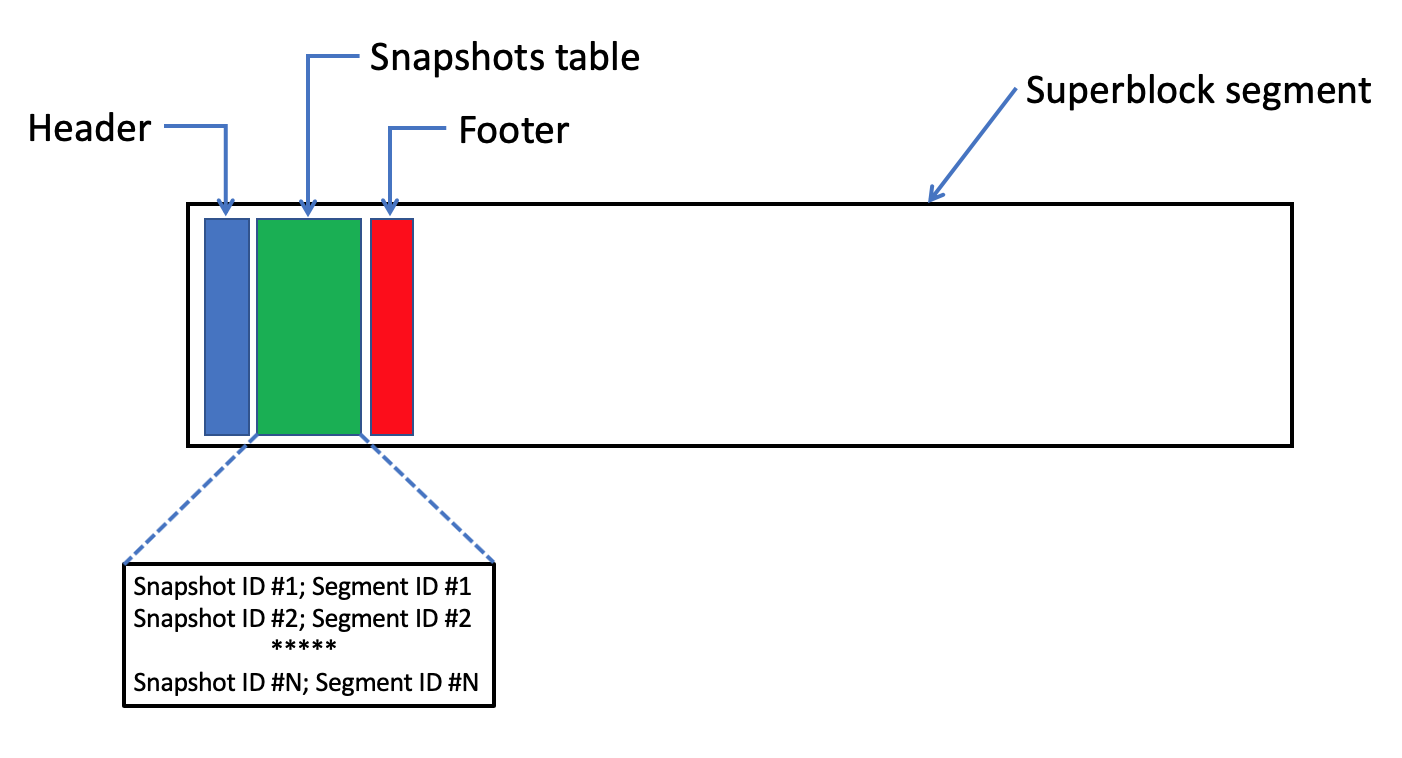}
\caption{Snapshots table concept.}
\label{fig:fig018}
\end{figure}

\textbf{Snapshot segment}. SSDFS file system is Log-structured File System (LFS) with using of Copy-On-Write (COW) policy for data updates. It means that SSDFS provides the rich basis for the concept of snapshots of file system's volume's states. Generally speaking, every log represents a checkpoint of user data's or metadata's state. Such checkpoint is accessible until the applying on a PEB the next erase operation. It means that the checkpoint has to be converted into the snapshot for the long-term keeping the state of this checkpoint. As a result, it is possible to state that snapshot is the long-term storage of the file system's state or the namespace's portion for some timestamp. The snapshot is able to play the role of starting point for the evolution of some version of a file system's state. Generally speaking, the snapshot's state or some version of the file system's state is able to be accessed by means of the mount operation for a snapshot's ID. The superblock segment is able to store in the log's payload the content of some critical metadata structures. As a result, the snapshots table is capable to be stored into the superblock segment (Fig. \ref{fig:fig018}). Generally speaking, the snapshots table is the array of records that keep the snapshot ID and the corresponding number of snapshot segment. Finally, this table provides the mechanism to find the snapshot segment number(s) for the case of a snapshot ID.

\begin{figure}[h]
\centering
 
 \includegraphics[width=0.80\columnwidth,keepaspectratio]{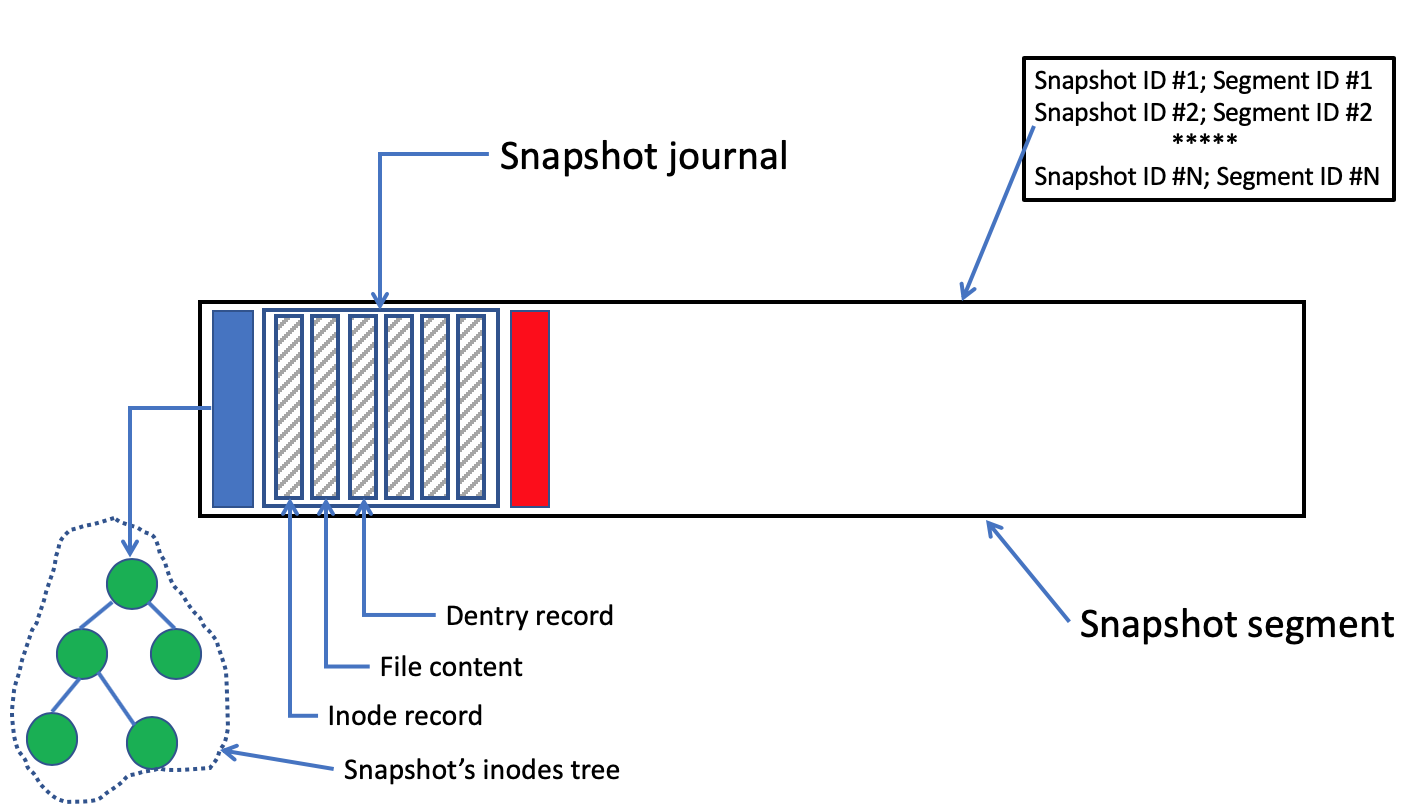}
\caption{Snapshot segment concept.}
\label{fig:fig019}
\end{figure}

SSDFS file system introduces the concept of specialized snapshot segment (Fig. \ref{fig:fig019}). The snapshot segment is dedicated to the long-term storing of a checkpoint's state or some portion of file system's namespace (that could play the role of parent snapshot for a version of file system's state). Generally speaking, the snapshot segment contains a sequence of logs that keep a special journal (Fig. \ref{fig:fig019}). This journal is simply aggregation of records that keep the state of files' content or metadata. Moreover, log's header or footer is capable to store the root node of inodes tree that represents the initial state of namespace for the particular snapshot. This root node points out on the index/leaf nodes that will be stored inside of the regular, specialized segments. Generally speaking, only inodes tree needs to be defined explicitly in the snapshot segment because the rest of b-trees (extents, dentries, xattr b-trees) will be defined by means of root node(s) in the particular inode records. Finally, the nodes of these child b-tree will be stored in the regular, specialized segments that are dedicated to keep the index/leaf nodes of b-trees.

\begin{figure}[h]
\centering
 
 \includegraphics[width=0.80\columnwidth,keepaspectratio]{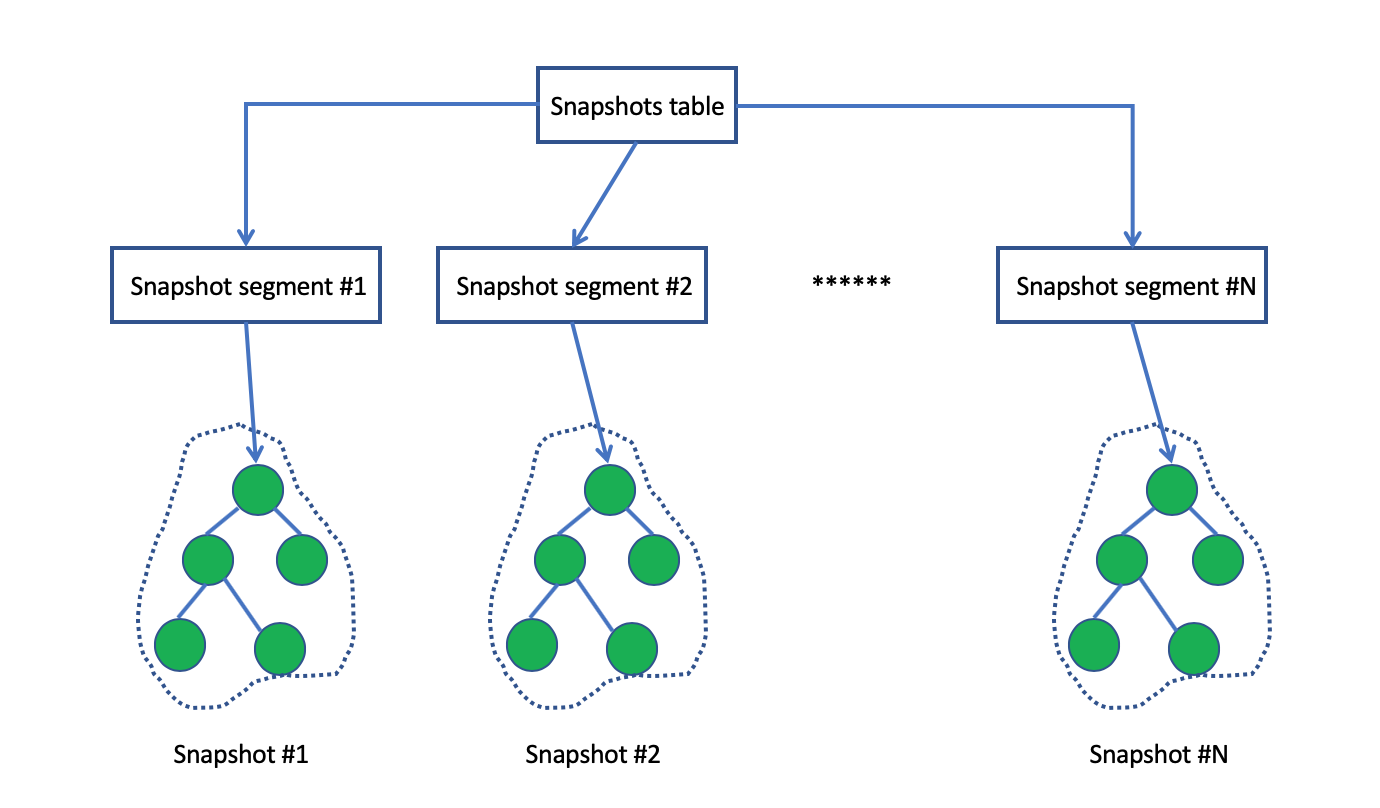}
\caption{Snapshots concept.}
\label{fig:fig020}
\end{figure}

Finally, snapshot table in the superblock segment is capable to associate the snapshot IDs with segment numbers (Fig. \ref{fig:fig020}). Every segment number in this table defines the specialized snapshot segment that stores a snapshot of some state of file system’s volume. Also snapshot segment contains the root node of inodes tree in the log's segment header that plays the role of starting point for evolving the file system's state by means of adding index and leaf nodes into the inodes, extents, dentries, and xattr b-trees. The snapshot state is able to evolve in the case when the file system's volume is mounted with using some snapshot ID.

\textbf{PEB mapping table}. SSDFS file system is based on the concept of logical segment that is the aggregation of LEBs. Moreover, initially, LEB hasn't association with a particular PEB. It means that segment could have the association not for all LEBs or, even, to have no association at all with any PEB (for example, in the case of clean segment). Generally speaking, SSDFS file system needs in special metadata structure (PEB mapping table) that is capable to associate any LEB with any PEB. The PEB mapping table is the crucial metadata structure that has several goals: (1) mapping LEB to PEB, (2) implementation the logical extent concept, (3) implementation the concept of PEB migration, (4) implementation the delayed erase operation by specialized thread, (5) implementation the approach of bad erase block recovering.

\begin{figure}[h]
\centering
 
 \includegraphics[width=0.80\columnwidth,keepaspectratio]{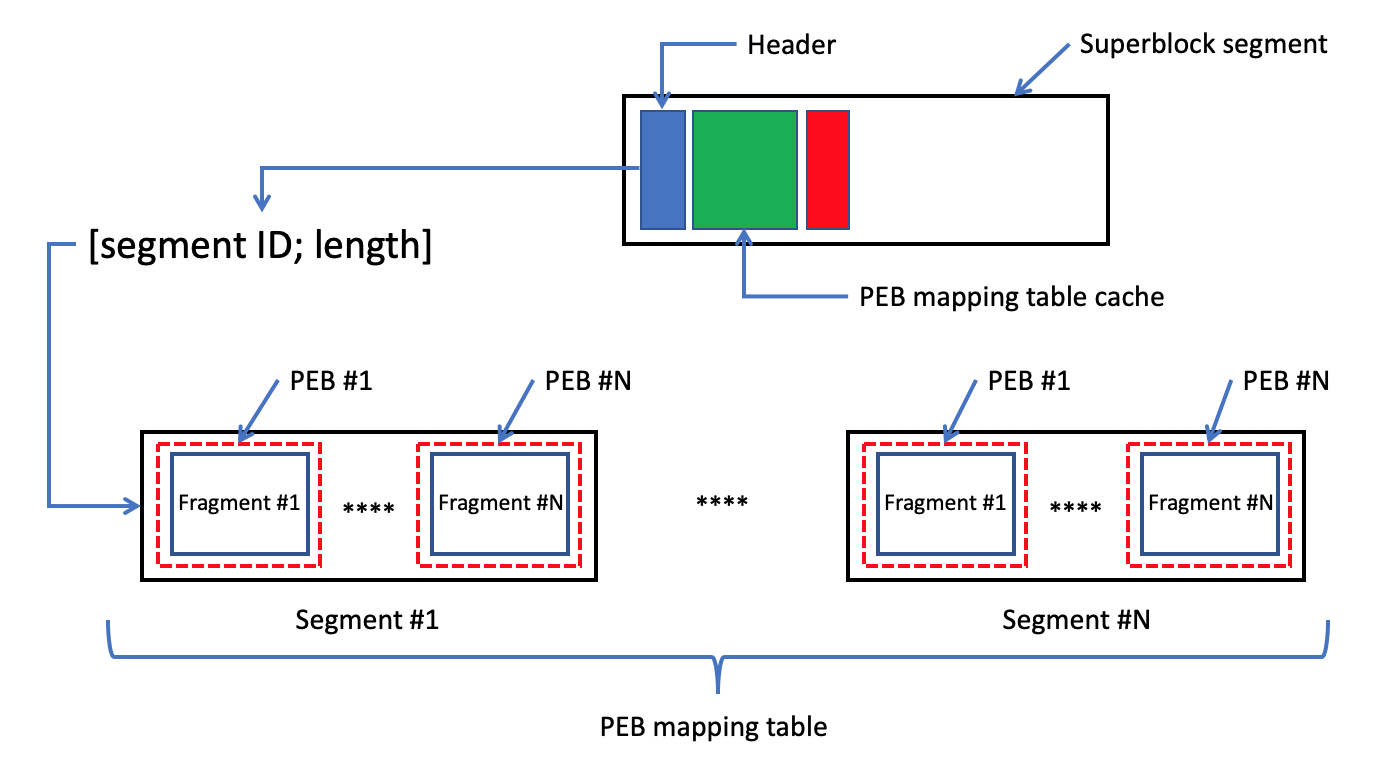}
\caption{PEB mapping table architecture.}
\label{fig:fig022}
\end{figure}

Generally speaking, PEB mapping table describes the state of all PEBs on a particular SSDFS file system's volume. These descriptors are split on several fragments that are distributed amongst PEBs of specialized segments (Fig. \ref{fig:fig022}). Numbers of these specialized segments are stored in the segment headers of every log and it describes the reserved space of a SSDFS file system's volume for the PEB mapping table. Because SSDFS file system employs the concept of logical segment then the reserved numbers of specialized segments remain the same for the volume's lifetime. But if some PEB achieves the exhausted state then it triggers the migration mechanism of moving the exhausted PEB into another one. Also PEB mapping table is enhanced by special cache is stored in the payload of superblock segment's log (Fig. \ref{fig:fig022}). Generally speaking, the cache stores the copy of records of PEBs' state. The goal of PEB mapping table's cache is to resolve the case when a PEB's descriptor is associated with a LEB of PEB mapping table itself, for example. If unmount operation triggers the flush of PEB mapping table then there are the cases when the PEB mapping table could be modified during the flush operation's activity. As a result, actual PEB's state is stored only into PEB mapping table's cache. Such record is marked as inconsistent and the inconsistency has to be resolved during the next mount operation by means of storing the actual PEB's state into the PEB mapping table by specialized thread. Moreover, the cache plays another very important role. Namely, PEB mapping table's cache is used for conversion the LEB ID into PEB ID for the case of basic metadata structures (PEB mapping table, segment bitmap, for example) before the finishing of PEB mapping table initialization during the mount operation.

\begin{figure}[h]
\centering
 
 \includegraphics[width=0.80\columnwidth,keepaspectratio]{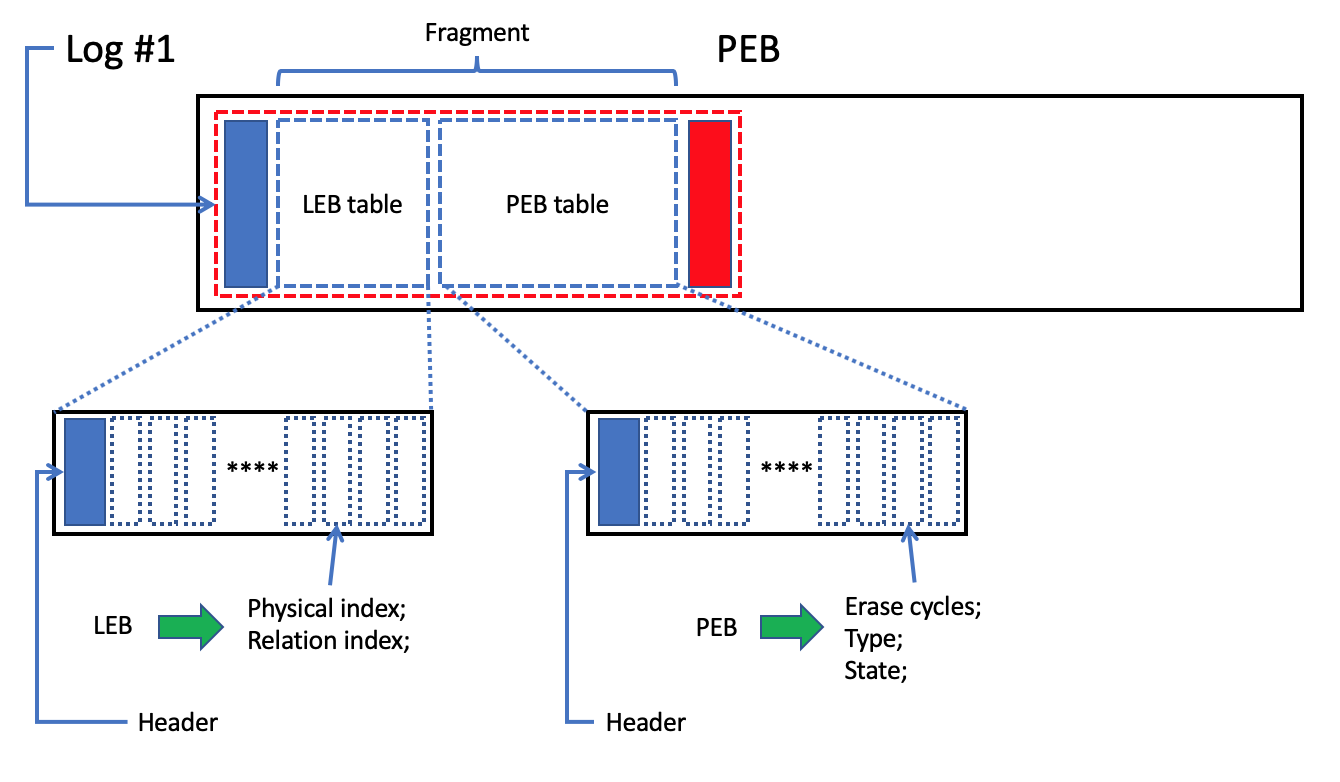}
\caption{PEB mapping table's fragment structure.}
\label{fig:fig023}
\end{figure}

Every fragment of PEB mapping table represents the log's payload in a specialized segment (Fig. \ref{fig:fig023}). Generally speaking, the payload's content is split on: (1) LEB table, and (2) PEB table. The LEB table starts from the header and it contains the array of records are ordered by LEB IDs. It means that LEB ID plays the role of index in the array of records. As a result, the responsibility of LEB table is to define an index inside of PEB table. Moreover, every LEB table's record defines two indexes. The first index (physical index) associates the LEB ID with some PEB ID. Additionally, the second index (relation index) is able to define a PEB ID that plays the role of destination PEB during the migration process from the exhausted PEB into a new one. It is possible to see (Fig. \ref{fig:fig023}) that PEB table starts from the header and it contains the array of PEB's state records is ordered by PEB ID. The most important fields of the PEB's state record are: (1) erase cycles, (2) PEB type, (3) PEB state.

\begin{figure}[h]
\centering
 
 \includegraphics[width=0.80\columnwidth,keepaspectratio]{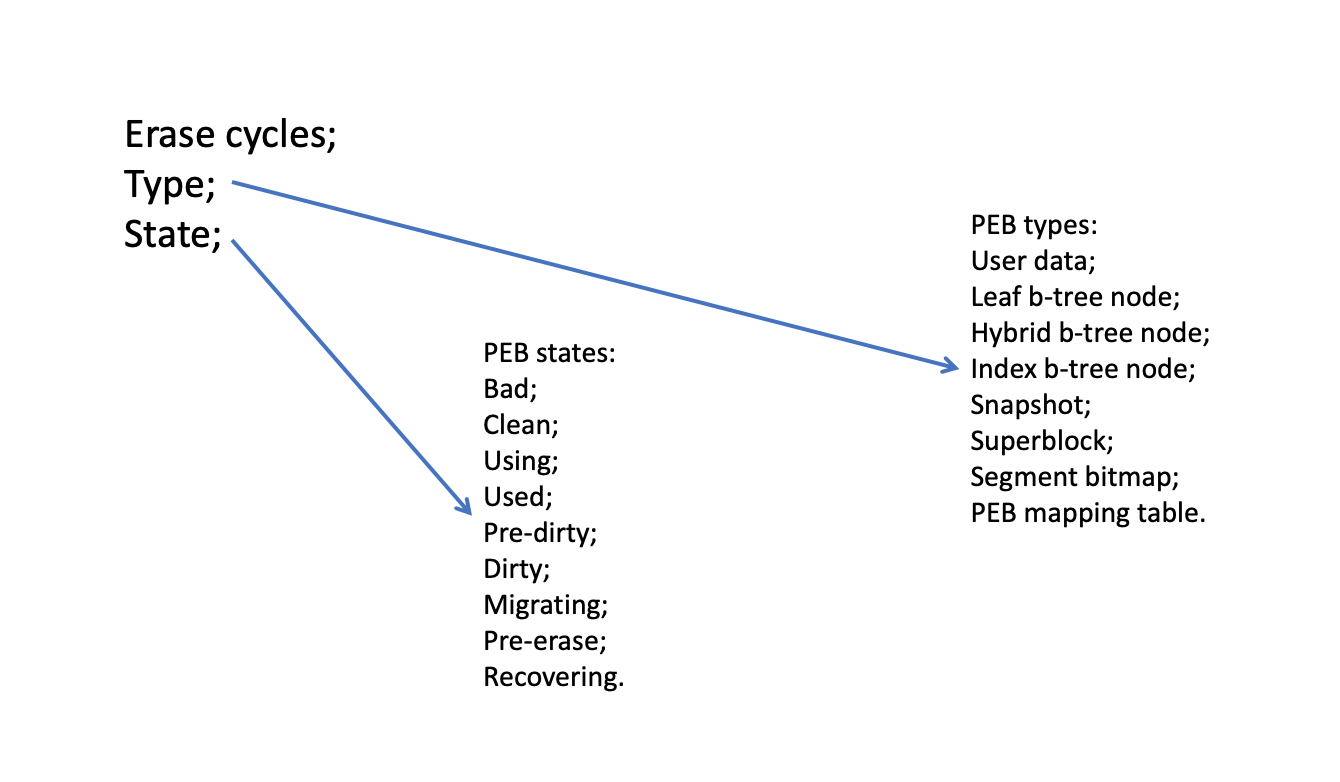}
\caption{Possible PEB's types and states.}
\label{fig:fig024}
\end{figure}

PEB type (Fig. \ref{fig:fig024}) describes possible types of data that PEB could contain: (1) user data, (2) leaf b-tree node, (3) hybrid b-tree node, (4) index b-tree node, (5) snapshot, (6) superblock, (7) segment bitmap, (8) PEB mapping table. PEB state (Fig. \ref{fig:fig024}) describes possible states of PEB during the lifecycle: (1) clean state means that PEB contains only free NAND flash pages are ready for write operations, (2) using state means that PEB could contain valid, invalid, and free pages, (3) used state means that PEB contains only valid pages, (4) pre-dirty state means that PEB contains as valid as invalid pages only, (5) dirty state means that PEB contains only invalid pages, (6) migrating state means that PEB is under migration, (7) pre-erase state means that PEB is added into the queue of PEBs are waiting the erase operation, (8) recovering state means that PEB will be untouched during some amount of time with the goal to recover the ability to fulfill the erase operation, (9) bad state means that PEB is unable to be used for storing the data. Generally speaking, the responsibility of PEB state is to track the passing of PEBs through various phases of their lifetime with the goal to manage the PEBs' pool of the file system's volume efficiently.

\begin{figure}[h]
\centering
 
 \includegraphics[width=0.80\columnwidth,keepaspectratio]{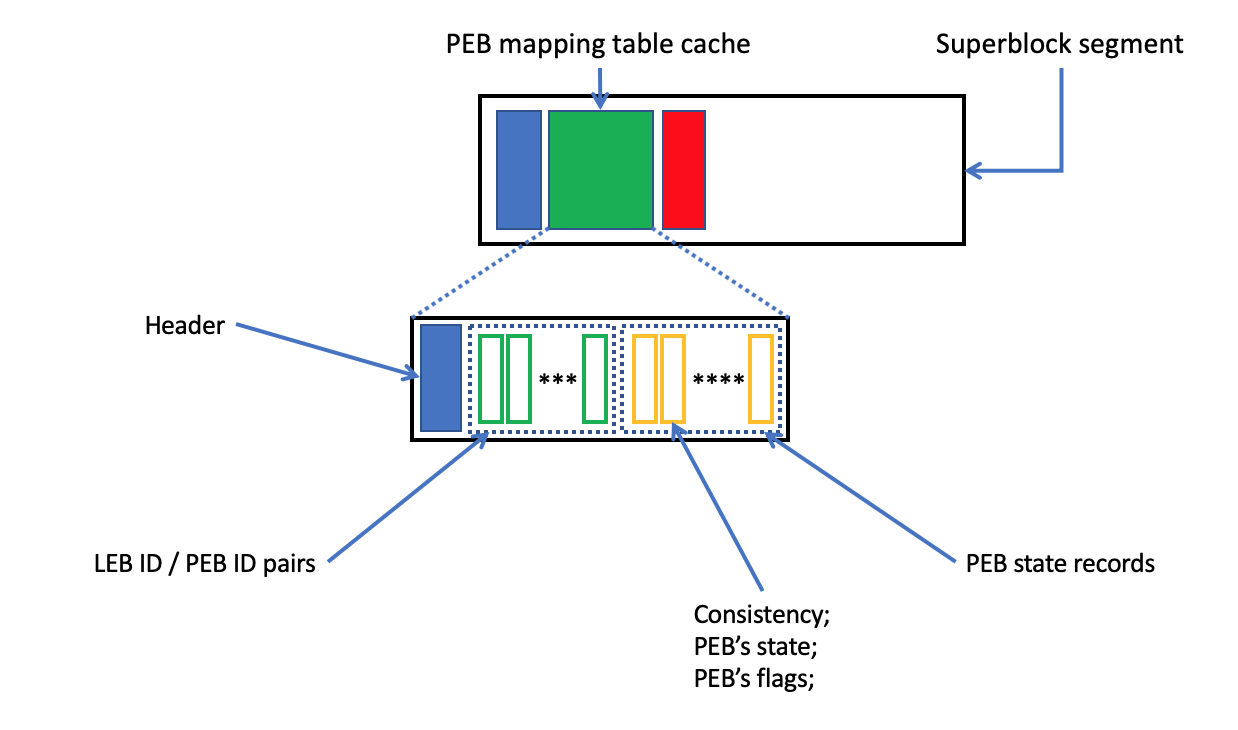}
\caption{PEB mapping table's cache.}
\label{fig:fig025}
\end{figure}

PEB mapping table's cache (Fig. \ref{fig:fig025}) starts from the header that precedes to: (1) LEB ID / PEB ID pairs, (2) PEB state records. The pairs' area associates the LEB IDs with PEB IDs. Additionally, PEB state records' area contains information about the last actual state of PEBs for every record in the pairs' area. It makes sense to point out that the most important fields in PEB state area are: (1) consistency, (2) PEB state, and (3) PEB flags. Generally speaking, the consistency field simply shows that a record in the cache and mapping table is identical or not. If some record in the cache has marked as inconsistent then it means that the PEB mapping table has to be modified with the goal to keep the actual value of the cache. As a result, finally, the value in the table and the cache will be consistent.

The PEB migration approach is the key technique of SSDFS file system. Generally speaking, the migration mechanism implements the logical segment and logical extent concepts with the goal to decrease or completely eliminate the write amplification issue. Moreover, SSDFS file system is widely using the data compression, delta-encoding technique, and small files compaction technique that provides the opportunity to employ the PEB migration mechanism without the necessity to use the additional overprovisioning. PEB mapping table plays the critical role in the implementation of PEB migration technique by means of relation index in the LEB table (Fig. \ref{fig:fig023}). Finally, this index creates the relation between two PEBs that defines the source and destination points for data migration.

\begin{figure}[h]
\centering
 
 \includegraphics[width=0.90\columnwidth,keepaspectratio]{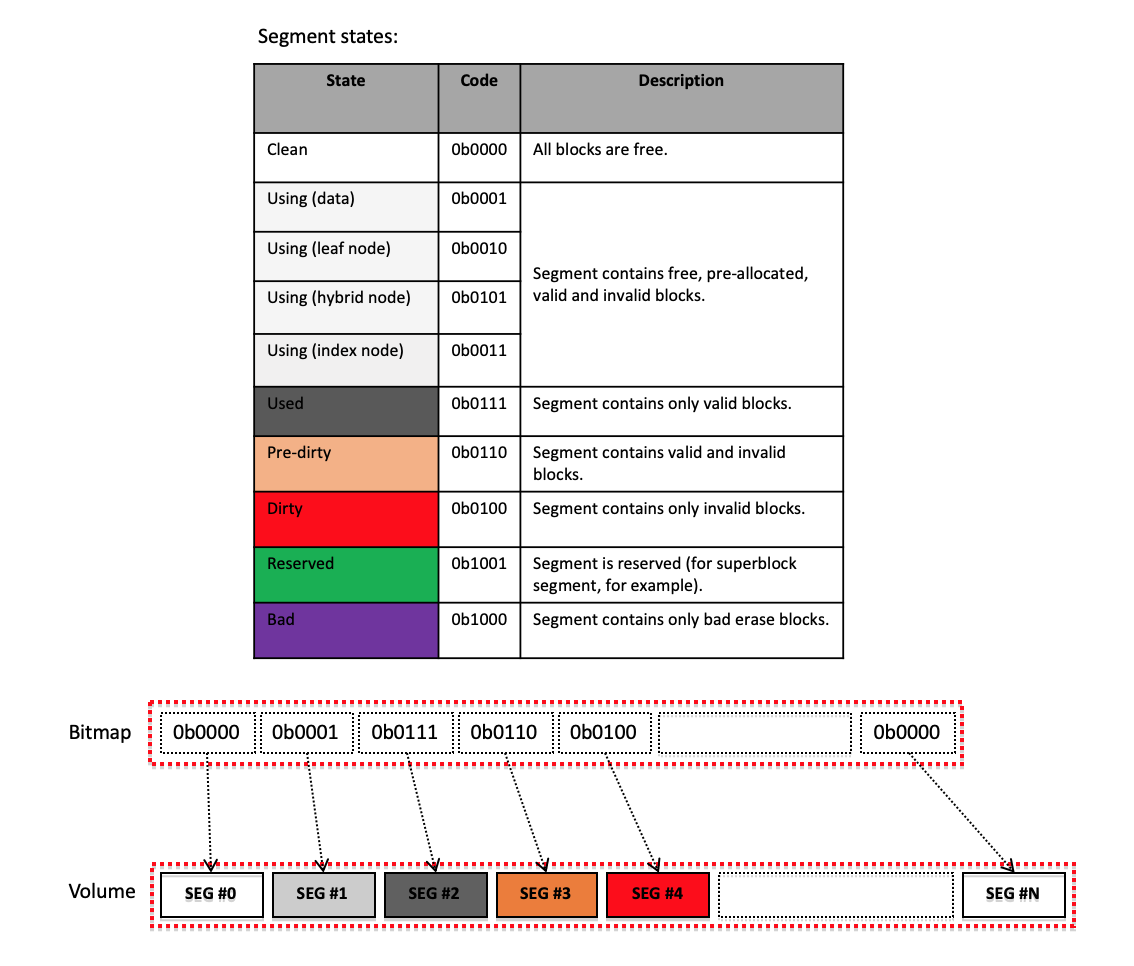}
\caption{Segment bitmap concept.}
\label{fig:fig026}
\end{figure}

\textbf{Segment bitmap} (Fig. \ref{fig:fig026}) is the critical metadata structure in SSDFS file system that implements several goals: (1) searching a candidate for a current segment is capable to store a new data, (2) searching by GC subsystem a most optimal segment (pre-dirty state, for example) with the goal to prepare the segment in background for storing a new data. Segment bitmap (Fig. \ref{fig:fig026}) is able to represent such set of states: (1) clean state means that a segment contains the free logical blocks only, (2) using state means that a segment could contain valid, invalid, and free logical blocks, (3) used state means that a segment contains the valid logical blocks only, (4) pre-dirty state means that a segment contains valid and invalid logical blocks, (5) dirty state means that a segment contains only invalid blocks, (6) reserved state is used for reservation the segment numbers for some metadata structures (for example, for the case of superblock segment), (7) bad state means that a segment is excluded from the usage because a file system's volume hasn't enough valid erase blocks (PEBs).

Generally speaking, PEB migration scheme implies that segments are able to migrate from one state to another one without the explicit using of GC subsystem. For example, if some segment receives enough truncate operations (data invalidation) then the segment could change the used state on pre-dirty state. Additionally, the segment is able to migrate from pre-dirty into using state by means of PEBs migration in the case of receiving enough data update requests. As a result, the segment in using state could be selected like the current segment without any GC-related activity. However, a segment is able to stick in pre-dirty state in the case of absence the update requests. Finally, such situation can be resolved by GC subsystem by means of migration in the background of pre-dirty segments into the using state if a SSDFS file system's volume hasn't enough segments in the clean or using state.

\begin{figure}[h]
\centering
 
 \includegraphics[width=0.80\columnwidth,keepaspectratio]{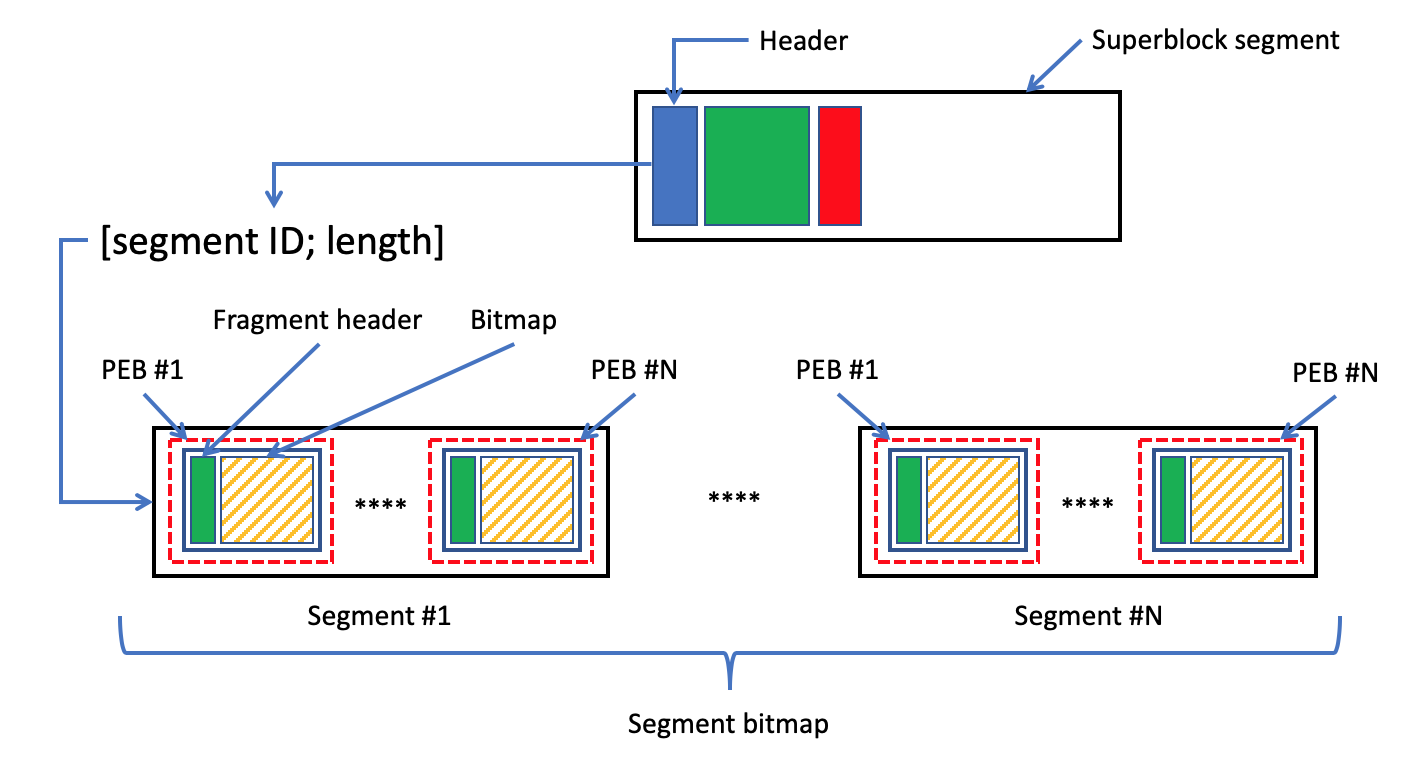}
\caption{Segment bitmap architecture.}
\label{fig:fig027}
\end{figure}

Segment bitmap is implemented like the bitmap metadata structure that is split on several fragments (Fig. \ref{fig:fig027}). Every fragment is stored into a log of specialized PEB. As a result, the full size of segment bitmap and PEB's capacity define the number of fragments. The mkfs utility reserves the necessary number of segments for storing the segment bitmap's fragments during a SSDFS file system's volume creation. Finally, the numbers of reserved segments are stored into the segment headers of every log on the volume. The segment bitmap "lives" in the same set of reserved segments during the whole lifetime of the volume. However, the update operations of segment bitmap could trigger the PEBs migration in the case of exhaustion of any PEB is used for keeping the segment bitmap's content.

\begin{figure}[h]
\centering
 
 \includegraphics[width=0.80\columnwidth,keepaspectratio]{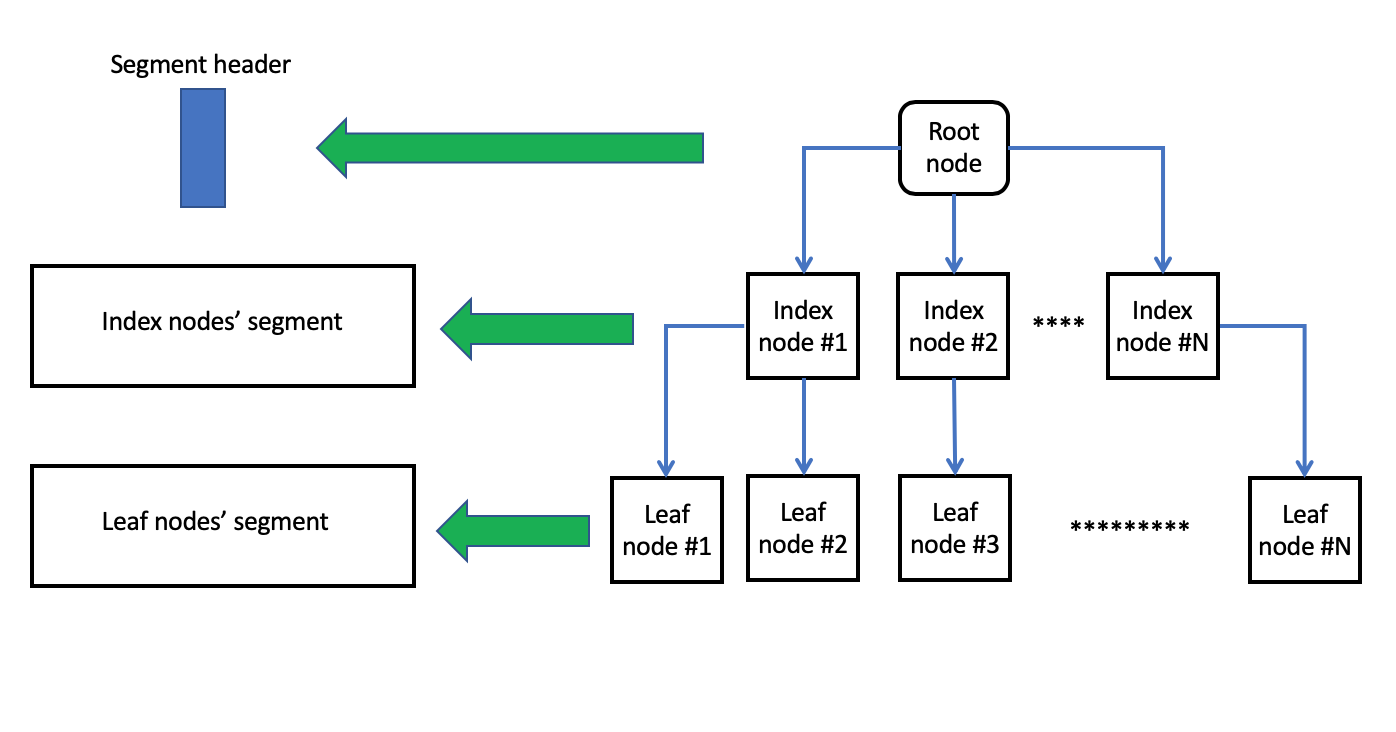}
\caption{B-tree concept.}
\label{fig:fig028}
\end{figure}

\textbf{B-trees} (Fig. \ref{fig:fig028}) represent efficient and compact metadata structure for storing and representing metadata on a file system's volume. Also b-tree is able to provide the efficient and the fast way of searching items. Moreover, another important feature of the b-tree is the compact representation of sparse metadata and easy increasing the reserved capacity of metadata space. Especially, such feature could be very important for the case of NAND flash because it could be expensive and useless way, for example, to store a huge and flat array of inodes in the case of absence of any file in the namespace of a file system. However, usually, b-tree is treated like not suitable solution for the case of flash-oriented file systems because of wandering tree and excessive write amplification issues. But SSDFS file system is based on logical segment concept, logical extent concept, and delta-encoding technique that completely exclude the wandering tree issue. Also these concepts are the basis for decreasing (or complete elimination) the write amplification issue. Moreover, b-tree metadata structure provides the way not to keep an unnecessary reserve of metadata space on the volume. As a result, it means the exclusion of management operations of reserved metadata space (moving from a PEB to another one) with the goal to support it in the valid state. Generally speaking, it is the way to decrease the amount of PEBs' erase and write operations.

\begin{figure}[h]
\centering
 
 \includegraphics[width=0.80\columnwidth,keepaspectratio]{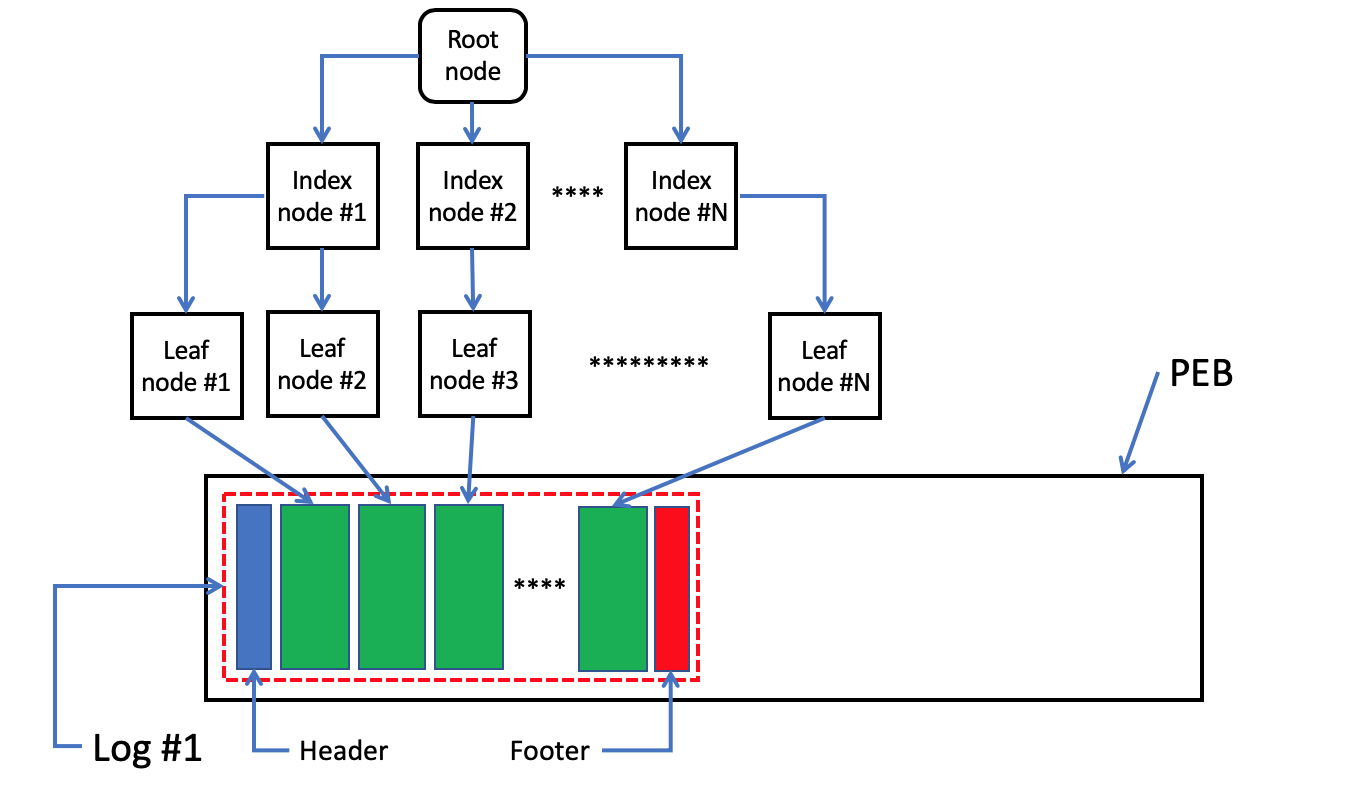}
\caption{B-tree segment type.}
\label{fig:fig029}
\end{figure}

Root node of the key b-trees (inodes b-tree, shared extents b-tree, shared dictionary b-tree) is stored into segment header or/and log footer (Fig. \ref{fig:fig028}). However, oppositely, root node of extents b-tree, dentries b-tree, and xattr b-tree is stored into an inode record. Generally speaking, if the root node is capable to store some amount of metadata records or b-tree hasn't any metadata records at all then a SSDFS file system's volume will not have any node on the volume. SSDFS file system is using specialized types of segments. It means that b-tree's leaf nodes will be stored in a segment is dedicated and reserved for the leaf nodes but index nodes will be stored into a segment is containing the index nodes only (Fig. \ref{fig:fig029}). Moreover, nodes of different b-trees can be stored in the same segment. Generally speaking, it is expecting that workload type of b-tree's nodes of the same type (leaf nodes, for example) could be the same and it is the basis for grouping the nodes of different b-trees into one segment. Also it could create more compact representation of the b-trees on the volume. Every b-tree's node could use one or several logical blocks. As a result, any log of segment with b-trees' nodes (Fig. \ref{fig:fig029}) contains: (1) segment header, (2) log footer, and (3) payload that contains b-trees' nodes. Finally, b-tree's nodes' content is distributed among the main, diff updates, and journal areas of the log.

\begin{figure}[h]
\centering
 
 \includegraphics[width=0.80\columnwidth,keepaspectratio]{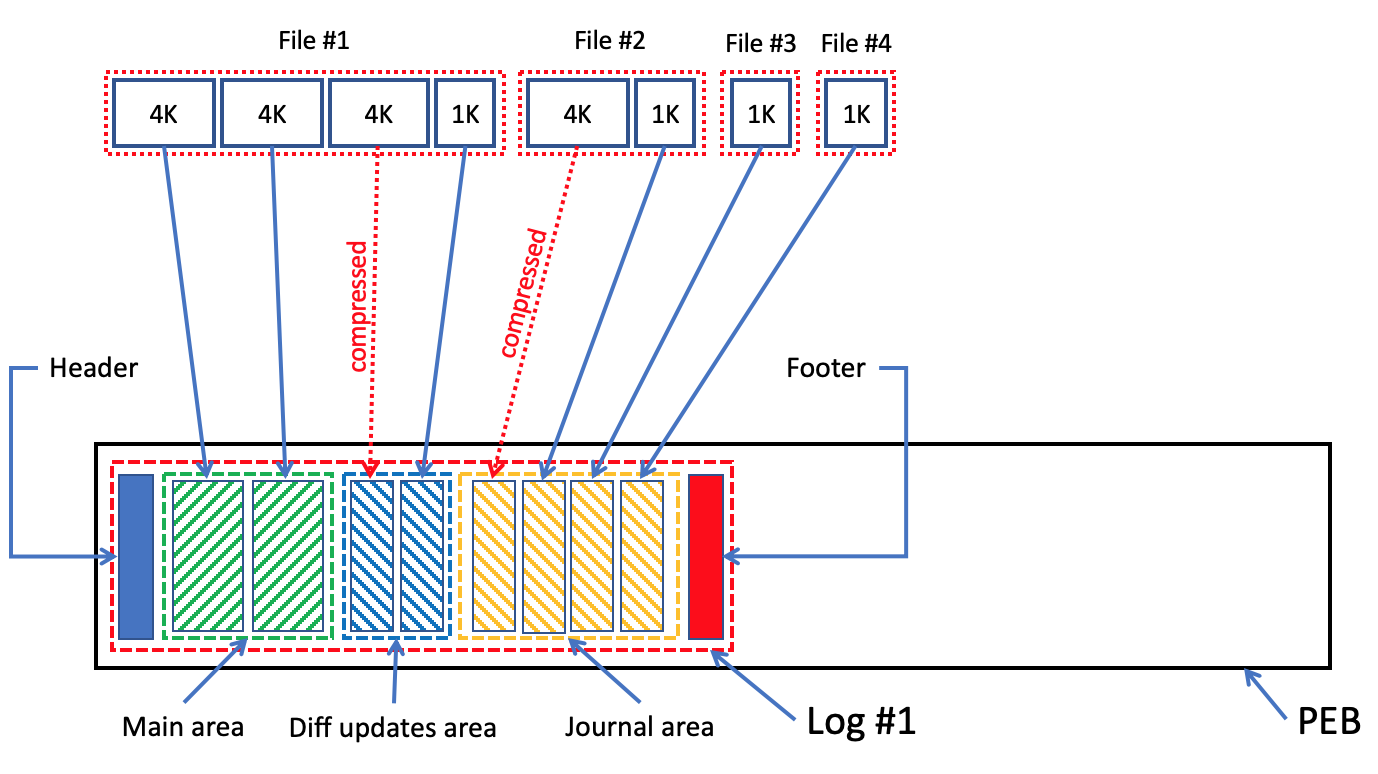}
\caption{User data segment type.}
\label{fig:fig030}
\end{figure}

\textbf{User data segment}. SSDFS file system aggregates user data inside of segments are dedicated to user data's type (Fig. \ref{fig:fig030}). It needs to point out that the user data could live under different workloads and to have the various features. Theoretically, it is possible to introduce the various types of segments for the user data. However, nevertheless, it is used only one type of segment for the user data.

Generally speaking, the main, diff updates, and journal areas are the key technique of user data's management on a SSDFS file system's volume (Fig. \ref{fig:fig030}). The main area plays the role of cold data because it is dedicated to store the plain, full logical blocks. Moreover, if a logical block is stored into the main area then all subsequent updates of this logical block are directed to the diff updates or journal areas. As a result, this is the key technique to achieve the cold nature of data into the main area of log. Finally, the data in main area can be moved from the initial PEB into a new one by means of PEB migration scheme (in the case of update operation is directed to the exhausted PEB) or by GC subsystem (in the case of absence of update operation and exhausted state of the PEB). The moving operation of data in the main area could be accompanied by applying the associated updates are stored into the diff updates or/and journal areas.

The diff updates area is able to store into one NAND flash page the compressed blocks, delta-encoded portions of data, and tail of the same file (Fig. \ref{fig:fig030}). It is possible to expect that frequency of updates for different portions of the same file could be lower than frequency of updates for different files. As a result, diff updates area is treated like area with warm data. It is important to point out that data in the diff updates area could be invalidated partially or completely by update operations. Also, the valid data of diff updates area is used during migration of the main area's content with the goal to prepare the actual state of logical block(s). Generally speaking, it is not necessary to reserve any space for the diff updates area's content in a destination PEB during the migration operation.

The responsibility of journal area is to gather into one NAND flash page the small files, the tails of different files, compressed updates or delta-encoded data of different files (Fig. \ref{fig:fig030}). Generally speaking, the different files are able to grow or to be updated with higher frequency than content of one file. As a result, gathering the content of different files into the one NAND flash page increases the probability of updates in journal area. Finally, the journal area is treated like area with hot data. Moreover, the amount of updates in journal area is expected to be high enough to achieve the complete invalidation of the journal area by means of natural migration of data between logs and PEBs (by means of migration scheme) without any GC subsystem's activity.

However, some small files could be completely "cold" or to be updated rarely. As a result, it means the necessity to employ the logic of GC subsystem or PEB migration scheme to process some data in the journal area. However, the scheme of compaction several small files into one NAND flash page decreases the write amplification issue by virtue of the opportunity to move the several small files into one page by single copy operation.

\begin{figure}[h]
\centering
 
 \includegraphics[width=0.80\columnwidth,keepaspectratio]{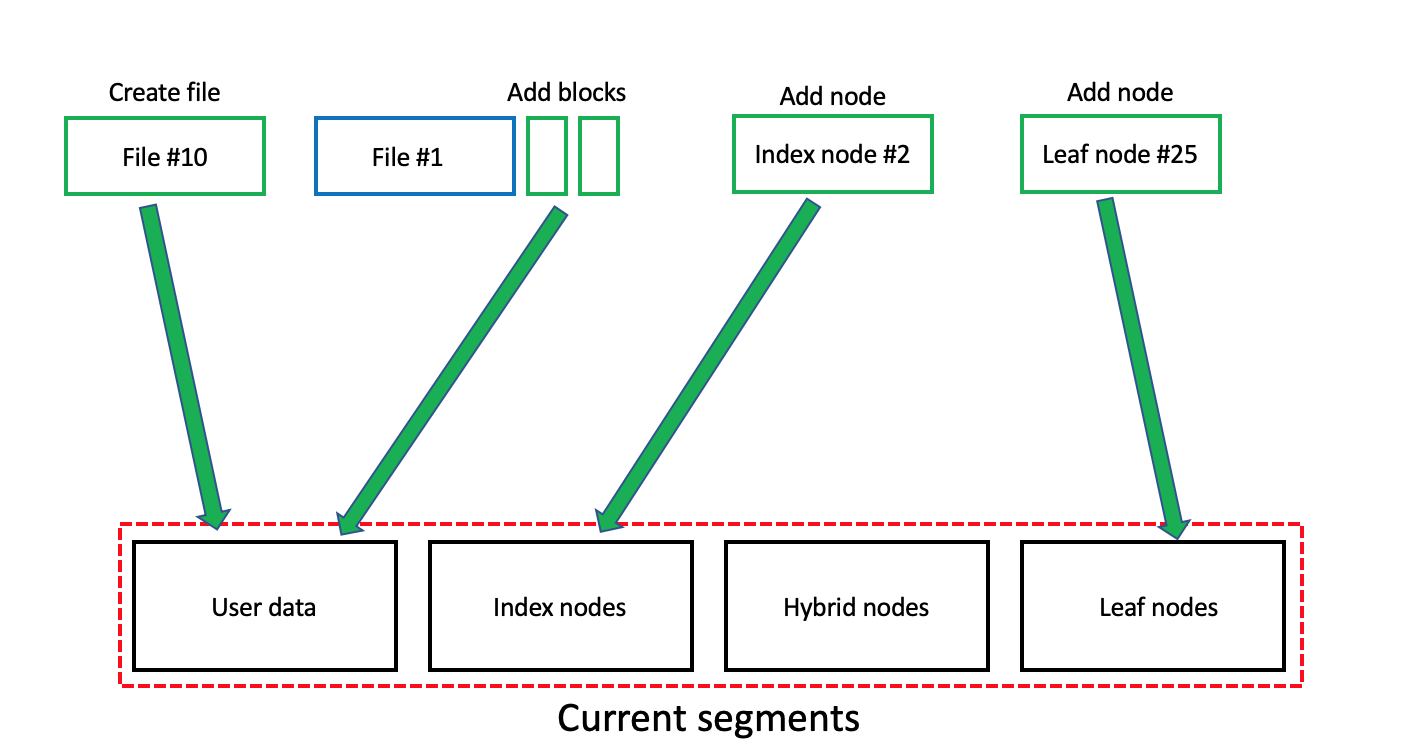}
\caption{Current segment concept.}
\label{fig:fig031}
\end{figure}

\textbf{Current segment}. SSDFS file system employs the concept of current segments (Fig. \ref{fig:fig031}). Generally speaking, if it is necessary to add some new data on the volume (new file, new logical blocks of the existing file, new b-tree's node) then it needs to use a segment that has the free logical blocks. Only segment in clean or using state can be used for adding a new data. As a result, SSDFS file system's driver has the set of current segments (user data, index node, hybrid node, leaf node) because of the policy of grouping different types of data into different type of segments (Fig. \ref{fig:fig031}). The segment can be used as current until the complete exhaustion of the free logical blocks' pool in the particular segment (used or pre-dirty state). In the case of absence the free logical blocks in the current segment, the file system driver tries to find in the segment bitmap a new segment in the clean or using state. If the driver is unable to find any clean or using segment then it will trigger the GC logic for searching the segments in pre-dirty or dirty state with the goal to transform the pre-dirty segment into using state and dirty segment into clean state. Finally, if no pre-dirty or dirty segment can be processed or transformed into clean or using state then the driver will need to inform the user about absence of free space on the volume. Generally speaking, GC subsystem has to track the state of segment bitmap in the background and to prepare enough number of clean or using segments (in the case of enough free space on the volume). However, GC subsystem's activity in the background should not affect the whole performance of the file system driver and not to increase the write amplification issue.

From one point of view, the several types of current segments create the several independent threads of processing a new data. Additionally, the segment object in file system driver is implemented in such way that every segment has a queue for requests with the new data. Generally speaking, it means that a thread adds some new data into a current segment by means of simple insertion of the request into the tail of queue. The rest processing of the requests in the queue will be executed in the background by specialized PEBs' flush threads. Finally, the whole architecture creates the fast, efficient, simple, and multi-threaded mechanism of the new data processing. Of course, if a volume was mounted in synchronous mode then a thread needs to wait the finishing of request processing that was added into the queue. And it means the inevitable degradation of file system performance in one thread. However, if the several threads add the requests into the queue then the whole file system's performance could not degrade dramatically even for the case of synchronous mode. Finally, it is possible to state that SSDFS file system has flexible and efficient subsystem of the current segments is capable to provide the good performance of data processing.

\subsection{B-tree Architecture}

\begin{figure}[h]
\centering
 
 \includegraphics[width=0.90\columnwidth,keepaspectratio]{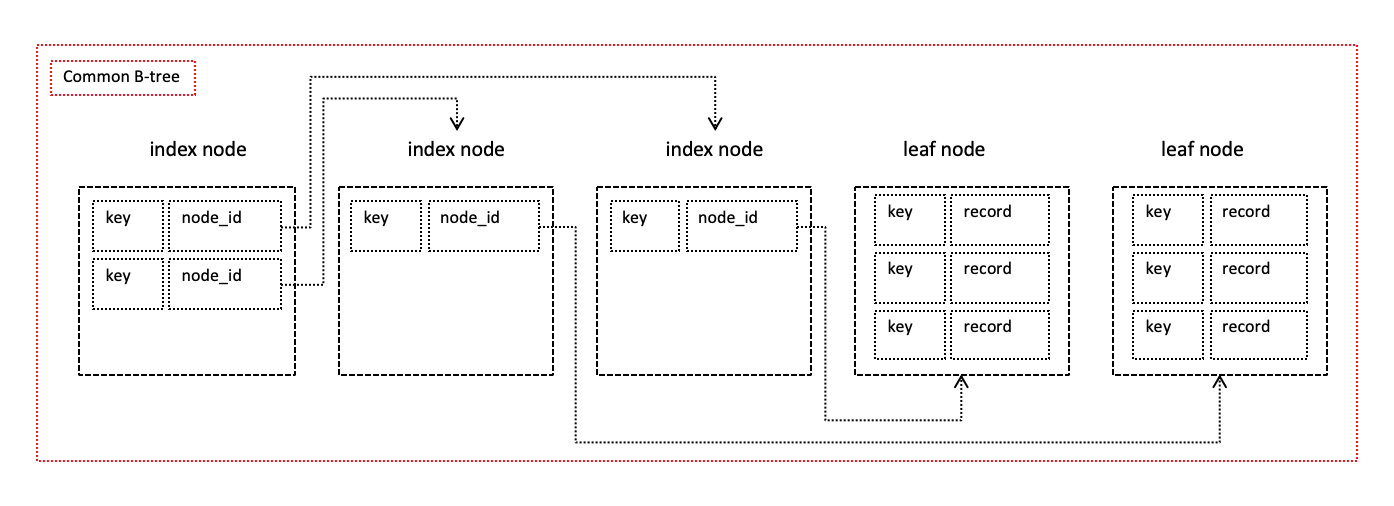}
\caption{Common b-tree architecture.}
\label{fig:fig032}
\end{figure}

\textbf{B-tree} is widely used metadata structure has been proven to be efficient for the case of various file systems. For example, XFS, btrfs, jfs, reiserfs, HFS+ are using different b-tree's implementation. Usually, b-tree provides the opportunity to have multiple child nodes for the same parent node. A regular b-tree (Fig. \ref{fig:fig032}) contains a root node, index nodes, and leaf nodes. Generally speaking, the key advantage of any b-tree is a capability to store data in the form of nodes with the goal to provide an efficient data extraction in the case of block-oriented storage device. Because any b-tree's node is capable to include multiple data items and, as a result, to decrease the number of I/O operations for the case of searching any item in the b-tree. Mostly, b-tree is used for storing and representation various file system's metadata types (for example, inodes or extents). Usually, root node (Fig. \ref{fig:fig032}) represents a starting point of a b-tree. It keeps index records that contain a key and a pointer on a child node. Any index node contains the same index records because the responsibility of the index node is to provide the way to find a leaf node with the data records. The data record is the pair of a key with some associated value (for example, extent or inode). A hash, an ID or any other value could play a role of the key that would be a field of the data record itself. Moreover, key values are the basis for data records ordering in the b-tree. Generally speaking, any lookup operation starts from the root node, passing by through the index levels, and finding some data record on the key basis in the found leaf node.

\textbf{Why b-tree for LFS file system?} Usually, b-tree is considered like not very good choice for the case of flash-oriented and flash-friendly file systems by virtue of wandering tree issue and high value of write amplification. However, b-tree architecture implements very important advantages: (1) efficient search mechanism, (2) compact storage of sparse data, (3) flexible technique of capacity increasing and shrinking. Generally speaking, the key reason of possible b-tree's inefficiency for the case of LFS file system is the Copy-On-Write (COW) policy. The COW policy means the necessity to store an updated logical block into some new and free position on the file system's volume. As a result, it initiates a lot of metadata updates that, finally, results in significant increasing of write amplification in the case of b-tree using.

However, SSDFS file system is using the logical segment, logical extent concepts, and the PEBs migration scheme. Generally speaking, these techniques provide the opportunity to exclude completely the wandering tree issue and to decrease significantly the write amplification. SSDFS file system introduces the technique of storing the data on the basis of logical extent that describes this data's position by means of segment ID and logical block number. Finally, PEBs migration technique guarantee that data will be described by the same logical extent until the direct change of segment ID or logical block number. As a result, it means that logical extent will be the same if data is sitting in the same logical segment. The responsibility of PEBs migration technique is to implement the continuous migration of data between PEBs inside of the logical segment for the case of data updates and GC activity. Generally speaking, SSDFS file system's internal techniques guarantee that COW policy will not update the content of b-tree. But content of b-tree will be updated only by regular operations of end-user with the file system.

SSDFS file system uses b-tree architecture for metadata representation (for example, inodes tree, extents tree, dentries tree, xattr tree) because it provides the compact way of reserving the metadata space without the necessity to use the excessive overprovisioning of metadata reservation (for example, in the case of plain table or array). The excessive overprovisioning of metadata reservation dictates the necessity to support the reserved space in valid state by means of migration among PEBs because of continuous increasing NAND unrecoverable bit error rate (UBER) for stored data. As a result, such migration activity (on SSD or file system side) increases the write amplification issue.

Moreover, b-tree provides the efficient technique of items lookup, especially, for the case of aged or sparse b-tree that is capable to contain the mixture of used and deleted (or freed) items. Such b-tree's feature could be very useful for the case of extent invalidation, for example. Also SSDFS file system aggregates the b-tree's root node in the superblock (for example, inodes tree case) or in the inode (for example, extents tree case). As a result, it means that an empty b-tree will contain only the root node without the necessity to reserve any b-tree's node on the file system's volume. Moreover, if a b-tree needs to contain only several items (two items, for example) then the root node's space can be used to store these items inline without the necessity to create the full-featured b-tree's node.

One of the fundamental mechanism of SSDFS file system is the current segments approach. This approach is used for aggregation of data living under similar workloads in the current segment of some type. For example, there are current segments for different b-tree's node types (index, hybrid, leaf nodes). Generally speaking, it means that the current segment for leaf nodes aggregates the leaf nodes of different b-trees (inodes, extents, dentries b-trees, for example). Every current segment allocates the logical blocks till the complete exhaustion of this segment. Finally, SSDFS file system driver needs to allocate a new current segment in the case of complete exhaustion of the free space of previous current segment. Moreover, SSDFS file system driver uses the compression and the delta-encoding techniques that is the way to achieve the compact representation of b-tree's nodes in the PEB's space.

As a result, SSDFS uses b-trees with the goal to achieve the compact representation of metadata, the flexible way to expend or to shrink the b-tree's space capacity, and the efficient mechanism of items' lookup.

\textbf{Hybrid b-tree architecture}. Regular b-tree contains two types of nodes: index and leaf ones. The index node keeps the pointers on other nodes with the goal to implement mechanism of fast lookup operation in the b-tree. Oppositely, the leaf node keeps items of real data that are stored in the b-tree. Moreover, a node creation means the reservation of 4-64 KB of file system's volume's space. However, usually, b-tree is metadata structure that is not receiving a lot of items at once. As a result, it means that growing b-tree could contain some number of empty or semi-empty index nodes. These index nodes could be empty or semi-empty significant amount of time that results in increasing of number of I/O operations during the search in b-tree and the flush of the b-tree. Generally speaking, this side effect could increase the write amplification for the case of flash-friendly file systems.

\begin{figure}[h]
\centering
 
 \includegraphics[width=0.90\columnwidth,keepaspectratio]{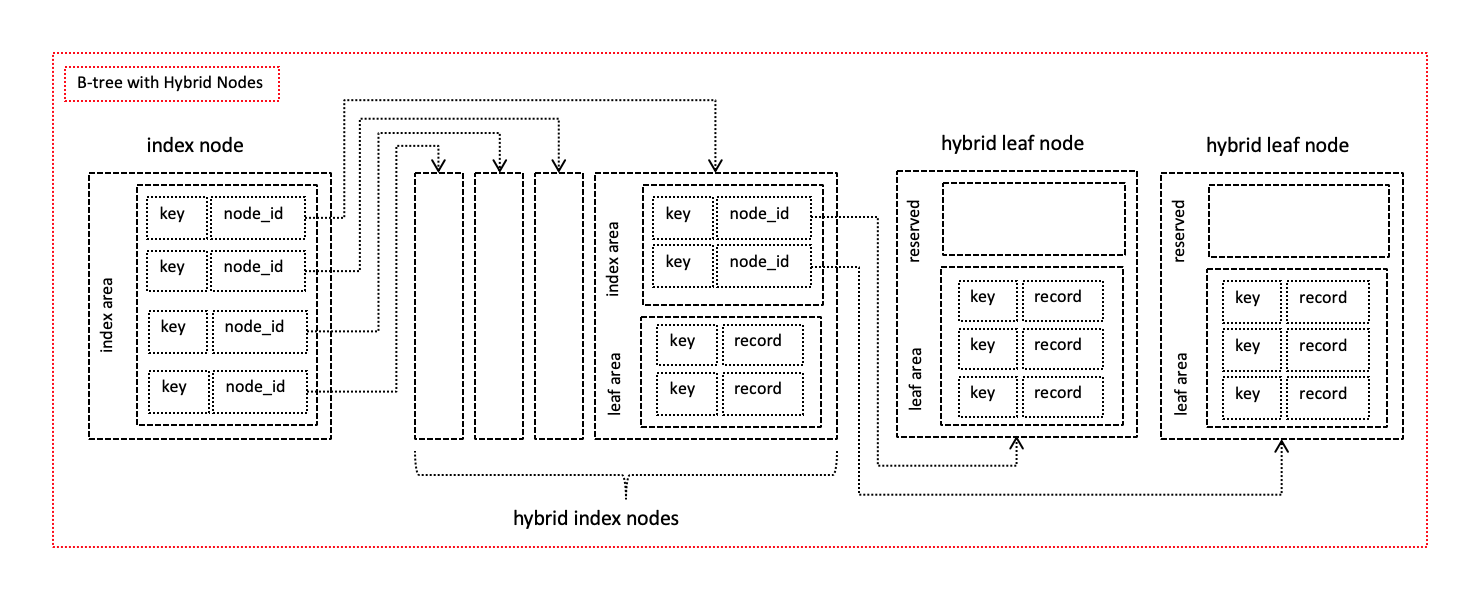}
\caption{B-tree architecture with hybrid nodes.}
\label{fig:fig033}
\end{figure}

\begin{figure}[h]
\centering
 
 \includegraphics[width=0.90\columnwidth,keepaspectratio]{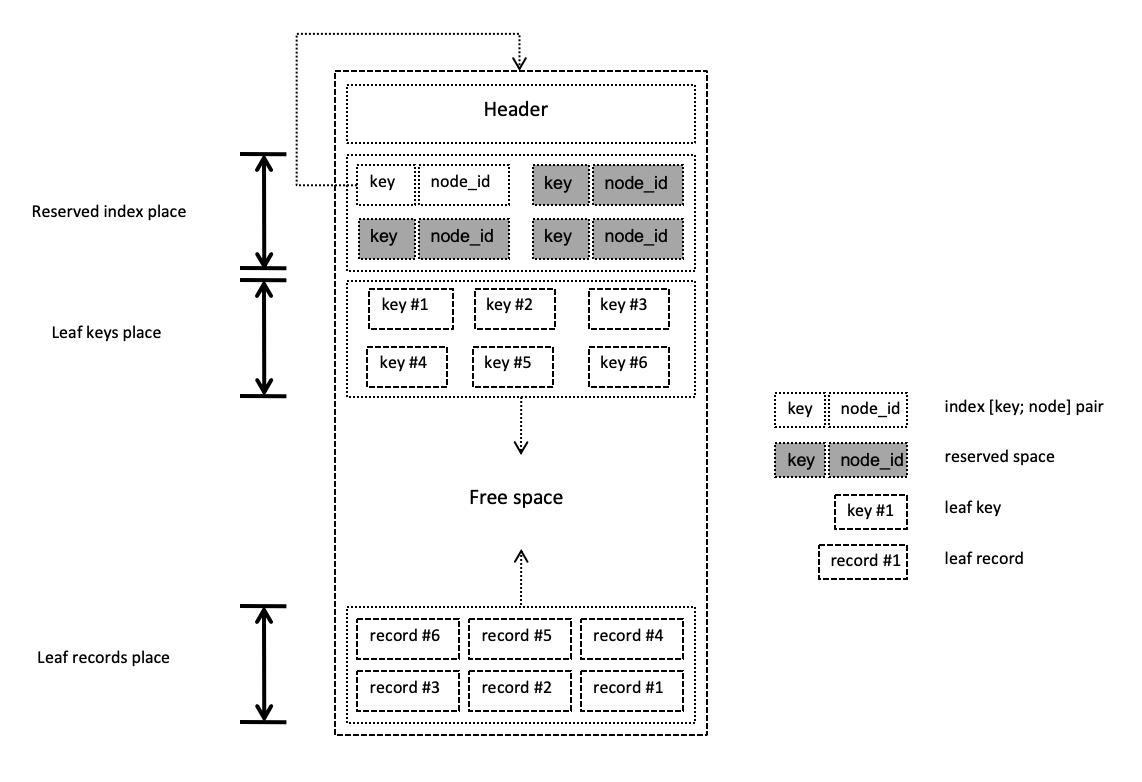}
\caption{Hybrid node architecture.}
\label{fig:fig034}
\end{figure}

\begin{figure}[h]
\centering
 
 \includegraphics[width=0.80\columnwidth,keepaspectratio]{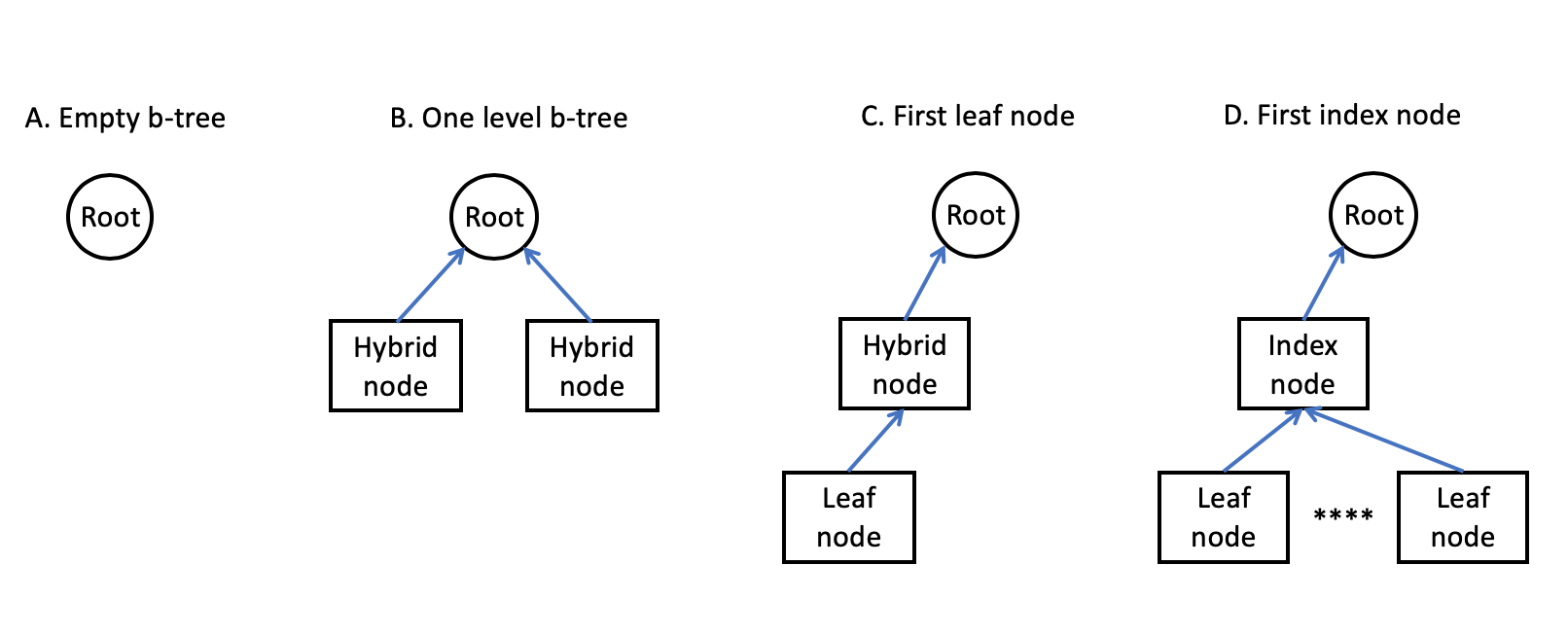}
\caption{Hybrid b-tree evolution.}
\label{fig:fig035}
\end{figure}

SSDFS file system uses a hybrid b-tree architecture (Fig. \ref{fig:fig033}) with the goal to eliminate the index nodes' side effect. The hybrid b-tree operates by three node types: (1) index node, (2) hybrid node, (3) leaf node. Generally speaking, the peculiarity of hybrid node (Fig. \ref{fig:fig034}) is the mixture as index as data records into one node. Hybrid b-tree starts with root node (Fig. \ref{fig:fig035} case A) that is capable to keep the two index records or two data records inline (if size of data record is equal or lesser than size of index record). If the b-tree needs to contain more than two items then it should be added the first hybrid node into the b-tree. The first level of b-tree is able to contain only two nodes (Fig. \ref{fig:fig035} case B) because the root node is capable to store only two index records. Generally speaking, the initial goal of hybrid node is to store the data records in the presence of reserved index area (Fig. \ref{fig:fig034}).

\begin{figure}[h]
\centering
 
 \includegraphics[width=0.80\columnwidth,keepaspectratio]{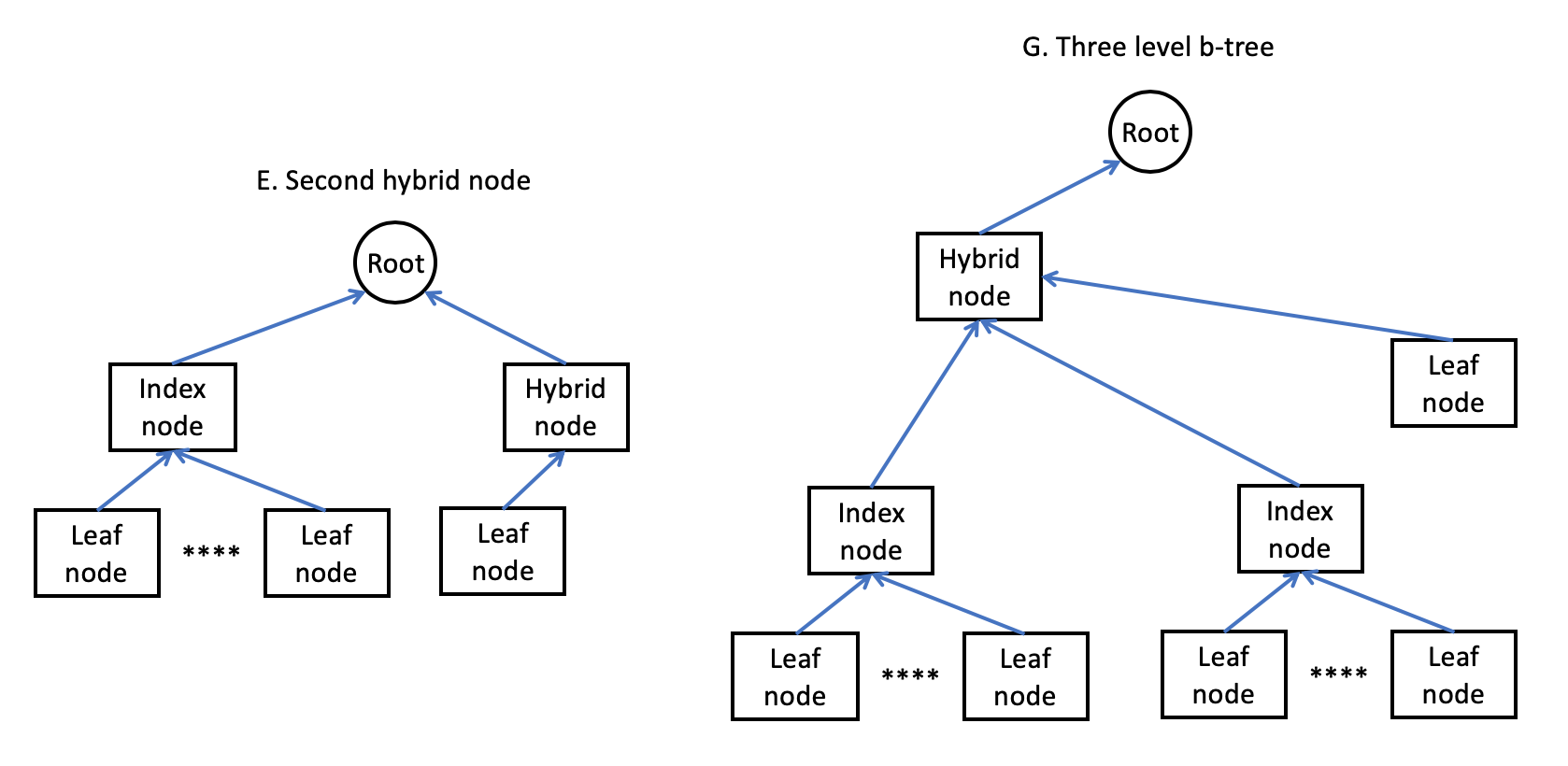}
\caption{Hybrid b-tree evolution.}
\label{fig:fig036}
\end{figure}

The exhaustion of the data area's space of the first hybrid node triggers addition of the second hybrid node on the first level of the b-tree (Fig. \ref{fig:fig035} case B). If both hybrid nodes of the first level are completely exhausted then it takes place a special transformation of the b-tree (Fig. \ref{fig:fig035} case C). First of all, the left hybrid node is transformed into the leaf node. The portion of data records are moved from the right hybrid node to the newly made leaf node. Also, the index area of the right hybrid node is increased in size after the move operation of data records. The next step is to move the index record of the left node from the root node into the index area of the right node. Finally, root node will keep the index record on the hybrid node but the hybrid node will keep the index record on the leaf node (Fig. \ref{fig:fig035} case C). As a result, hybrid b-tree will contain the completely full leaf node and the hybrid node with the free space for the new data records.

The next step of hybrid b-tree's evolution is the adding data records in the hybrid b-tree node till the complete exhaustion of the data area of the node. Exhaustion of the data area of hybrid node triggers: (1) creation of another leaf node, (2) moving all data records from the hybrid node into the newly created leaf node, (3) adding index record for the leaf node into index area of hybrid node. Usually, data area of hybrid node is lesser than the capacity of a leaf node. It means that data records will be added into the newly created leaf node at first. Finally, the hybrid node will gather the data records in the case of exhaustion of leaf node's capacity. It means that data area of hybrid node plays the role of temporary buffer that aggregates enough data records before a leaf node creation. Generally speaking, this sequence of leaf nodes creation takes place before the exhaustion of index area of hybrid node. Moreover, the index area's exhaustion triggers the increasing of index area's capacity. As a result, it means decreasing the capacity of data area in hybrid node. If the index area of hybrid node extends on the whole node's space then such node becomes to be the index node (Fig. \ref{fig:fig035} case D). Finally, the index node needs to be completely filled by index records.

The exhaustion of index node (Fig. \ref{fig:fig035} case D) implies the addition of right hybrid node on the same level (Fig. \ref{fig:fig036} case E). It means that root node will point out on the newly added hybrid node. As a result, this hybrid node plays the role of initial point for evolving the right branch of the b-tree by means of adding the new leaf nodes. Generally speaking, the evolution implies addition of data records until the state when the right hybrid node will be transformed into the index node. If both index node becomes exhausted by index records then it needs to add the hybrid node. This hybrid node will contain the pointers on exhausted index nodes and root node will point out on the newly added hybrid node (Fig. \ref{fig:fig036} case G). Finally, the hybrid node plays the essential role in the hybrid b-tree's evolution.

\begin{figure}[h]
\centering
 
 \includegraphics[width=0.80\columnwidth,keepaspectratio]{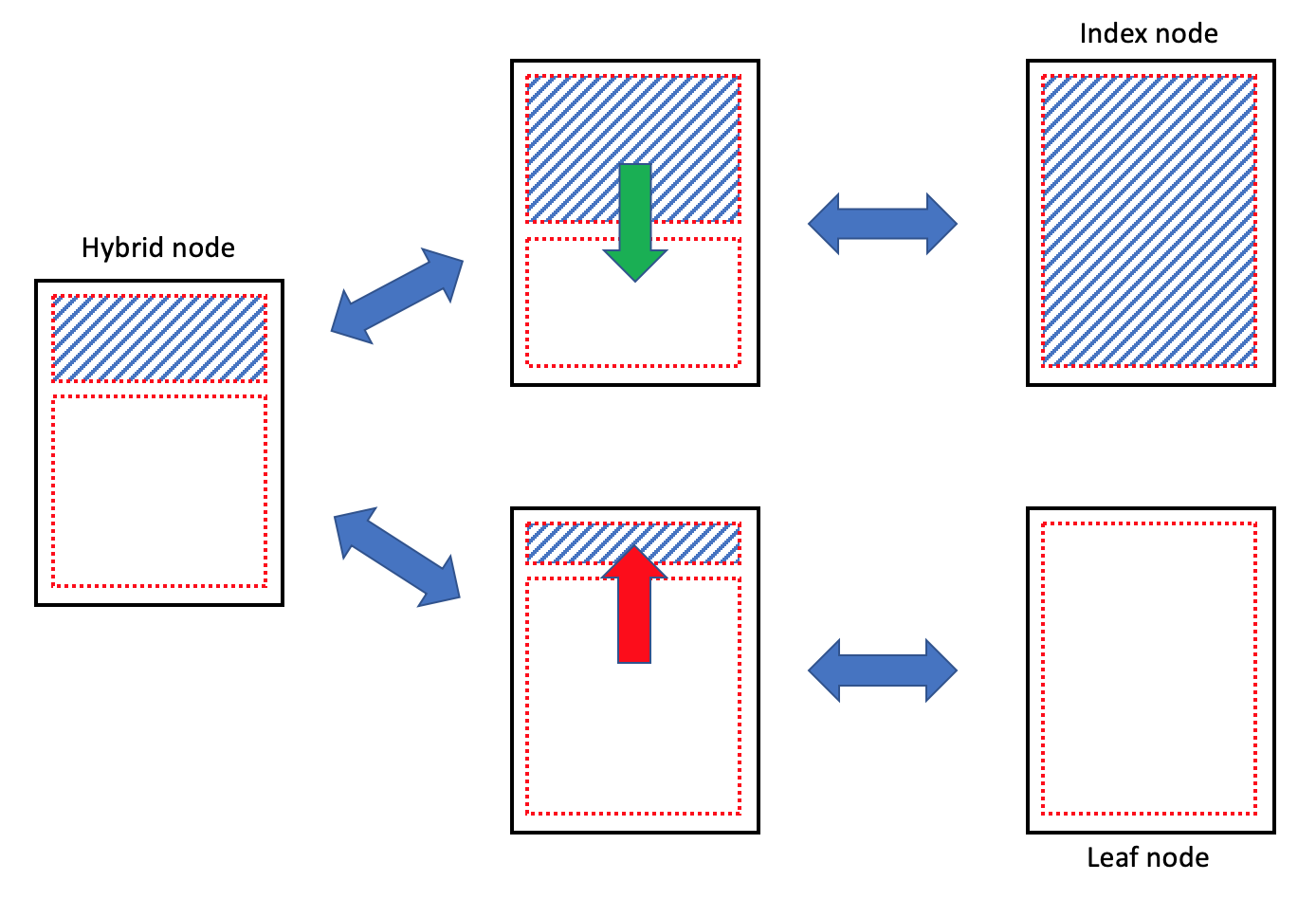}
\caption{Node type migration scheme.}
\label{fig:fig037}
\end{figure}

Operation of deletion of data records could initiate the transformation of index node(s) into the hybrid ones (Fig. \ref{fig:fig037}). Such transformation of node's type could take place many times by virtue of mixture of addition and deletion operations. Also it is possible to imagine the situation of necessity to split one index node on two hybrid ones in the case of inserting some data record in the middle of an exhausted leaf record. Moreover, such splitting operation could be resulted in the addition of hybrid node on the next upper level of the b-tree.

\textbf{B-tree delayed invalidation}. SSDFS file system processes the delete operations in hybrid b-trees by means of special techniques. For example, deletion of any inode from the inodes b-tree is treated like freeing of corresponding item in a particular node. Generally speaking, it means that deleted inodes can be allocated again and the volume space is used by inodes b-tree's nodes remains the reserved space.

If anyone considers the case of deletion of all items in a leaf node then such node can be transformed into the pre-allocated state instead of real deletion of the node from the b-tree structure. Moreover, the pre-allocated state means that it doesn't need to keep the allocated space in a PEB for this node but the index and/or hybrid nodes continue to keep the same index records for the node in the pre-allocated state. Finally, it decreases the write amplification because it doesn't need to update the index/hybrid nodes by means of deletion of index records that point out on the leaf node. However, if a b-tree becomes completely empty then it is the case of the real destruction of b-tree structure. 

But SSDFS file system's driver uses the technique of delayed b-tree's nodes or sub-trees invalidation/destruction. Especially, this technique could make the operation of big files truncation or deletion more fast and efficient. Generally speaking, SSDFS file system's driver has special invalidation queue for the index records (that point out on a node or a sub-tree) and a dedicated thread that has goal to invalidate the data records in the leaf/hybrid nodes and to destroy the sub-tree structure in the background. Finally, it means the necessity to place the index record on node or sub-tree into the invalidation queue during the truncate or delete operation but the real processing of the node or sub-tree will take place in the background. Interesting side effect of such approach is the opportunity to fulfill this background activity in the idle state of file system driver, for example.

\subsection{Inodes B-tree}

\textbf{Inode} is the cornerstone metadata structure of any Linux file system that keeps all information about a file excluding the file's name and content (user data, for example). Generally speaking, this metadata structure is the critical one that requires as high reliability of storing as high efficiency of access and modification operations. The creation of file results in the association of name and inode ID with newly created file. Moreover, inode ID is unique number in the scope of particular file system's volume. The name of file and inode ID are stored as an item of folder. Namely folder associates file names and inode instances. As a result, if end-user or application try to access a file by means of the name then OS employs this file's name for an inode ID lookup. The found inode ID is used by file system driver for retrieving the inode instance.

\begin{figure}[h]
\centering
 
 \includegraphics[width=0.80\columnwidth,keepaspectratio]{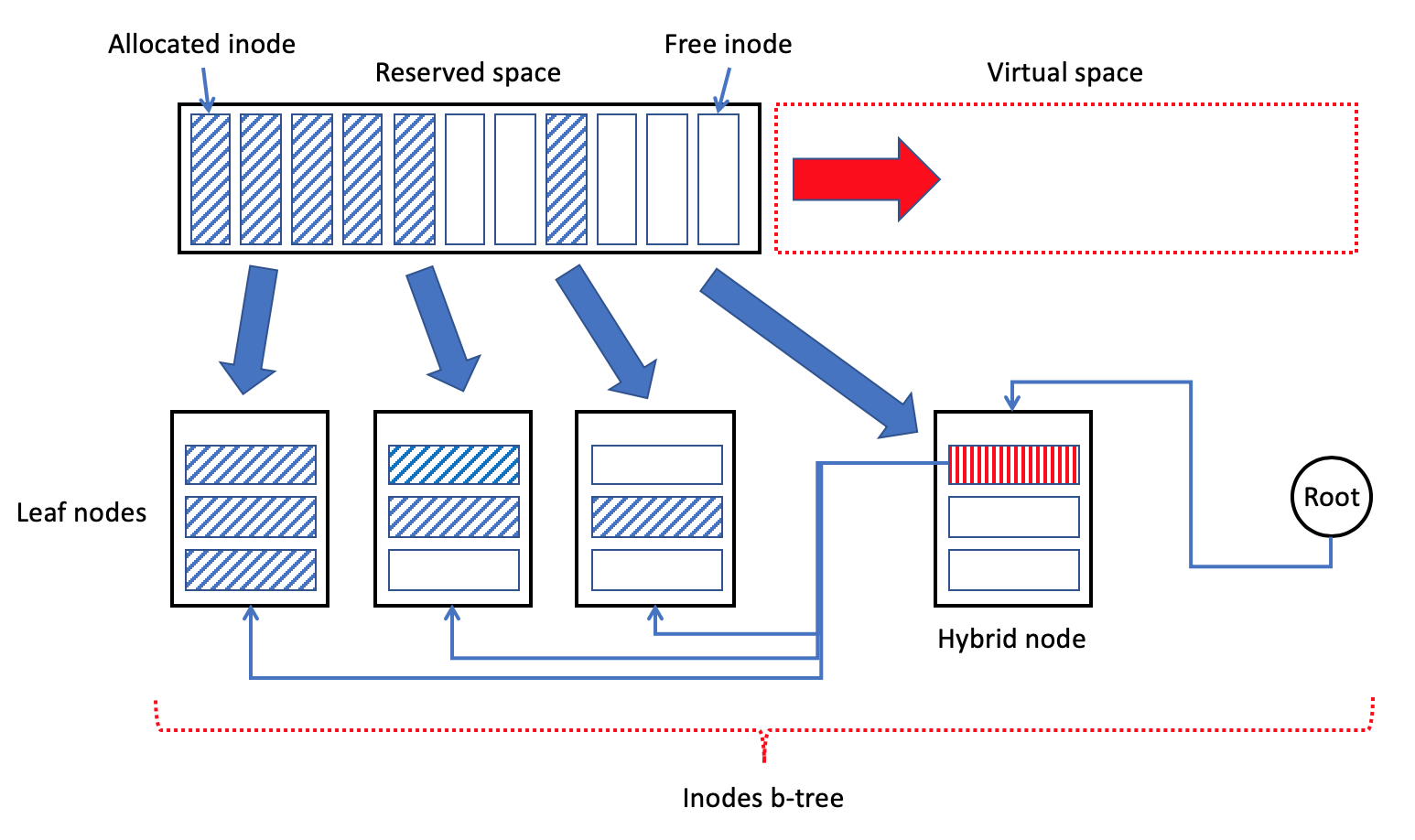}
\caption{Inodes b-tree architecture.}
\label{fig:fig038}
\end{figure}

Generally speaking, inode table can be imagined like a generalized array of inode instances (Fig. \ref{fig:fig038}) because every inode is identified by integer value (inode ID). However, huge capacity of the modern storage devices (HDD, SSD) and highly intensive operations of creation/deletion of files makes the efficient management of inode table by very complex problem. Moreover, the using of simple array or table for inode instances reserves a big space for such table. And such reservation could increase the write amplification because of necessity to keep the reserved space in the valid state. Another possible issue could be the easy exhaustion of the reserved space without the flexible way to extend the reserved space. Oppositely, b-tree provides the easy way of compact representation the small and sparse set of items. Moreover, b-tree is easily extendable metadata structure with the flexible mechanism as increasing as shrinking the nodes' space. Finally, the efficient and fast lookup technique is the another advantage of any b-tree. These points were the steady basis for selection b-tree as the basic metadata structure for inodes tree in SSDFS file system.

\begin{figure}[h]
\centering
 
 \includegraphics[width=0.80\columnwidth,keepaspectratio]{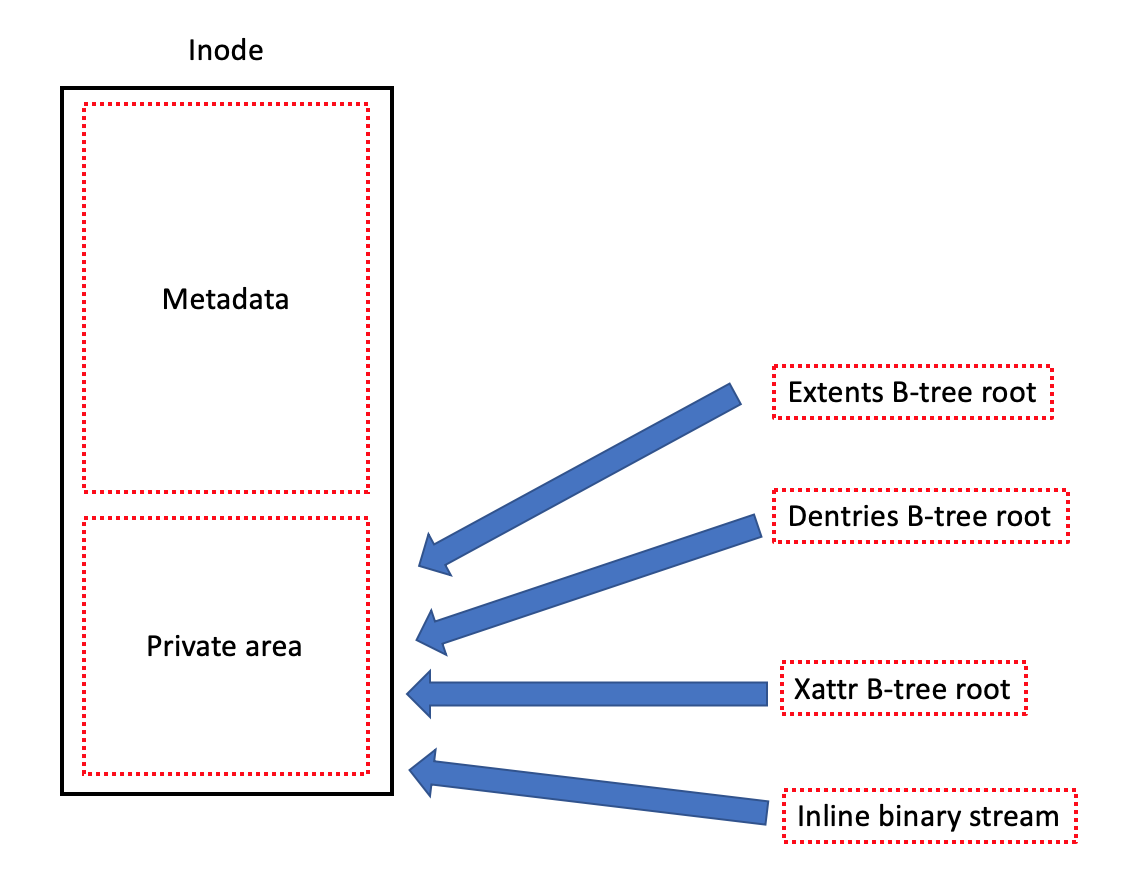}
\caption{Raw inode structure.}
\label{fig:fig039}
\end{figure}

\textbf{SSDFS raw inode} (Fig. \ref{fig:fig039}) is the metadata structure of fixed size that can vary from 256 bytes to several KBs. The size of inode is defined during the file system's volume creation. Usually, inode object includes file mode; file attributes; user/group ID; access, change, modification time; file size in bytes and blocks; links count. The most special part of the SSDFS raw inode is the private area that is used for storing: (1) small file inline, (2) root node of extents, dentries, and/or xattr b-tree.

\textbf{SSDFS inodes b-tree} is the hybrid b-tree that includes the hybrid nodes with the goal to use the node's space in more efficient way by means of combination the index and data records inside of the node. Root node of inodes b-tree is stored into the log footer or partial log header of every log. Generally speaking, it means that SSDFS file system is using the massive replication of the root node of inodes b-tree. Actually, inodes b-tree node's space includes header, index area (in the case of hybrid node), and array of inodes are ordered by ID values. If a node has 8 KB in size and inode structure is 256 bytes in size then the maximum capacity of one inodes b-tree's node is 32 inodes.

Generally speaking, inodes table can be imagined like an imaginary array that is extended by means of adding the new inodes into the tail of the array (Fig. \ref{fig:fig038}). However, inode can be allocated or deleted by virtue of create file or delete file operations, for example. As a result, every b-tree node has an allocation bitmap that is tracking the state (used or free) of every inode in the b-tree node. The allocation bitmap provides the mechanism of fast lookup a free inode with the goal to reuse the inodes of deleted files. Also inodes b-tree uses the special technique of processing the completely empty leaf nodes that could achieve the empty state after deletion the all inodes in this node. This technique is based on the conversion an empty b-tree node into the pre-allocated state. Generally speaking, the pre-allocated state means that the logical extent continues to be reserved for this b-tree node but no space is allocated in segment's PEBs. The important point of such technique is the opportunity not to update the index records in index/hybrid b-tree nodes that point out on the leaf node has converted into pre-allocated state. Also it means that the leaf node's space continues to be reserved on the file system volume.

Additionally, every b-tree node has a dirty bitmap that has goal to track modification of inodes. Generally speaking, the dirty bitmap provides the opportunity to flush not the whole node but the modified inodes only. As a result, such bitmap could play the cornerstone role in the delta-encoding or in the Diff-On-Write approach. Moreover, b-tree node has a lock bitmap that has responsibility to implement the mechanism of exclusive lock a particular inode without the necessity to lock exclusively the whole node. Generally speaking, the lock bitmap was introduced with the goal to improve the granularity of lock operation. As a result, it provides the way to modify the different inodes in the same b-tree node without the using of exclusive lock the whole b-tree node. However, the exclusive lock of the whole tree has to be used for the case of addition or deletion a b-tree node.

\subsection{Dentries B-tree}

Linux kernel identifies a file by means of inode that is unique for the file. However, the association of file name and inode's instance takes place by means of a directory entry. Moreover, different dentries in the same or different folder can identify the same file or inode. Dentries play an important role in the directory caching that contains metadata of frequently accessed files for the more efficient access operations. Another important role of dentries is the folders hierarchy traversing because the dentries connect folder with files.

\begin{figure}[h]
\centering
 
 \includegraphics[width=0.80\columnwidth,keepaspectratio]{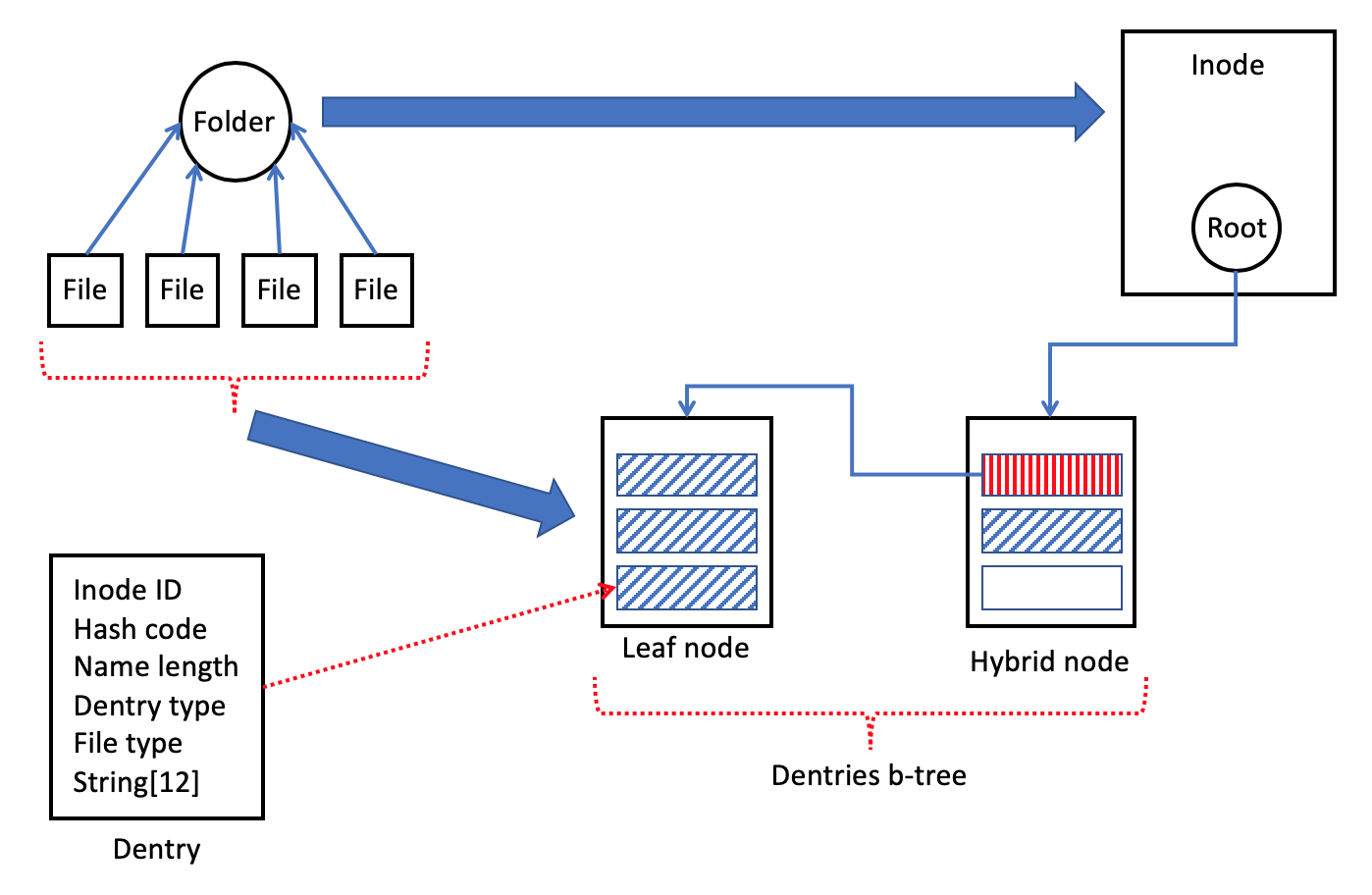}
\caption{Dentries b-tree architecture.}
\label{fig:fig040}
\end{figure}

\textbf{SSDFS dentry} (Fig. \ref{fig:fig040}) is the metadata structure of fixed size (32 bytes). It contains inode ID, name hash, name length, and inline string for 12 symbols. Generally speaking, the dentry is able to store 8.3 filename inline. If the name of file/folder has longer name (more than 12 symbols) then the dentry will keep only the portion of the name but the whole name will be stored into a shared dictionary. The goal of such approach is to represent the dentry by compact metadata structure of fixed size for the fast and efficient operations with the dentries. It is possible to point out that there are a lot of use-cases when the length of file or folder is not very long. As a result, dentry's inline string could be only storage for the file/folder name. Moreover, the goal of shared dictionary is to store the long names efficiently by means of using the deduplication mechanism.

\textbf{Dentries b-tree} is the hybrid b-tree (Fig. \ref{fig:fig040}) with the root node is stored into the private inode's area. By default, inode's private area has 128 bytes in size. Also SSDFS dentry has 32 bytes in size. As a result, inode's private area provides enough space for 4 inline dentries. Generally speaking, if a folder contains 4 or lesser files then the dentries can be stored into the inode's private area without the necessity to create the dentries b-tree. Otherwise, if a folder includes more than 4 files or folders then it needs to create the regular dentries b-tree with the root node is stored into the private area of inode. Actually, every node of dentries b-tree contains the header, index area (for the case of hybrid node), and array of dentries are ordered by hash value of filename. Moreover, if a b-tree node has 8 KB size then it is capable to contain maximum 256 dentries.

Generally speaking, the hybrid b-tree was opted for the dentries metadata structure by virtue of compactness of metadata structure representation and efficient lookup mechanism. Dentries is ordered on the basis of name's hash. Every node of dentries b-tree has: (1) dirty bitmap - tracking modified dentries, (2) lock bitmap - exclusive locking of particular dentries without the necessity to lock the whole b-tree node. Actually, it is expected that dentries b-tree could contain not many nodes in average because the two nodes (8K in size) of dentries b-tree is capable to store about 400 dentries.

\subsection{Extents B-tree}

\begin{figure}[h]
\centering
 
 \includegraphics[width=0.80\columnwidth,keepaspectratio]{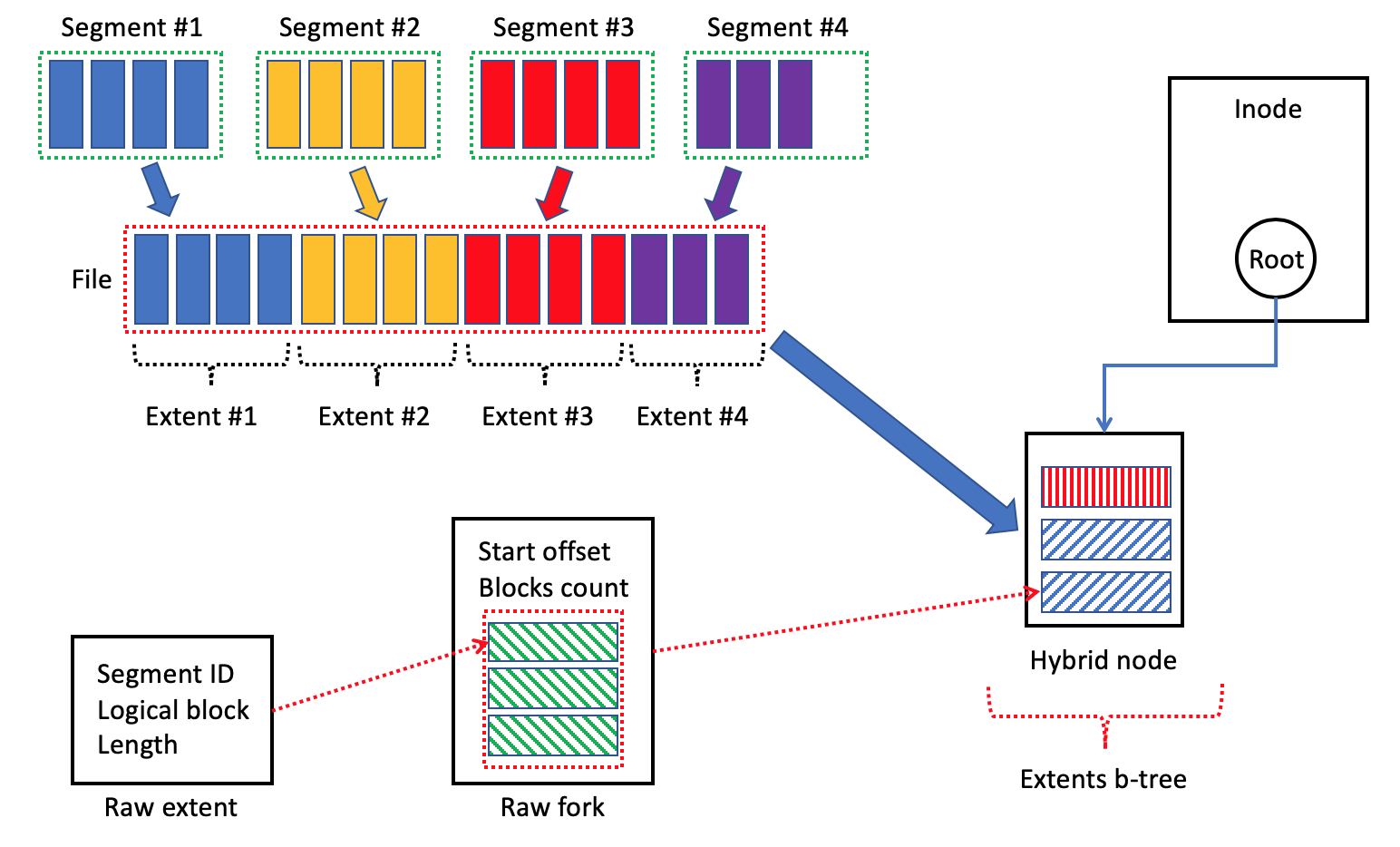}
\caption{Extents b-tree architecture.}
\label{fig:fig041}
\end{figure}

Any file system is dedicated to store the user data in the form of files. Various files could have different length and inode stores information about length of file in blocks and bytes. Also file system is responsible for logical blocks allocation in the case of adding a new data. Generally speaking, file system driver is always trying to allocate a contiguous sequence of logical blocks for any file's content. The contiguous sequence of logical blocks can be described by extent (starting LBA and length) as the most compact descriptor of such sequence. However, it is not always possible to allocate a contiguous sequence of free logical blocks by virtue of possible fragmentation of the file system's volume space by delete and truncate operations. As a result, the allocation operation can be fulfilled by means of allocation the several smaller contiguous sequences of logical blocks in various locations on the volume. Moreover, SSDFS extent cannot be greater than segment size (Fig. \ref{fig:fig041}). Finally, all the mentioned factors result in description of any file's content by means of the set of extents.

\textbf{SSDFS raw extent} (Fig. \ref{fig:fig041}) describes a contiguous sequence of logical blocks by means of segment ID, logical block number of starting position, and length. By default, SSDFS inode has the private area of 128 bytes in size and SSDFS extent has 16 bytes in size. As a result, the inode's private area is capable to store not more than 8 raw extents. Generally speaking, hybrid b-tree was opted with the goal to store efficiently larger number of raw extents. First of all, it was taken into account that file sizes can vary a lot on the same file system's volume. Moreover, the size of the same file could vary significantly during its lifetime. Finally, b-tree is the really good mechanism for storing the extents compactly with very flexible way of increasing or shrinking the reserved space. Also b-tree provides very efficient technique of extents lookup. Additionally, SSDFS file system uses compression that guarantee the really compact storage of semi-empty b-tree nodes. Moreover, hybrid b-tree provides the way to mix as index as data records in the hybrid nodes with the goal to achieve much more compact representation of b-tree's content.

Moreover, it needs to point out that extents b-tree's nodes group the extent records into forks (Fig. \ref{fig:fig041}). Generally speaking, the raw extent describes a position on the volume of some contiguous sequence of logical blocks without any details about the offset of this extent from a file's beginning. As a result, the fork (Fig. \ref{fig:fig041}) describes an offset of some portion of file's content from the file's beginning and number of logical blocks in this portion. Also fork contains the space for three raw extents that are able to define the position of three contiguous sequences of logical blocks on the file system's volume. Finally, one fork has 64 bytes in size. If anybody considers a b-tree node of 4 KB in size then such node is capable to store about 64 forks with 192 extents in total. Generally speaking, even a small b-tree is able to store a significant number of extents and to determine the position of fragments of generally big file. If anybody imagines a b-tree with the two 4 KB nodes in total, every extent defines a position of 8 MB file's portion then such b-tree is able to describe a file of 3 GB in total.

\subsection{Shared Extents B-tree}

\textbf{Deduplication}. One of the known technique of decreasing the write amplification is the deduplication approach. Generally speaking, the key mechanism of deduplication is the determination of replication of the same data on the volume with the goal to store the found duplicated fragment only in one place. As a result, it means that all files contain such deduplicated data should store the same extent in the extents b-trees. SSDFS file system uses a shared extents b-tree for implementation the deduplication technique.

\begin{figure}[h]
\centering
 
 \includegraphics[width=0.80\columnwidth,keepaspectratio]{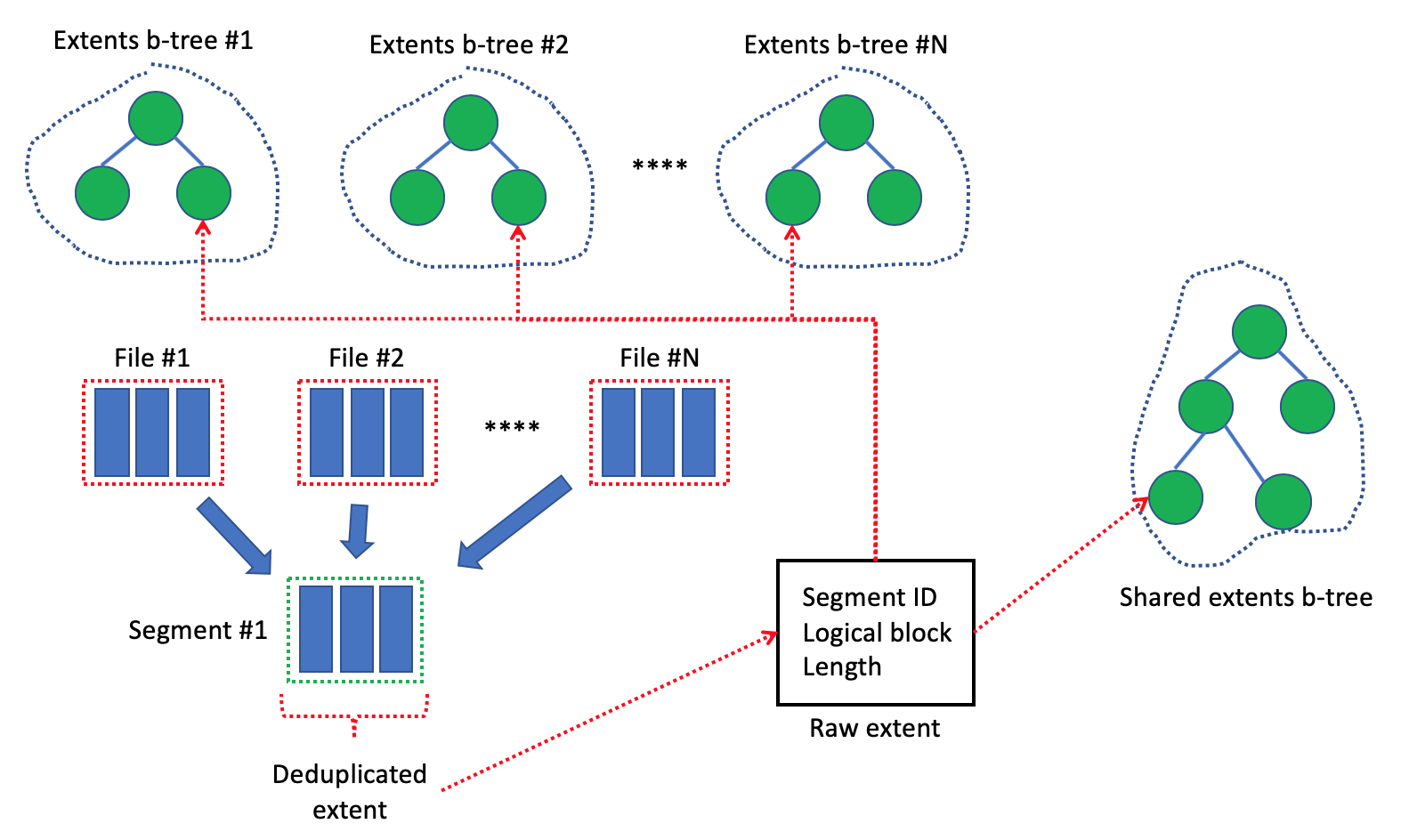}
\caption{Deduplication mechanism of shared extents b-tree.}
\label{fig:fig042}
\end{figure}

First of all, SSDFS file system driver takes into account the size of a file. If the size is smaller than some threshold (for example, 4 KB - 8 KB) then such file is not considered as a deduplication target. Otherwise, it needs to calculate the fingerprint of first 8 KB portion of a file (the size of initial portion can be defined by special threshold value). Then it needs to check the presence of calculated fingerprint in the shared extents b-tree. If no such fingerprint exists in the shared extents b-tree then the only calculated fingerprint has to be stored in the b-tree. Moreover, it doesn't need to calculate fingerprint(s) for the rest of the file in such case.

Oppositely, if there is the same fingerprint for the first 8 KB of the file in the shared extents b-tree then it needs to calculate the fingerprints for the rest of the file and to check the presence of these fingerprints in the shared extents b-tree. Again, it needs to store the calculated fingerprints in the shared extents b-tree if no such fingerprints were found. Otherwise, file system driver has to store extents of found deduplicated fragments into the extents b-trees of particular files (Fig. \ref{fig:fig042}).

Generally speaking, shared extents b-tree will keep only one fingerprint of the first 8 KB for all files that have unique content. Oppositely, the duplicated file's content will be detected during the trying to store a second copy of the same file. However, the detection of this duplication will be resulted in deduplication only first 8 KB of the file and in storing the fingerprints for the rest of duplicated file in the shared extents b-tree. Finally, the third (and next) try to store the duplicated file will be resulted in complete deduplication of the file's content.

\begin{figure}[h]
\centering
 
 \includegraphics[width=0.80\columnwidth,keepaspectratio]{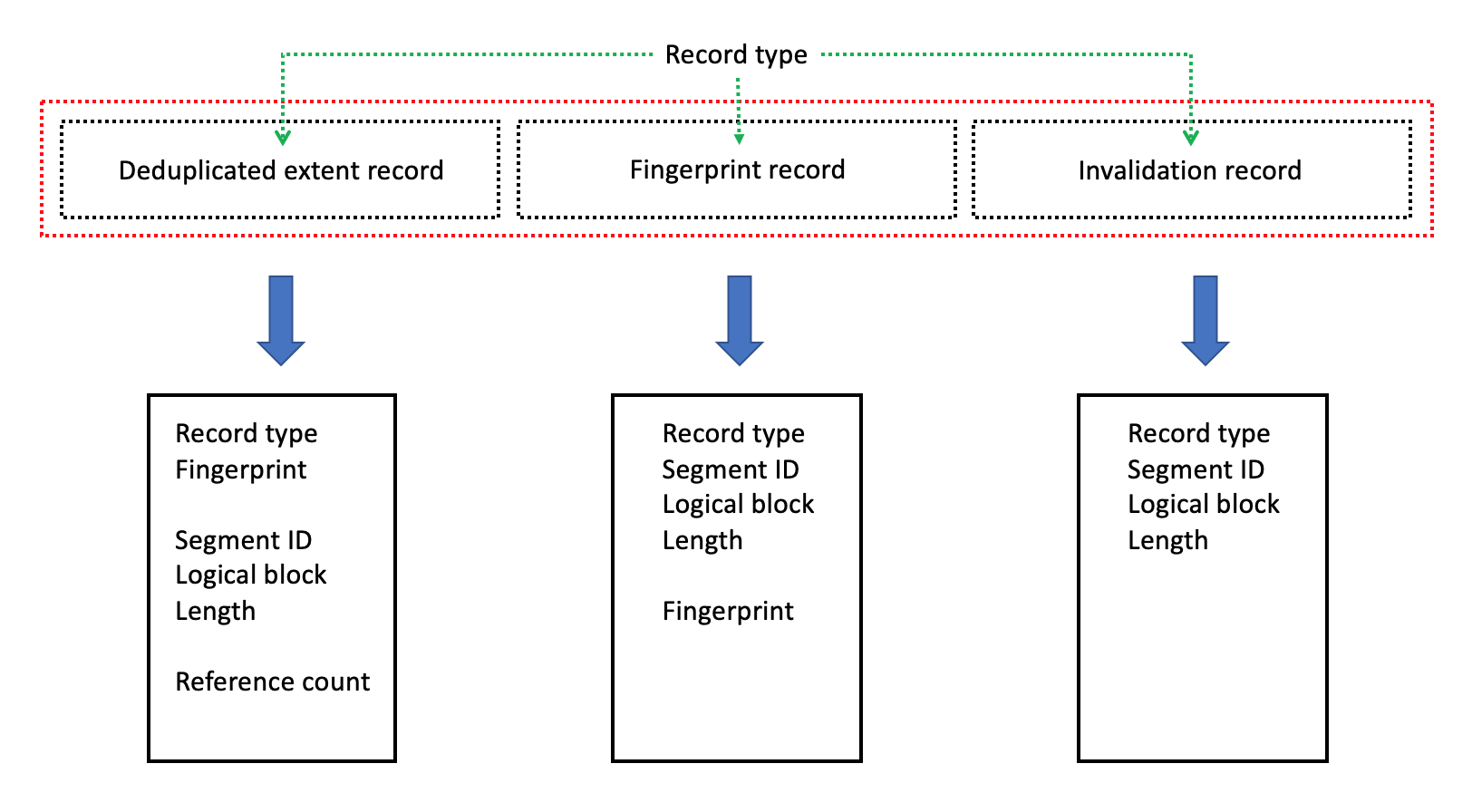}
\caption{Record types in shared extents b-tree.}
\label{fig:fig043}
\end{figure}

\begin{figure}[h]
\centering
 
 \includegraphics[width=0.80\columnwidth,keepaspectratio]{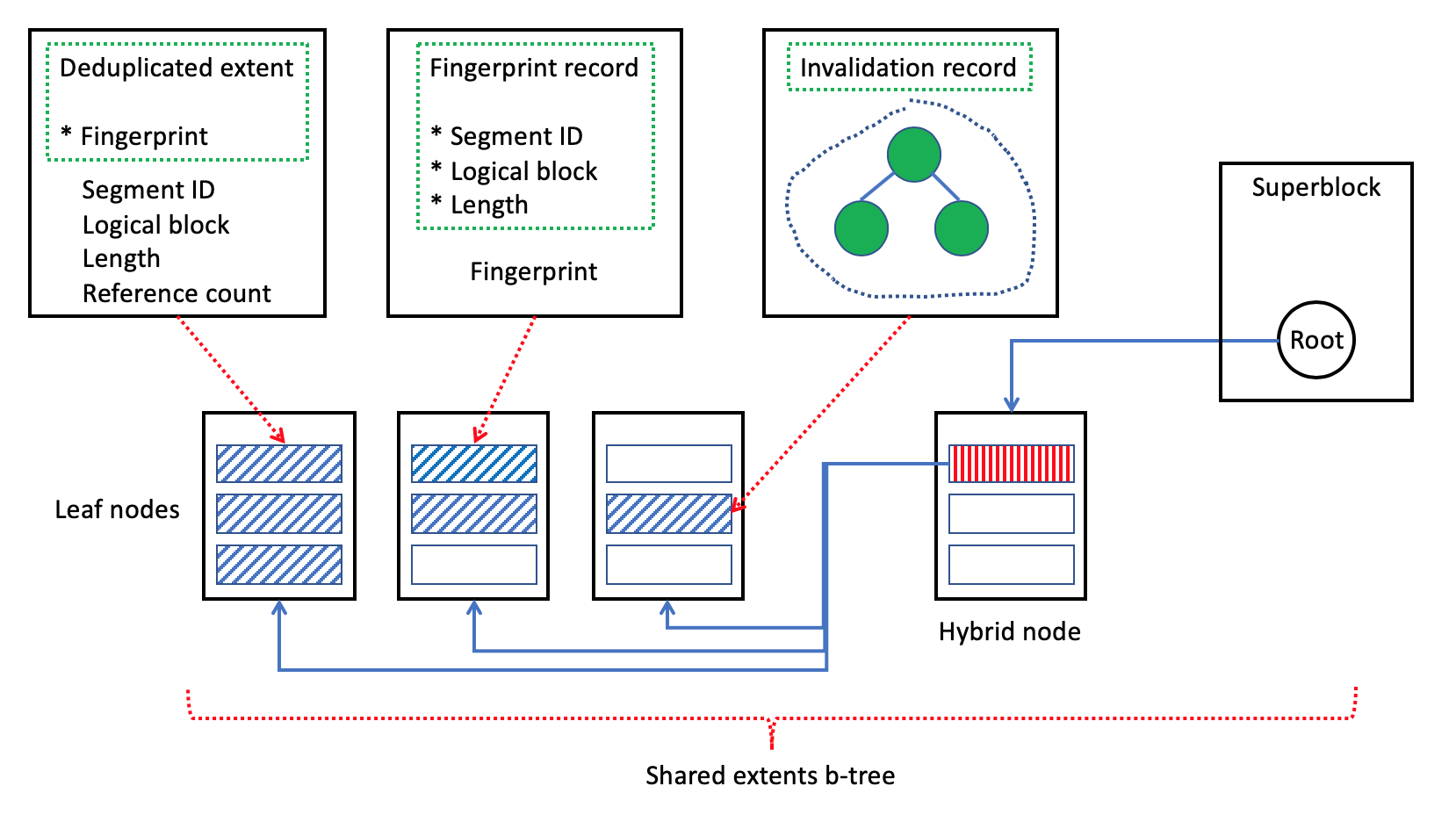}
\caption{Shared extents b-tree architecture.}
\label{fig:fig044}
\end{figure}

\textbf{SSDFS shared extents b-tree} is able to store several record types (Fig. \ref{fig:fig043} - \ref{fig:fig044}): (1) deduplicated extent record, (2) fingerprint record, (3) invalidation record. The deduplicated extent records are ordered by fingerprint value and it contains fingerprint, extent (segment ID, logical block, length), and reference counter values. Generally speaking, the goal of these records is to find the deduplicated extents on the basis of fingerprint value.

The fingerprint records are ordered by segment ID and logical block values and the responsibility of such records is to provide the way to find the fingerprint value on the basis of knowledge of segment ID and logical block values. Every time when it needs to add the information about a deduplicated extent then it needs to insert into the shared extents b-tree as deduplicated extent record as fingerprint record. Moreover, the reason to have two types of the record is the necessity to use the fingerprint record in the case of file deletion or truncation. Generally speaking, only extent data (segment ID, logical block, length) is available in the beginning of the delete or truncate operation. It means that extent data can be used for searching the fingerprint value. Finally, the found fingerprint value can be used for the searching a deduplicated extent record that has to be found with the goal to decrement the reference counter (or completely remove the record if the reference counter is equal to zero).

The third record type is the invalidation records that implement a mechanism of delayed invalidation of extents. Generally speaking, it means that it doesn't need to delete (or truncate) a big file immediately but it is possible to create the invalidation record(s) with the pointer on the whole (or sub-tree) extents b-tree and to store the invalidation record(s) into the shared extents b-tree at first. The processing of invalidation records takes place in the background by a dedicated thread (in the idle state of file system driver, for example). First of all, the thread has to extract an invalidation record and to check the presence of a deduplicated extent record for the extent under invalidation. If shared extents b-tree contains the deduplicated extent record for this extent then it needs to decrement the reference counter only. Otherwise, if shared extents b-tree hasn't deduplicated extent record or the reference counter achieved the nil value then it needs to invalidate the requested extent. Moreover, the corresponding deduplicated extent and fingerprint records have to be deleted from the shared extents b-tree in the case of zeroed reference counter. Finally, invalidation record has to be deleted from the shared extents b-tree also.

\subsection{Shared Dictionary B-tree}

SSDFS file system introduces dentry metadata structure of fixed size that is able to store only 12 inline symbols (8.3 filename) with the goal to achieve the efficient operations with dentries b-tree. However, it means that dentry itself is capable to store the short names only. From one viewpoint, files/folders have short names very frequently. As a result, it implies the high frequency to store the names in dentries only. Moreover, the fixed size of dentry provides simple and fast way to search a particular dentry in the b-tree node. Oppositely, varied size of dentry makes the searching algorithm more complex and inefficient and it require to add some additional metadata in the node.

As a result, SSDFS file system stores the short names only in the dentries and to use the shared dictionary for storing the long names. The shared dictionary's responsibility is to gather the long names are created on the file system's volume. Generally speaking, the gathering names in one place means that shared dictionary keeps only one copy of the name that can be used for different files. Also shared dictionary provides the basis for using the technique of substrings deduplication. Finally, shared dictionary provides the way to keep the names in very compact representation.

Moreover, one of the possible strategy of shared dictionary is not to delete the names at all. From one point of view, it means that such strategy is able to decrease the number of update operations for shared dictionary. From another point of view, if end-user will try to use the name of deleted file for a newly created one then such name doesn't need to be added in the shared dictionary because it will be there already. However, it needs to point out that strategy not to use the delete operation could have some side effect. Generally speaking, the malicious activity of names generation is able to result in unmanageable growing of shared dictionary. However, substring deduplication technique is able to manage such malicious activity efficiently.

\begin{figure}[h]
\centering
 
 \includegraphics[width=0.80\columnwidth,keepaspectratio]{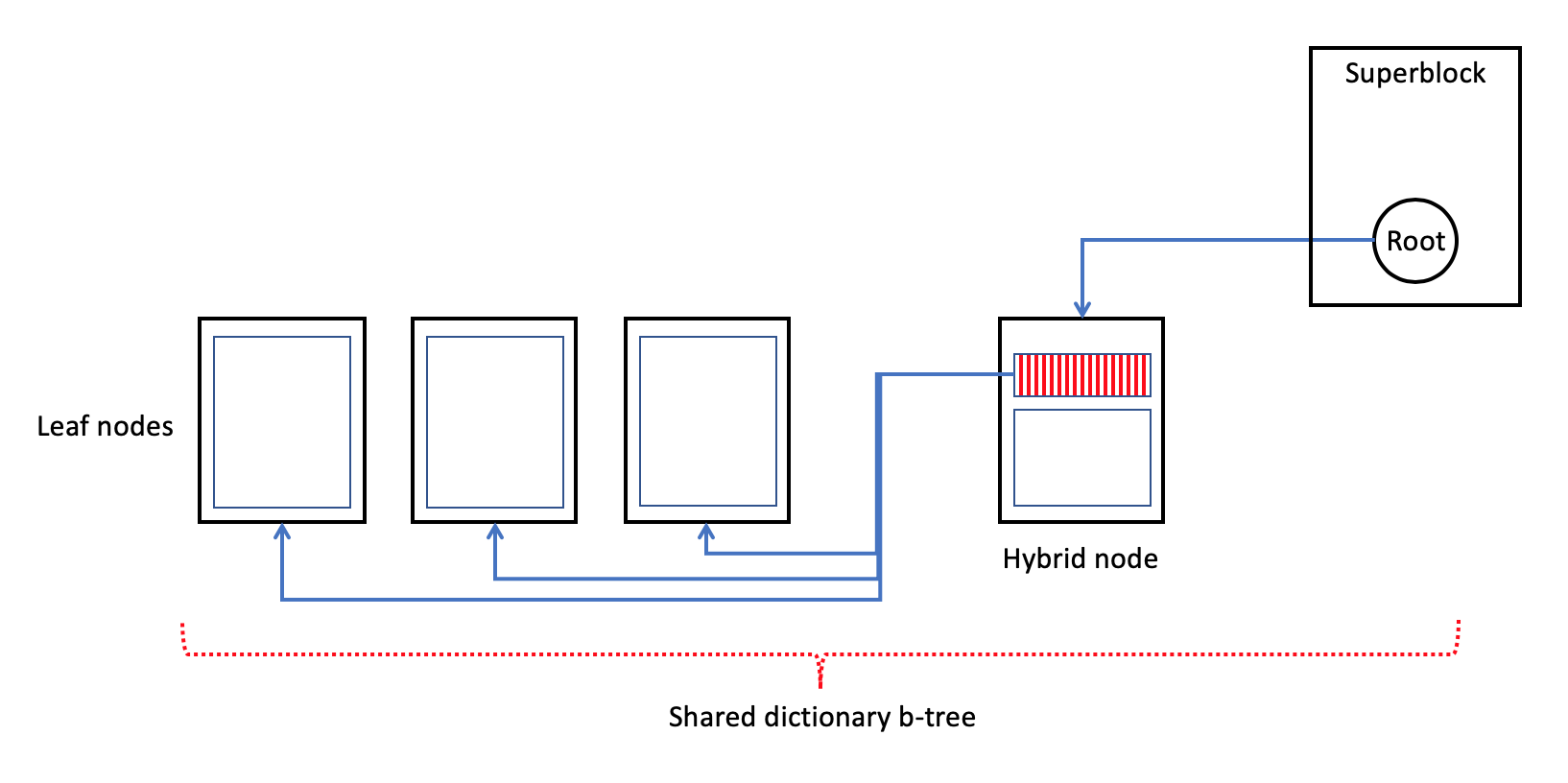}
\caption{Shared dictionary b-tree architecture.}
\label{fig:fig045}
\end{figure}

\begin{figure}[h]
\centering
 
 \includegraphics[width=0.80\columnwidth,keepaspectratio]{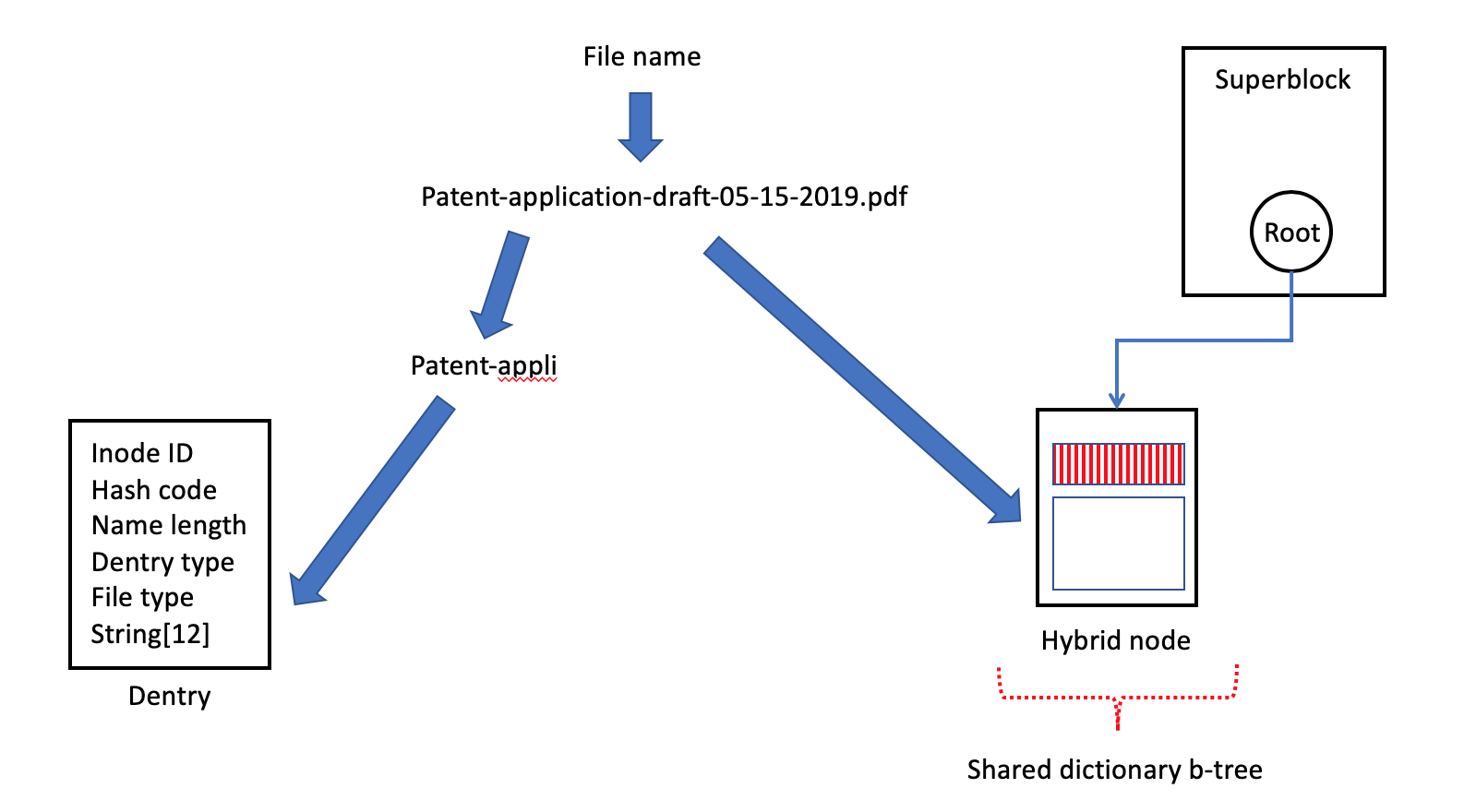}
\caption{Names deduplication mechanism.}
\label{fig:fig046}
\end{figure}

\begin{figure}[h]
\centering
 
 \includegraphics[width=0.80\columnwidth,keepaspectratio]{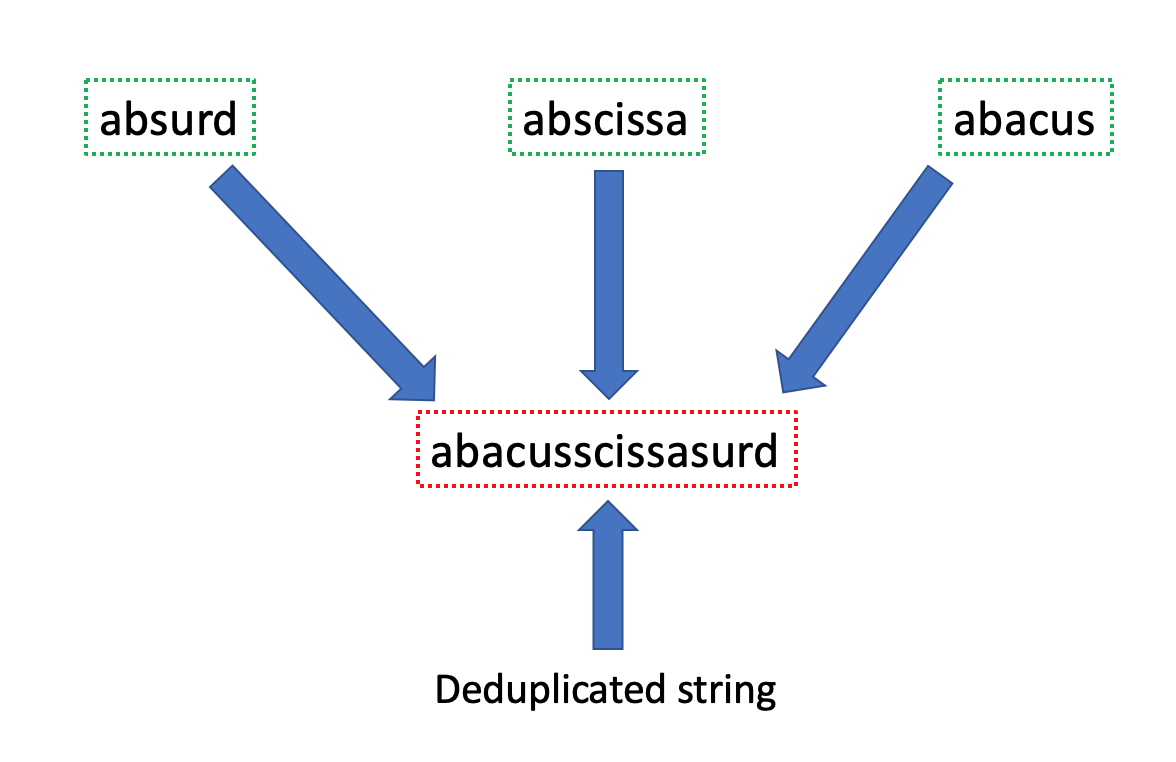}
\caption{Deduplicated strings representation.}
\label{fig:fig047}
\end{figure}

\begin{figure}[h]
\centering
 
 \includegraphics[width=0.80\columnwidth,keepaspectratio]{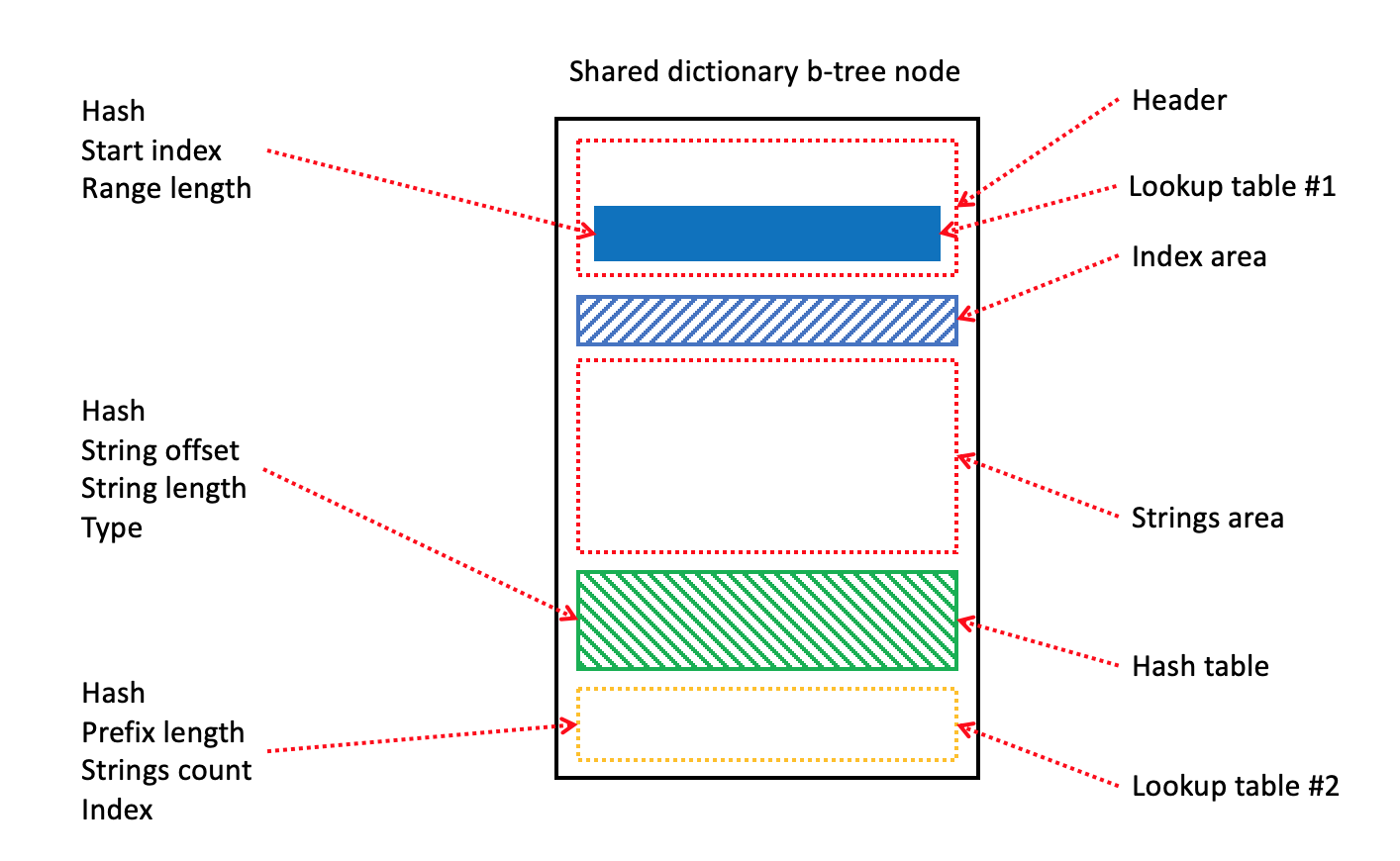}
\caption{Shared dictionary b-tree's node structure.}
\label{fig:fig048}
\end{figure}

\textbf{Shared dictionary} is the hybrid b-tree with root node is stored into the superblock (Fig. \ref{fig:fig045}). Every hybrid or leaf node of shared dictionary b-tree includes: (1) lookup table1, (2) lookup table2, (3) hash table, and (4) strings area (Fig. 48).

The lookup table1 is located into the node's header and it implements clustering or grouping the items of lookup table2. By design, lookup table1 is capable to keep only 20 items. Every item (Fig. \ref{fig:fig048}) contains: (1) hash value, (2) starting index in the lookup table2, and (3) number of items in the group. Generally speaking, the responsibility of lookup table1 is to provide the mechanism of fast search of some items' cluster in the lookup table2 on the basis of hash value.

As a result, the found item in the lookup table1 is the basis for further search in the lookup table2. This table (lookup table2) is located in the bottom of node (Fig. \ref{fig:fig048}) and it has goal to provide the mechanism for the search a position of name's prefix (or starting keyword). Every item of lookup table2 (Fig. \ref{fig:fig048}) contains: (1) hash value, (2) prefix length, (3) number of deduplicated names, and (4) index in hash table. Generally speaking, the lookup table2 describes positions of names' prefixes in strings area.

Finally, hash table (Fig. \ref{fig:fig048}) is located upper the lookup table2. It is responsible to describe every name in the strings area. Every item of hash table contains: (1) hash value, (2) name offset, (3) name length, and (4) name type. Generally speaking, hash table implements the mechanism to define the position and the length of a suffix of deduplicated name because the full name is constructed from the prefix and the suffix (Fig. \ref{fig:fig046} - \ref{fig:fig047}). Finally, it needs to find the prefix from the lookup table2 and the suffix from the hash table for the extraction of a full name. The last item of the node is strings area that keeps the full and deduplicated names. Generally speaking, b-tree is efficient mechanism for storing and searching the strings of variable length.

\subsection{Extended Attributes B-tree}

\textbf{Extended attribute} represents the pair of name and value is associated with a file or a folder. It is possible to say that extended attributes play the role of extension of regular attributes that are associated with inodes. Frequently, extended attributes are used with the goal to provide an additional functionality in file system, for example, additional security features - Access Control Lists (ACL). Name of extended attribute is the null-terminated string and it is defined in the fully qualified namespace form (for example, security.selinux). Currently, it exists the security, system, trusted, and user classes of extended attributes. Usually, VFS limits the length of xattr's name by 255 bytes and size of the value by 64 KB.

\begin{figure}[h]
\centering
 
 \includegraphics[width=0.80\columnwidth,keepaspectratio]{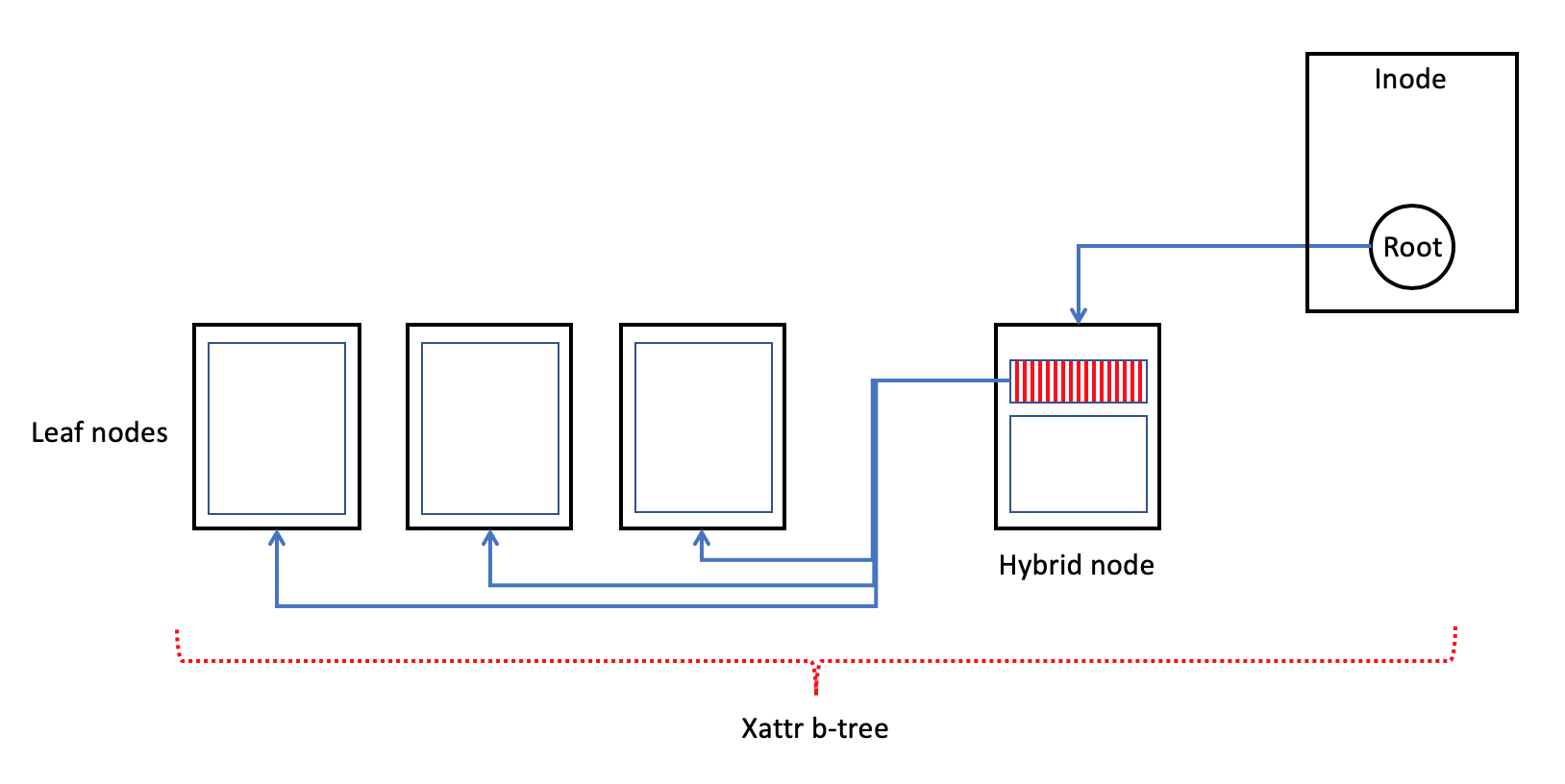}
\caption{Extended attributes (xattr) b-tree architecture.}
\label{fig:fig049}
\end{figure}

\begin{figure}[h]
\centering
 
 \includegraphics[width=0.80\columnwidth,keepaspectratio]{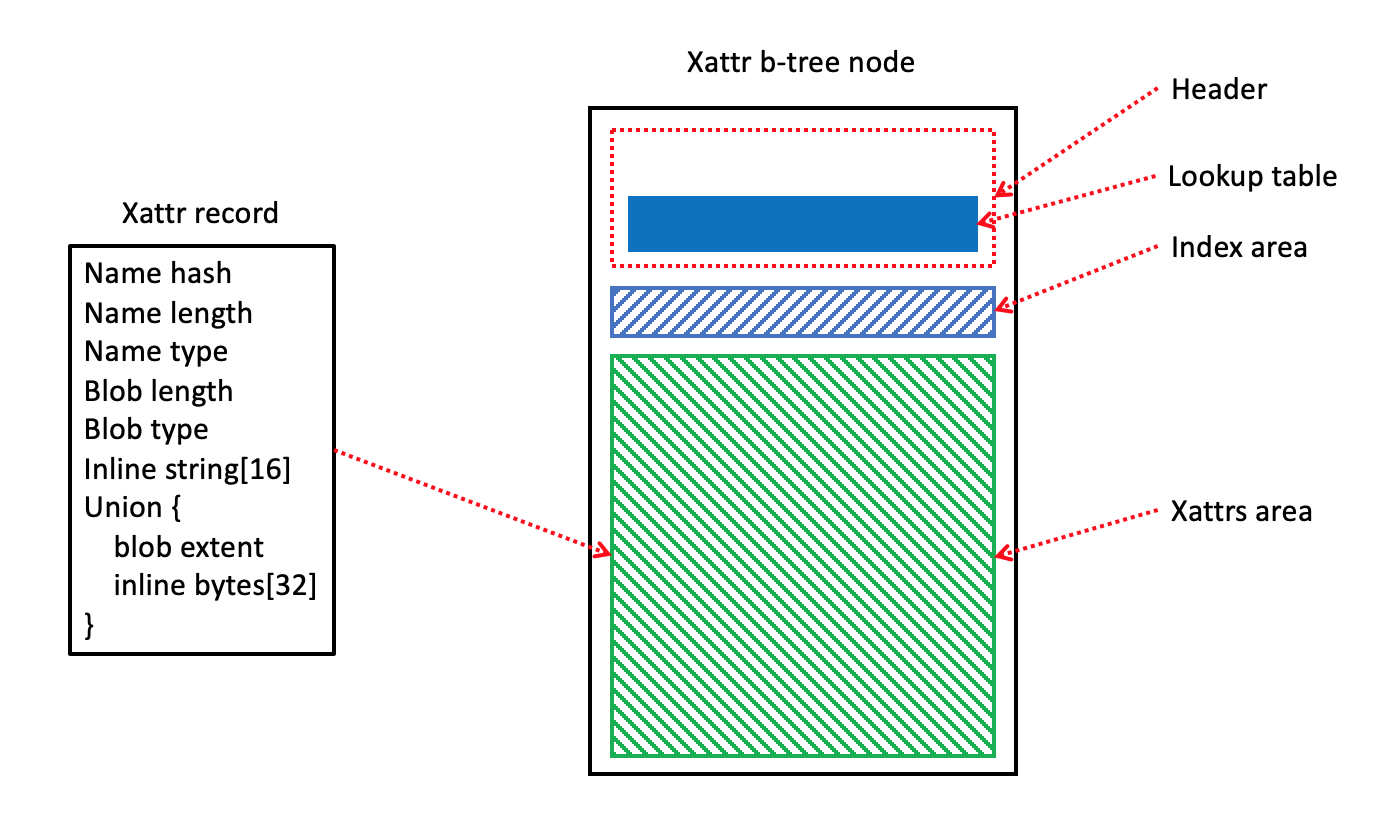}
\caption{Extended attributes b-tree's node structure.}
\label{fig:fig050}
\end{figure}

SSDFS file system uses a metadata structure of fixed size (64 bytes) for representation and storing the xattr record on a file system's volume. Moreover, this metadata structure is capable to keep the 16 symbols inline and value of 32 bytes (Fig. \ref{fig:fig050}). However, namespace class is represented not by string itself but by means of special field of name type. Generally speaking, it means that if the name or the value is lesser than declared limit then it can be stored inline in the xattr record. Otherwise, if a name is longer than 16 symbols then initial portion of the name will be stored inline in the xattr record but the whole name has to be stored into the shared dictionary. Also, if a value is bigger than 32 bytes then the blob has to be stored in some logical block(s) of the volume but the xattr record will keep the extent that describes the position of this blob. Moreover, it is possible to employ the shared extents b-tree for storing the xattr's blobs in the range from 32 bytes to 4 KB. Additionally, shared extents b-tree is able to deduplicate the blobs with identical content.

\textbf{SSDFS xattr tree} (Fig. \ref{fig:fig049}) is implemented as hybrid b-tree with root node is stored in the inode's private area. By default, the private area of inode has 128 bytes in size. Usually, file owns the extents b-tree but folder has dentries b-tree. Finally, it means that the first 64 bytes of private area will be used by the root node of extents or dentries b-tree but the rest 64 bytes can be used for root node of xattr b-tree. Also, if file or folder has only one extended attribute then the xattr record (64 bytes) can be stored inline in the second half of private area.

\textbf{The xattr record} is the metadata structure of fixed size. Generally speaking, the goal of such approach is to keep in the node an array of xattr records because the fixed size of every item in the array provides a very efficient mechanism of lookup, access and modification operations. Moreover, the header of b-tree's node contains a lookup table (Fig. \ref{fig:fig050}) is capable to store 22 records. The goal of such lookup table is the clustering of xattr records in the main area for implementing the efficient mechanism of searching operation. Every item in the lookup table is a hash value of an extended attribute's name. Generally speaking, the hash value identifies the position of starting xattr record in a group (or cluster) of xattr records. Every such starting record is located on a fixed position in the main area of node. As a result, the lookup table provides the way to restrict the search by some cluster in the main area.

Generally speaking, the case of significant number of extended attributes for the same file/folder is very rare. It means that it makes sense to consider the xattr record of bigger size (128 bytes, for example) with the goal to optimize the operations with xattr records by increasing inline area of value (blob). Moreover, it is possible to consider the inode's record of bigger size (512 bytes, for example). Such inode will be able to keep about 5 inline xattr records. Additionally, it is possible to implement a shared xattrs b-tree that will be able to store xattr records of different files/folders into the one b-tree. However, even if anybody considers only dedicated xattrs b-tree then the b-tree with 2 nodes of 4 KB in size is capable to store about 128 xattr records in total.

\subsection{Write Amplification Management}

The write amplification issue is the crucial problem for the case of flash-oriented and flash-friendly file systems. It is possible to state that this issue is the key reason of SSD lifetime shortening. Every particular file system has unique reasons of the write amplification issue and it contains some techniques to decrease or to eliminate this problem. SSDFS file system uses such techniques for resolving the problem of write amplification issue: (1) compression, (2) small files compaction scheme, (3) logical extent concept, (4) Diff-On-Write approach, (5) deduplication, (6) inline files.

\textbf{Compression}. SSDFS file system widely uses compression as for user data as for metadata. Current file system driver implementation supports zlib and LZO compression. Moreover, SSDFS file system uses a special compaction scheme which gathers several compressed fragments (even for different files) into one NAND flash page inside of special log's area (diff update or journal areas). Generally speaking, this compaction technique provides the opportunity to use only one NAND flash page for several compressed fragments of different files instead of several ones. As a result, the decreasing number of used NAND flash pages decreases number of I/O operations and it creates the opportunity to reduce the write amplification issue.

\textbf{Small files compaction}. It took place some number of research papers with the goal to investigate the aged file system volumes' state and to elaborate some vision of distribution of data amongst various types. As a result, it has been found that many file system volumes contain significant number of small files. Some researchers estimate the number of small files as 61\% of total number of files on the volume. SSDFS file system introduces a special compaction scheme for the case of small files. Generally speaking, PEB's log can contain a special journal area that is used for gathering into one NAND flash page the several small files. As a result, this compaction technique reduces the number of I/O operations and is able to decrease the factor of write amplification issue.

\textbf{Inline content}. SSDFS file system has inode's format with reservation of 128 bytes for private area (by default). Moreover, increasing the size of inode transforms the private area to bigger size. Generally speaking, private area can be used for keeping inline the content of small files, extent, dentry or xattr records. As a result, it means that keeping data inline in the inode's private area creates the opportunity not to allocate the logical blocks (NAND flash pages) for storing these data or metadata. Finally, mechanism of keeping data inline is the way to reduce the write amplification issue and to improve the file system's performance.

\textbf{Logical extent concept}. SSDFS file system implements the logical extent concept as the additional mechanism of decreasing the write amplification issue. Generally speaking, Copy-On-Write policy is the main technique of data updates in the scope of any LFS file system. It means that the necessity to update some data on the volume results in writing the actual state of data in a new position (logical block) on the file system's volume. As a result, the main problem of such approach is the necessity to update a metadata (block mapping table, for example) for any of such update with the goal to track the position of actual state of data. Finally, it results in increasing the number of I/O operations and making the write amplification issue like more severe problem.

But SSDFS file system tracks the position of any data on the volume by means of logical extent. The logical extent structure includes: (1) segment ID, (2) logical block number inside of this segment, (3) number of logical blocks in the extent. Moreover, SSDFS file system implements PEBs migration technique. Finally, it means that if any logical block is stored into some segment then the logical extent remains the same during any update or modification operations with data inside of this logical extent. Generally speaking, the logical extent will have the same value until the data will be moved into another segment. As a result, the nature of logical extent provides the opportunity not to update the metadata structure that tracks the position of data on the volume by means of logical extents. Moreover, this technique reduces the write amplification issue.

\begin{figure}[h]
\centering
 
 \includegraphics[width=0.90\columnwidth,keepaspectratio]{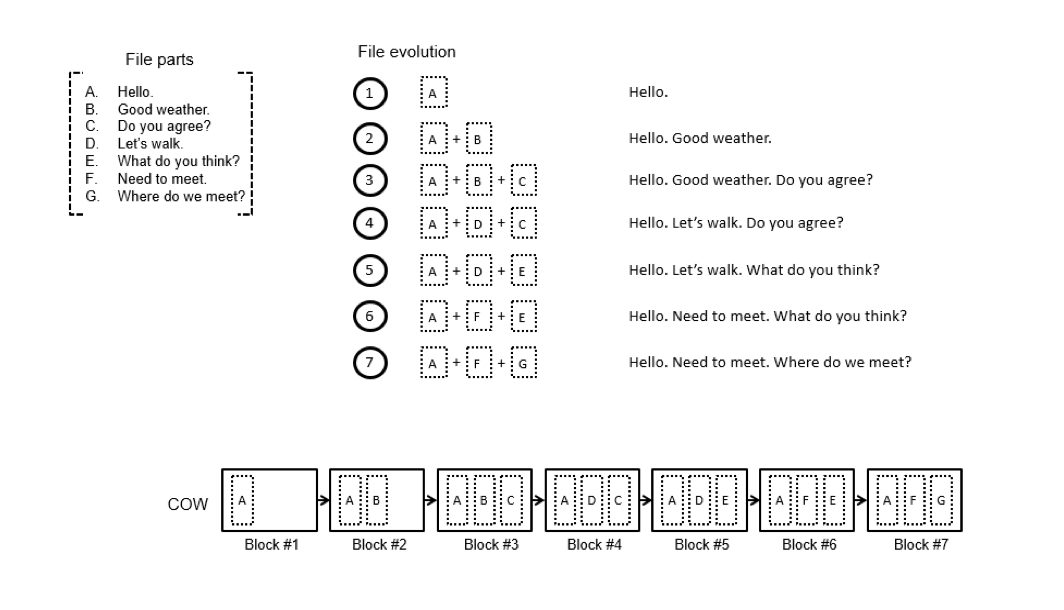}
\caption{Copy-On-Write policy side effect.}
\label{fig:fig051}
\end{figure}

\textbf{Diff-On-Write approach}. The Copy-On-Write (COW) policy is the central technique of Log-structured file system. The goal of this policy is to overcome peculiarity of NAND flash. Namely, clean physical page of NAND chip can be written once. And it needs to erase a whole physical erase block for operation of re-writing the page. Usually, physical erase block includes a bunch of pages. But, from another point of view, the COW policy can be treated as a reason of write amplification issue. Because every update of file's data results in moving updated block of file into new physical page of NAND flash (Fig. \ref{fig:fig051}).

\begin{figure}[h]
\centering
 
 \includegraphics[width=0.90\columnwidth,keepaspectratio]{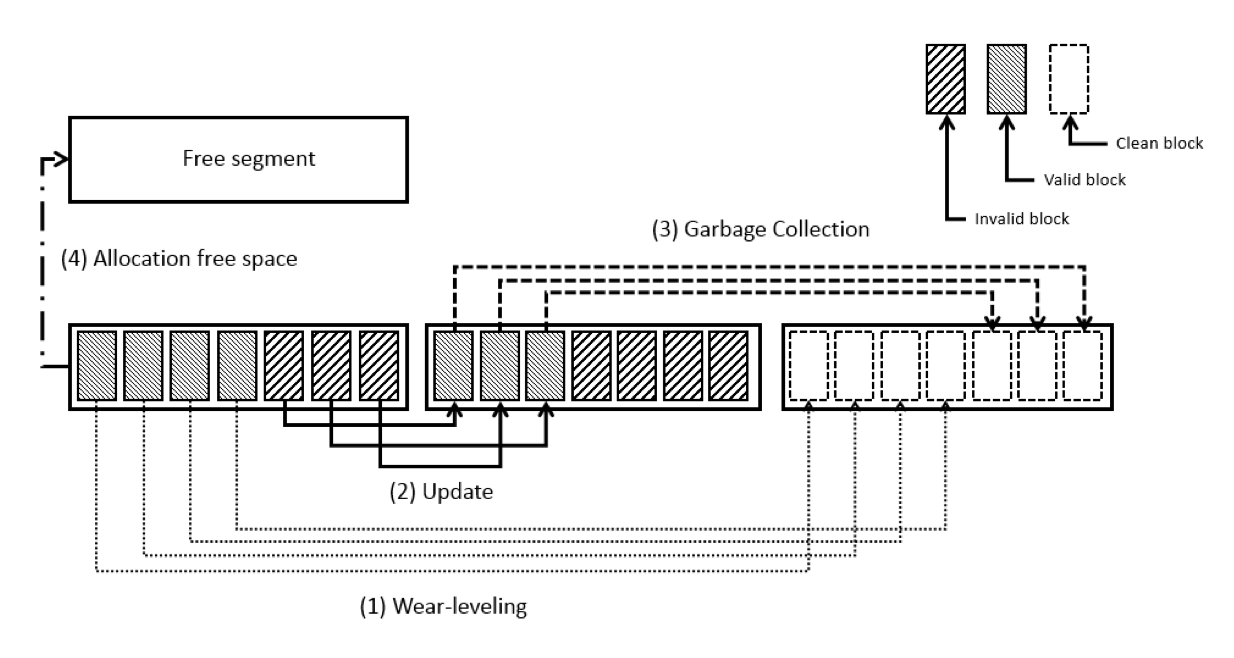}
\caption{Write amplification issue.}
\label{fig:fig052}
\end{figure}

Write amplification issue (Fig. \ref{fig:fig052}) has several reasons. First of all, necessity to overcome write and read disturbance effects of NAND flash and necessity to wear NAND flash erase blocks uniformly are resulted in wear-leveling policy. This policy dictates regular moving of user data from aged segment into new one. The COW policy as basic technique of Log-structured file system can be treated as another reason of write amplification issue. And final reason of write amplification issue could be an inefficient Garbage Collection policy.

\begin{figure}[h]
\centering
 
 \includegraphics[width=0.90\columnwidth,keepaspectratio]{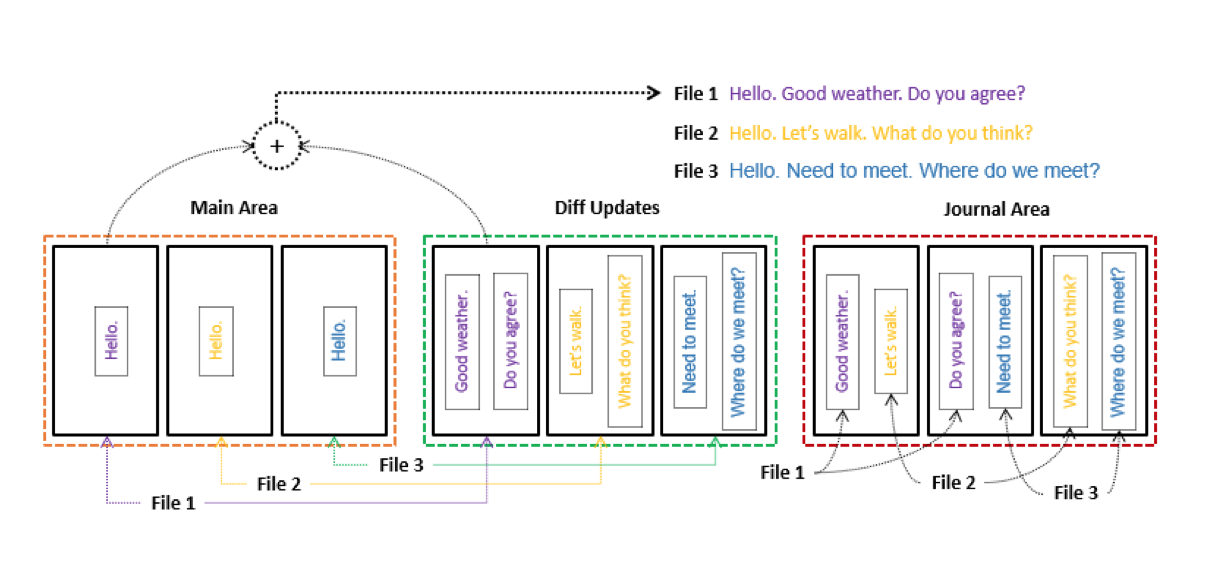}
\caption{Diff-On-Write approach.}
\label{fig:fig053}
\end{figure}

Main, Diff Updates and Journal areas are foundation for Diff-On-Write approach (Fig. \ref{fig:fig053}). This approach distinguishes main, unchangeable ("cold") part of file's data. These data are stored in Main area. For example, a file's contiguous 4 KB binary stream can be treated as "cold" data. Such piece of data can be saved into one physical page of Main area. And read-only nature of this physical page can be provided by means of saving of all updates of this page into another area (Diff Updates area). For example, File 1 has string "Hello" as "cold" data on Fig. \ref{fig:fig053}. The Journal area provides shared space for gathering updates of different files. Joining of all current updates in one area looks like as journal and to provide gathering all "hot" data in one area. For example, Fig. \ref{fig:fig053} shows situation when one block of Journal area contains updates for File 1 (string "Good weather.") and for File 2 (string "Let's walk."). Journal area can be imagined as mixed sequence of updates for different files. As a result, if Journal area in one or several logs has been gathered updates of one file with accumulated size equals to physical page size then it makes sense to join these updates in one block of Diff Updates or Main areas. It needs to store updates in the Diff Updates area for the case of presence updates from different file's parts. And, finally, it needs to store a sequence of contiguous updates into one block of Main area.

\begin{figure}[h]
\centering
 
 \includegraphics[width=0.90\columnwidth,keepaspectratio]{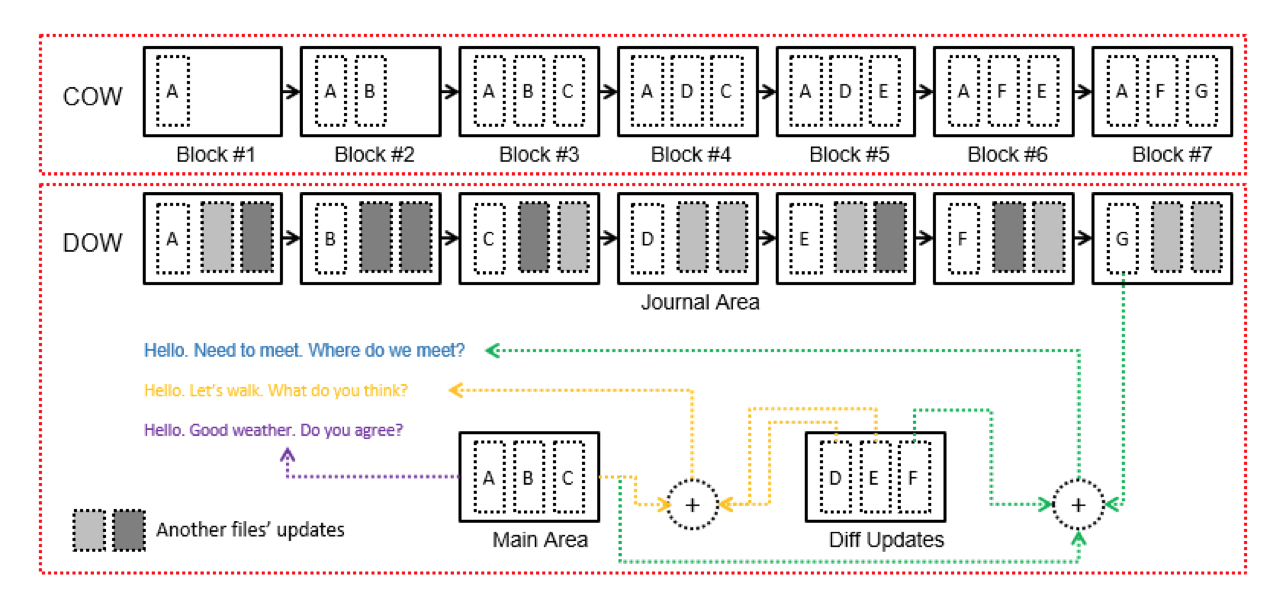}
\caption{Copy-On-Write vs. Diff-On-Write.}
\label{fig:fig054}
\end{figure}

The Copy-On-Write (COW) policy means that every updated block should be copied in a new place. The Diff-On-Write approach suggests to store only diff between initial and updated state of data for every update (Fig. \ref{fig:fig054}).

\begin{figure}[h]
\centering
 
 \includegraphics[width=0.90\columnwidth,keepaspectratio]{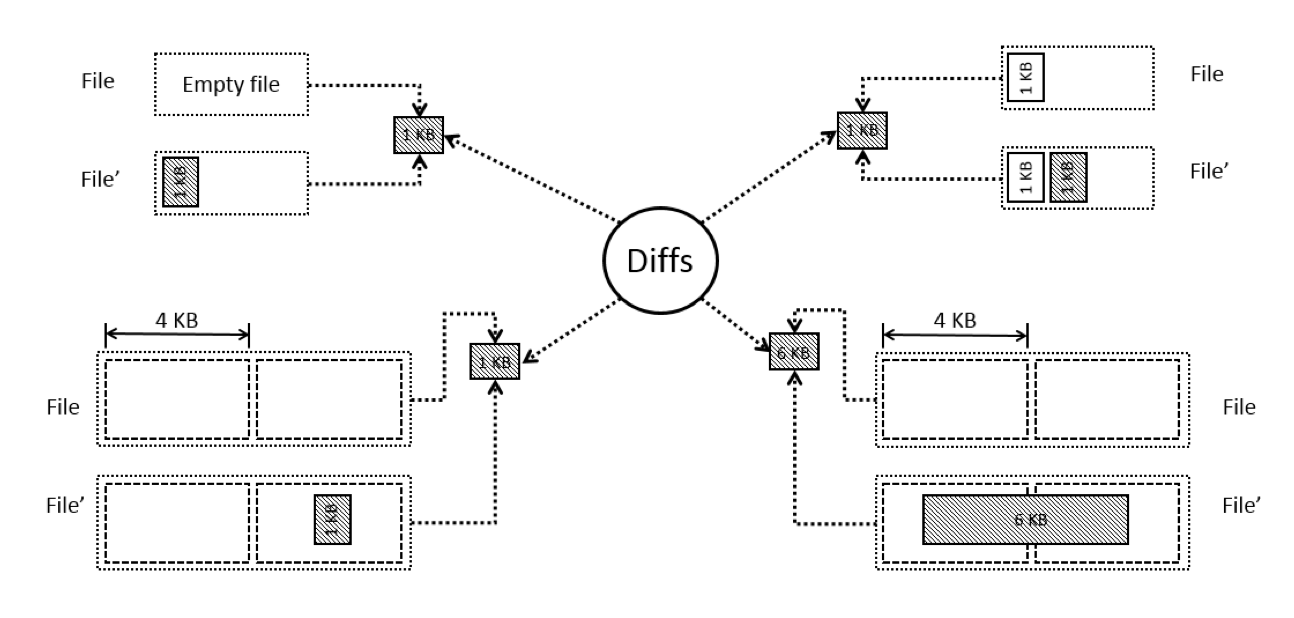}
\caption{Diff concept.}
\label{fig:fig055}
\end{figure}

Fig. \ref{fig:fig055} shows different examples of diff. The diff can be result of: (1) file creation; (2) adding data into existing file; (3) update of some file's part.

\begin{figure}[h]
\centering
 
 \includegraphics[width=0.90\columnwidth,keepaspectratio]{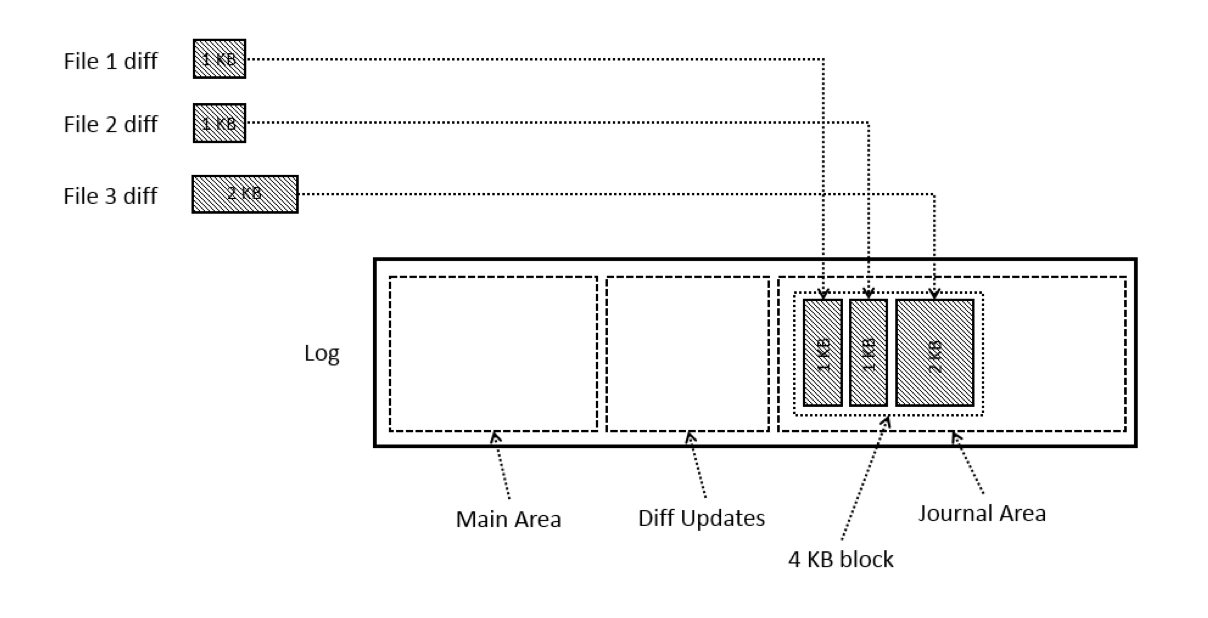}
\caption{Technique of joining files' diffs in journal area.}
\label{fig:fig056}
\end{figure}

Diff-On-Write approach suggests to gather small parts or small updates of different files in one block of Journal area (Fig. \ref{fig:fig056}). It is well known fact that about 61\% of all files on a volume are smaller than 10KB. Such technique suggest the way of decreasing write amplification factor and decreasing over-provisioning for the case of small files. Moreover, such approach gathers frequent updates in dedicated "hot" area. As a result, it can improve efficiency of GC policy.

\begin{figure}[h]
\centering
 
 \includegraphics[width=0.90\columnwidth,keepaspectratio]{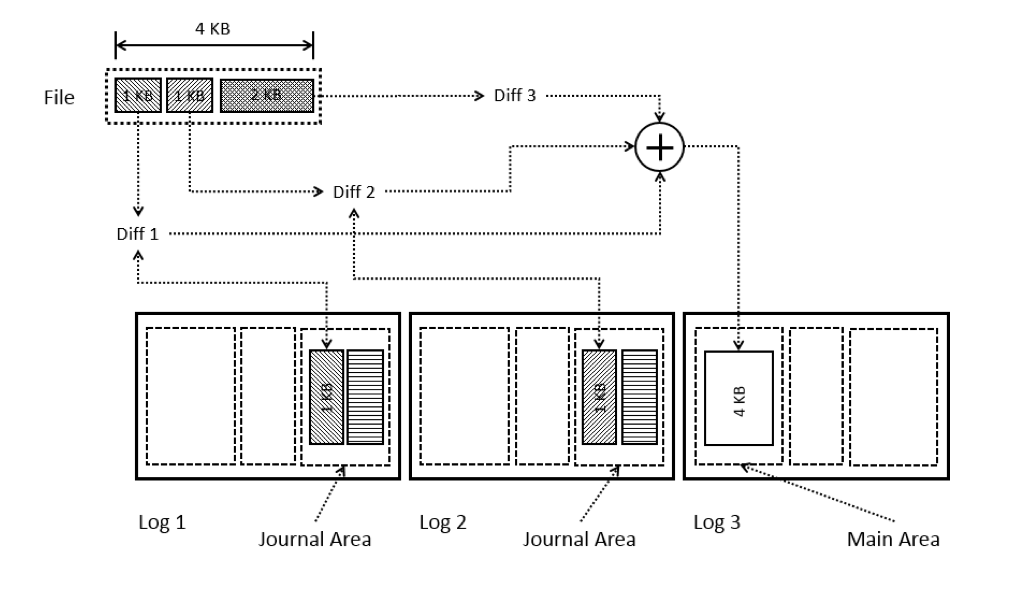}
\caption{Technique of main and journal areas interaction in Diff-On-Write approach.}
\label{fig:fig057}
\end{figure}

Diff-On-Write approach provides basis for decreasing write amplification factor in the case of gradual growing of file's content (Fig. \ref{fig:fig057}). Let's suppose that file contains 1 KB data after creation. Then additional 1 KB will be added on another day, for example. And, finally, 2 KB of data will be added after several days. First two 1 KB diffs can be stored in Journal areas of different logs. Every diff will share space of physical page with updates of another files. Finally, file content will be saved into Main area of a log with joining of all available updates.

\begin{figure}[h]
\centering
 
 \includegraphics[width=0.90\columnwidth,keepaspectratio]{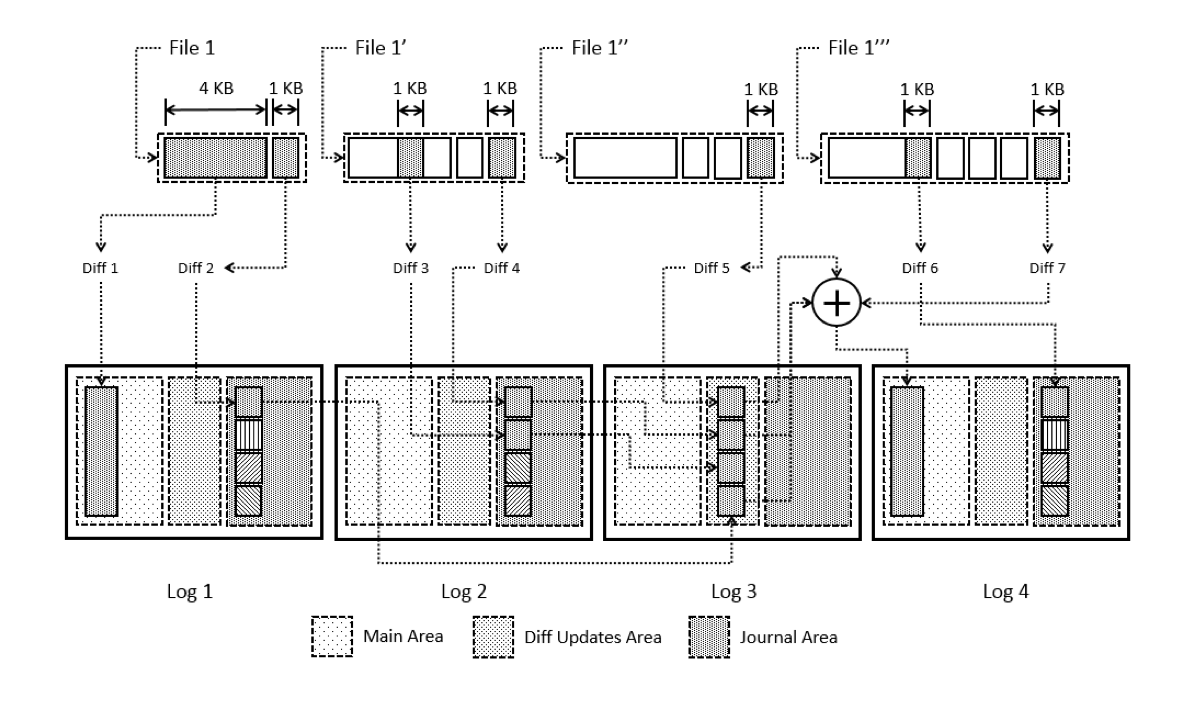}
\caption{Technique of journal and diff updates areas interaction in Diff-On-Write approach.}
\label{fig:fig058}
\end{figure}

Diff-On-Write approach provides especially good basis for decreasing write amplification factor in the case of mixed workloads. Let's assume that workload contains as adding data to the end of file as updating of internal areas of file (Fig. \ref{fig:fig058}). First of all, diffs can be stored into Journal area of different logs. Then diffs of one file can be moved into Diff Updates area with the goal to join updates of different areas of the file into one block. And, finally, a sequence of contiguous diffs from Diff Updates and Journal areas can be joined into one block of Main area.

\textbf{Deduplication}. Technique of deduplication is the well known and proven mechanism of exclusion of the duplicated content of files. The essence of this technique is the detection of data duplication on the basis of fingerprint calculation and comparison the calculated fingerprint value with the hash table of existing fingerprints. Generally speaking, the deduplication technique is very efficient mechanism of reducing the write amplification factor by virtue of the opportunity to share the same deduplicated content amongst the several files.

SSDFS file system uses the shared extents b-tree as the key mechanism of deduplication implementation. Generally speaking, the shared extents b-tree has the goal to keep a fingerprint value and an associated extent structure. The fingerprint value is used for comparison and detection of the duplication event but the extent structure is used for sharing the deduplicated data fragment amongst the different files. However, deduplication technique could be a compute-intensive task because a file system's volume could contain the small number of duplicated fragments or to have no duplications at all. Also the calculated fingerprint values need to keep in some metadata structure that has to be stored on file system's volume. As a result, the deduplication subsystem is capable to decrease the file system driver performance.

The architecture of SSDFS's deduplication subsystem is designed with taking into account the possible drawbacks. First of all, SSDFS file system driver calculates fingerprint value of the first 8 KB of the file only if the file is bigger than some threshold value (for example, 8 KB in total). The next step is searching the identical fingerprint value in shared extents b-tree.  If no fingerprint value has been found then the calculated fingerprint value should be stored into the shared extents b-tree. Moreover, the rest of the file is simply ignored by means of skipping the calculation of fingerprints. Oppositely, if it was found the identical fingerprint value for the first 8 KB of the file in shared extents b-tree then it needs to calculate the fingerprint values for the rest of file and to try to find the identical fingerprints in the tree. Again, if no identical fingerprints were found then it needs to store the calculated fingerprint values into the shared extents b-tree. But it needs to use the associated extent structures for the file's content deduplication in the case of detection the identical fingerprint values in shared extents b-tree. Generally speaking, it means that shared extents b-tree is ready to deduplicate the file's content only in the case of detection of third case of data duplication. Moreover, it means that file system volume will have two copies of identical data on the volume that could increase the reliability of data storing.

\subsection{GC Overhead Management}

\textbf{Garbage Collector (GC)} is inevitable subsystem of any LFS file system because of Copy-On-Write (COW) policy. Generally speaking, the simplified way of thinking about a volume of LFS file system is to imagine the volume like a sequence of logs are filling the volume's space sequentially. Moreover, the data update operations create the volume's state when old logs are the mixture of valid and invalid data (or completely invalid data). It means that the responsibility of GC subsystem is the moving valid data from old logs into the new ones and to erase completely invalid erase blocks (segments) with the goal to prepare the completely clear erase blocks for allocation for the new logs. Generally speaking, GC activity is the vital but auxiliary action that could compete with the regular file system's I/O operations. Finally, GC activity degrades the file system's performance dramatically and in completely unpredictable way. It is possible to say that GC overhead management problem is the crucial and the key problem of any LFS file system and it needs to be taken into account on initial stage of a file system's architecture design.

\textbf{Segment bitmap}. Any classic GC subsystem of LFS file system is implemented like a thread that selects in the background the aged segments with the goal to move valid data into a new log(s) and to apply the erase operation for these segments. Generally speaking, the important problem of such approach is to find an aged segment with as minimum as possible number of valid blocks because this is the possible strategy to manage the GC overhead efficiently. It needs to point out that SSDFS file system doesn't use this classical way of GC overhead management as the basic and fundamental mechanism of GC operations. However, the technique of searching of segment with minimal number of valid blocks can be used in the environment of critical lack of free space on the volume.

SSDFS file system uses segment bitmap as the basic metadata structure for searching the segments with minimum overhead for GC activity. The responsibility of segment bitmap is the tracking of segments' state (clean, using, used, pre-dirty, dirty). However, the key responsibility of a main GC thread is: (1) detecting the idle state of file system driver, (2) defining the total I/O budget that can be employed by GC subsystem, (3) selecting segments for processing by GC subsystem on the segment bitmap basis, (4) distribution of total I/O budget between the GC threads of particular PEBs. Finally, GC thread of particular PEB has to move gradually in the background the cold data on the basis of determined I/O budget.

The using state means that segment has free logical blocks. This state of segments doesn't need to be processed by GC subsystem. The used state means that the whole segment is filled by valid blocks. Finally, it means that such segment contains the cold data and it is the most expensive type of segments for processing by GC subsystem. However, flash-friendly file system could delegate the migration of such cold data on SSD side and not to process it by GC subsystem.

The dirty state means that no valid blocks exist in such segment and GC subsystem needs to apply only erase operation for all erase blocks in dirty segment. Generally speaking, it is the cheapest case of segment processing by GC subsystem and the dirty segments are the key target for the GC subsystem. The pre-dirty state means that segment contains as valid as invalid logical blocks. This segment's state has the lower priority for GC subsystem and this state will be used only in the case of complete absence of dirty segments. Finally, the key technique of processing the pre-dirty segment is to create the gradual migration of cold data by means of adding of GC operations to the regular I/O operations with data in the segment.

\textbf{PEBs migration scheme}. Migration scheme is the fundamental technique of GC overhead management in the SSDFS file system. The key responsibility of the migration scheme is to guarantee the presence of data in the same segment for any update operations. Generally speaking, the migration scheme's model is implemented on the basis of association an exhausted PEB with a clean one. The goal of such association of two PEBs is to implement the gradual migration of data by means of the update operations in the initial (exhausted) PEB. As a result, the old, exhausted PEB becomes invalidated after complete data migration and it will be possible to apply the erase operation to convert it in the clean state. Moreover, the destination PEB in the association changes the initial PEB for some index in the segment and, finally, it becomes the only PEB for this position. Namely such technique implements the concept of logical extent with the goal to decrease the write amplification issue and to manage the GC overhead. Because the logical extent concept excludes the necessity to update metadata is tracking the position of user data on the file system's volume. Generally speaking, the migration scheme is capable to decrease the GC activity significantly by means of the excluding the necessity to update metadata and by means of self-migration of data between of PEBs is triggered by regular update operations.

\textbf{Hot/warm data self-migration}. The important addition of migration scheme is a technique of hot/warm data self-migration. It means that any update operation in the environment of two PEBs' association results in the moving of data from the exhausted PEB into the new one. Finally, if a PEB contains only hot data then all data is able to migrate between PEBs by means of regular update operations without the necessity to employ the GC activity. Moreover, it is possible to delay the applying of erase operation to the completely invalidated PEB. However, the important peculiarity of such approach is to provide enough time for complete migration of valid data between PEBs. If a file system's volume contains enough clean PEBs then it will be possible to finish the data migration by means of regular update operations only (without using the GC service). However, if a PEB contains significant amount of cold valid blocks or volume hasn't enough clean PEBs then it needs to stimulate the migration process by means of GC activity. The key item of stimulation activity is the PEB's dedicated GC thread. Generally speaking, the responsibility of such GC thread is to orchestrate the gradual migration of cold data on the basis of allocated I/O budget for a particular GC thread. The goal of such approach is to minimize the GC threads' activity and to guarantee the stable file system driver's performance for regular I/O operations. Finally, this policy has to exclude the degradation of file system's performance because of GC threads' activity and to prepare enough free space for file system operations.

\subsection{Overprovisioning Management}

\textbf{Overprovisioning} is widely using technique of reservation some amount of SSD's erase blocks (for example, 20\% of the whole volume) with the goal to exchange the bad erase blocks on the good ones from the reserved pool. One of the critical reason of presence of bad erase blocks could be the high number of erase cycles because of write amplification issue and significant GC activity. Generally speaking, decreasing write amplification factor and elimination the GC activity is able to prolong the SSD lifetime because of capability to reduce the erase cycles number is used for auxiliary file system activity (for example, GC activity). Moreover, it also means the opportunity to prolong lifetime of the main pool of SSD's erase blocks. As a result, overprovisioning pool can be decreased or it could be used for prolongation of SSD lifetime.

\textbf{Pre-allocated state}. SSDFS file system introduces a special pre-allocated state of logical blocks. Generally speaking, the pre-allocated state defines the presence of some data portion or reservation without the allocation of the whole NAND flash page for the logical block. As a result, pre-allocated state provides the opportunity to reserve some space on the file system's volume without the real allocation (delayed allocation). The goal of pre-allocated state is not only to reserve some space (for metadata, for example) but it can be used for small files or compressed data portions. If some file or compressed data portion has lesser than 4 KB in size then such data portion can be marked as pre-allocated and the several data portions can be compacted or gathered into the one NAND flash page. Generally speaking, such compaction scheme is able to reduce the number of used NAND flash pages and, as a result, could decrease the overprovisioning and to prolong the SSD lifetime.

\textbf{Compression + delta-encoding}. SSDFS file system widely uses the compression for more compact representation of user data and metadata. Moreover, compression is added by compaction scheme with the goal to merge several compressed data fragments into one NAND flash page. Also SSDFS file system is trying to use a delta-encoding technique. This delta-encoding technique implies not to save the whole modified logical block (for example, 4 KB in size) but the extraction and saving only modified area (for example, 128 bytes) in this logical block. Finally, it means that it will be flushed on the volume only 128 bytes instead of 4 KB. SSDFS file system uses the delta-encoding technique with the compaction scheme for gathering several data portions into one NAND flash page. Finally, the goal of these techniques is to achieve the more compact representation of user data and metadata and to reduce the amount of write operations on the file system's volume. Generally speaking, it implies decreasing the number of erase cycles and the prolongation of SSD lifetime.

\subsection{Performance Management}

The whole SSDFS file system's design and architecture is trying to achieve the prolongation of SSD lifetime through decreasing the write amplification factor. Generally speaking, the suggested and implemented approaches are capable to improve the flush/write operations' performance. However, potential side effect of such efforts could be some reducing of read operations' performance. But asynchronous nature of read/write latency of NAND flash (read operations is faster and no seek operation penalties) gives a steady basis to expect a good performance of read operations for the case of SSDFS file system's architecture. Moreover, aggregation of several PEBs into one segment, PEB's dedicated threads model, GC I/O budget model provide the rich opportunities for achieving a good file system's performance.

Any SSD represents multi-die and multi-channel architecture. If some protocol of interaction with SSD (for example, open-channel SSD model) shares the knowledge of distribution erase blocks among NAND dies then file system is able to employ this knowledge for enhancing the I/O operations performance. SSDFS file system uses the technique of aggregation several PEBs inside of one segment. Generally speaking, if one segment aggregates several PEBs from different NAND dies then such approach provides the way to process the I/O requests in parallel by different NAND dies. As a result, it is capable to improve the performance of I/O requests in the scope of one segment significantly.

\section{DISCUSSION}

\textbf{SSDFS file system} has goals: (1) manage write amplification in smart way, (2) decrease GC overhead, (3) prolong SSD lifetime, and (4) provide predictable file system's performance. To implement these goals SSDFS file system introduces several authentic concepts and mechanisms: logical segment, logical extent, segment's PEBs pool, Main/Diff/Journal areas in the PEB's log, Diff-On-Write approach, PEBs migration scheme, hot/warm data self-migration, segment bitmap, hybrid b-tree, shared dictionary b-tree, shared extents b-tree.

\textbf{It has been shown that 80\% or more of the files are smaller than 32 KB}. To manage this peculiarity, SSDFS file system uses inode's private area to store the small files inline. Moreover, it was introduced a special compaction scheme that gathers several small files into one NAND flash page. Also, SSDFS file system uses the block-level compression with addition of the compaction scheme that keeps several compressed portions into one NAND flash page. Additionally, it is employed delta-encoding, compaction scheme, and deduplication for the case of big files.

\textbf{The vast majority of files are deleted within a few minutes of their creation}. One of the efficient technique of management such case is using inode's private area for keeping inline files. Default raw SSDFS inode is able to store about 128 bytes of file's content. But the bigger size of raw inode is able to provide more space for inline files. Moreover, SSDFS file system gathers content of files into specialized user data segment. As a result, deletion of files creates the self-invalidation effect for the case of user data segment that can decrease the GC activity or completely eliminate the GC overhead through PEB migration scheme.

\textbf{The median file age ranges between 80 and 160 days. 0.8\% of the files are used essentially every day}. Flash-friendly file system doesn't need to follow by strict wear-leveling scheme. It makes sense to delegate moving the cold data by SSD's FTL once in 3 months (90 days). Extensive using the GC operations and strict wear-leveling scheme for moving cold data by LFS increases the write amplification issue. PEB's migration scheme is able to provide free space in cost-efficient manner. Compression and delta-encoding is combined with PEB's migration scheme is able to provide easy and efficient mechanism for combining in one PEB as hot as cold data and to guarantee the space for gathering hot data updates with fast/easy migration between PEBs.

\textbf{Several research works showed the growing of files count per file system and directories count per file system. It needs to expect as minimum 30K - 90K files per file system and 1K - 4K directories per file system.} To manage the growing demands for number of files and folders on the file system volume, SSDFS file system employs inodes b-tree that provides the way for easy increasing number of files and efficient management for the case of frequent delete/remove operations. Also, inline files provides the way to store the small files in inode itself without allocation of volume's space.

\textbf{File name length falls in the range from 9 to 17 characters. The peak occurs for file names with length of 12 characters.} SSDFS's dentry is able to include 12 inline characters. The tail of longer name is stored into shared dictionary. The fixed size of dentry provides the efficient mechanism of dentries management. Mostly, file names will be stored into dentries only.

\textbf{23-25\% of directories contain no files. 65-67\% of directories contain no subdirectories. 46-49\% of directories contain two or fewer entries.} SSDFS raw inode is able to keep two inline dentries. It means that very frequently raw inode is able to store the content of dentries tree. Moreover, SSDFS b-tree node is stored compressed. As a result, it implies that small dentries b-tree could be represented in very efficient way on the file system's volume.

\textbf{There are many files deep in the namespace tree, especially at depth 7.} SSDFS b-tree node stores several raw inodes. It means that operation of reading one b-tree node is able to provide access to the whole or part of namespace tree. Also, SSDFS gathers b-tree node of the same type in one segment/PEB. As a result, the readahead operation is able to read several b-tree nodes. It implies that several contiguous b-tree nodes could contain the whole namespace tree. Finally,  SSDFS raw inode is able to keep  the content of dentries tree inline.

\textbf{Many end-users have file system volume is on average only half full.} PEBs migration technique is trying to employ this fact. It means that PEBs association during migration can be done without the affection of availability of free space on the volume. Moreover, SSDFS uses compression and delta-encoding. Finally, it provides good basis for the PEBs migration technique.

\textbf{On average, half of the files in a file system have been created by copying without subsequent writes.} It is possible to conclude that user data is mostly cold but metadata is mostly hot. SSDFS uses the model of current segments of different types. It means that user data is aggregated in one segment but metadata is aggregated into another one. As a result, user data segment will be managed under cold data policy but metadata segment will be managed under hot data policy. SSDFS uses three area types in the log (main area, diff updates area, journal area). It provides the efficient way to manage data for the case of mixed nature of data (cold and hot) into one log. SSDFS is flash-friendly file system and it doesn't move segment with cold data as part of GC activity. SSDFS delegates error correction and read block reclaiming on FTL side. Also, deduplication is able to exclude the replication of existing files on the volume.

\textbf{Modern applications manage large databases of information organized into complex directory trees (A File Is Not a File).} First of all, SSDFS is able to manage case of mixed nature of data efficiently by means of three area types in the log (main area, diff updates area, journal area). Also, using delta-encoding technique provides the way to store only updated portion(s) of data. Moreover, segment is able to contain several PEBs from different NAND dies. It means that different extents of a file can be stored into different PEBs and to implement the parallelism of operations. SSDFS associates read/write threads with PEBs that implements parallelism as on file system as on SSD level.

\textbf{Applications help users create, modify, and organize content, but user files represent a small fraction of the files touched by modern applications. Most files are helper files that applications use to provide a rich graphical experience, support multiple languages, and record history and other metadata.} Auxiliary files of the same application can be aggregated into one segment/PEB. It means that readahead operation will be able to extract the content of all auxiliary files from one segment/PEB. SSDFS's segment is able to contain several PEBs from different NAND dies. As a result, this approach is capable to implement parallelism of read operation as on file system as on SSD level.

\textbf{Most written data is explicitly forced to disk by the application; for example, iPhoto calls fsync thousands of times in even the simplest of tasks.} SSDFS supports partial logs. It means that file system driver tries to prepare the full log before flushing on the volume. However, the file system driver is able to prepare the partial logs in the case of fsync requests or synchronous mount. The partial logs could increase amount of metadata on the volume. SSDFS supports several types of current segments. It means that metadata and user data will be processed simultaneously in different threads. PEB has associated flush thread. As a result, the update operations in different PEBs will be processed by different threads in multi-threaded environment.

\textbf{It has been shown that applications create many temporary files.} From one point of view, it is possible to consider adding an additional type of current segment for storing the temporary files. Finally, it means that such type of segment will be invalidated completely. And it will make the GC activity for such type of segments very cheap. But it will be much efficient way to keep the temporary files in page cache without flushing on the volume.

\textbf{Home-user applications commonly use atomic operations, in particular rename, to present a consistent view of files to users.} As a result, it is possible to expect more frequent of metadata's update operations (hot data). Finally, PEBs migration technique could migrate updated metadata between PEBs without necessity to use the GC activity.

\textbf{Write amplification issue}. SSDFS file system uses such techniques for resolving the problem of write amplification issue: (1) compression, (2) small files compaction scheme, (3) logical extent concept, (4) Diff-On-Write approach, (5) deduplication, (6) inline files.

The logical extent concept is the technique of resolving the write amplification issue for the case of LFS file system. It means that any metadata structure keeping a logical extent doesn't need in updating the logical extent value in the case of data migration between the PEBs because the logical extent remains the same until the data is living in the same segment. The migration mechanism implements the logical segment and logical extent concepts with the goal to decrease or completely eliminate the write amplification issue. Moreover, SSDFS file system is widely using the data compression, delta-encoding technique, and small files compaction technique that provides the opportunity to employ the PEB migration mechanism without the necessity to use the additional overprovisioning.

Moreover, b-tree metadata structure provides the way not to keep an unnecessary reserve of metadata space on the volume. As a result, it means the exclusion of management operations of reserved metadata space (moving from a PEB to another one) with the goal to support it in the valid state. Generally speaking, it is the way to decrease the amount of PEBs' erase and write operations.

SSDFS file system uses a special compaction scheme which gathers several compressed fragments (even for different files) into one NAND flash page inside of special log's area (diff update or journal areas). Generally speaking, this compaction technique provides the opportunity to use only one NAND flash page for several compressed fragments of different files instead of several ones. As a result, the decreasing number of used NAND flash pages decreases number of I/O operations and it creates the opportunity to reduce the write amplification issue.

SSDFS file system introduces a special compaction scheme for the case of small files. Generally speaking, PEB's log can contain a special journal area that is used for gathering into one NAND flash page the several small files. As a result, this compaction technique reduces the number of I/O operations and is able to decrease the factor of write amplification issue. Mechanism of keeping data inline in inode's private area is the way to reduce the write amplification issue and to improve the file system's performance.

\textbf{GC overhead management}. There are several type of segments on any SSDFS file system's volume: (1) superblock segment, (2) snapshot segment, (3) PEB mapping table segment, (4) segment bitmap, (5) b-tree segment, (6) user data segment. Generally speaking, the goal to distinguish the different type of segments is to localize the peculiarities of different types of data (user data and metadata, for example) inside of specialized segments.

Migration scheme is the fundamental technique of GC overhead management in the SSDFS file system. The key responsibility of the migration scheme is to guarantee the presence of data in the same segment for any update operations. Generally speaking, the migration scheme is capable to decrease the GC activity significantly by means of the excluding the necessity to update metadata and by means of self-migration of data between of PEBs is triggered by regular update operations. The important addition of migration scheme is a technique of hot/warm data self-migration. It means that any update operation in the environment of two PEBs' association results in the moving of data from the exhausted PEB into the new one. Finally, if a PEB contains only hot data then all data is able to migrate between PEBs by means of regular update operations without the necessity to employ the GC activity.

However, if a PEB contains significant amount of cold valid blocks or volume hasn't enough clean PEBs then it needs to stimulate the migration process by means of GC activity. The key item of stimulation activity is the PEB's dedicated GC thread. Generally speaking, the responsibility of such GC thread is to orchestrate the gradual migration of cold data on the basis of allocated I/O budget for a particular GC thread. The goal of such approach is to minimize the GC threads' activity and to guarantee the stable file system driver's performance for regular I/O operations. Finally, this policy has to exclude the degradation of file system's performance because of GC threads' activity and to prepare enough free space for file system operations.

The compaction of several fragments of different logical blocks into one NAND flash page creates the capability to move more data for one GC operation. From another viewpoint, warm/hot areas introduce the areas with high frequency of update operations. Generally speaking, it is possible to expect that high frequency of update operations (in diff updates and journal areas) creates the natural migration of data between PEBs without the necessity to use the extensive GC operations.

\textbf{SSD lifetime}. Decreasing write amplification factor and elimination the GC activity is able to prolong the SSD lifetime because of capability to reduce the erase cycles number is used for auxiliary file system activity (for example, GC activity). Moreover, it also means the opportunity to prolong lifetime of the main pool of SSD's erase blocks. As a result, overprovisioning pool can be decreased or it could be used for prolongation of SSD lifetime. SSDFS file system uses the delta-encoding technique with the compaction scheme for gathering several data portions into one NAND flash page. Finally, the goal of these techniques is to achieve the more compact representation of user data and metadata and to reduce the amount of write operations on the file system's volume. Generally speaking, it implies decreasing the number of erase cycles and the prolongation of SSD lifetime.

\textbf{File system performance}. The whole SSDFS file system's design and architecture is trying to achieve the prolongation of SSD lifetime through decreasing the write amplification factor. Generally speaking, the suggested and implemented approaches are capable to improve the flush/write operations' performance. However, potential side effect of such efforts could be some reducing of read operations' performance. But asynchronous nature of read/write latency of NAND flash (read operations is faster and no seek operation penalties) gives a steady basis to expect a good performance of read operations for the case of SSDFS file system's architecture. Moreover, aggregation of several PEBs into one segment, PEB's dedicated threads model, GC I/O budget model provide the rich opportunities for achieving a good file system's performance. Any SSD represents multi-die and multi-channel architecture. If some protocol of interaction with SSD (for example, open-channel SSD model) shares the knowledge of distribution erase blocks among NAND dies then file system is able to employ this knowledge for enhancing the I/O operations performance. SSDFS file system uses the technique of aggregation several PEBs inside of one segment. Generally speaking, if one segment aggregates several PEBs from different NAND dies then such approach provides the way to process the I/O requests in parallel by different NAND dies. As a result, it is capable to improve the performance of I/O requests in the scope of one segment significantly.

\section{CONCLUSION}

Solid state drives have a number of interesting characteristics. However, there are numerous file system and storage design issues for SSDs that impact the performance and device endurance. Many flash-oriented and flash-friendly file systems introduce significant write amplification issue and GC overhead that results in shorter SSD lifetime and necessity to use the NAND flash overprovisioning. SSDFS file system introduces several authentic concepts and mechanisms: logical segment, logical extent, segment's PEBs pool, Main/Diff/Journal areas in the PEB's log, Diff-On-Write approach, PEBs migration scheme, hot/warm data self-migration, segment bitmap, hybrid b-tree, shared dictionary b-tree, shared extents b-tree. Combination of all suggested concepts are able: (1) manage write amplification in smart way, (2) decrease GC overhead, (3) prolong SSD lifetime, and (4) provide predictable file system's performance.

\section{FUTURE WORK}

Currently, SSDFS file system driver is not fully functional and is not completely implemented. It still needs in bug fix. Diff-On-Write approach is implemented only partially. Deduplication and snapshot support is not implemented yet. Additionally, SSDFS file system hasn't fsck tool.

\section{SOURCE CODE}

SSDFS project information is available online (\url{http://www.ssdfs.org/ssdfs.html}). Source code of user-space tools is available in \url{https://github.com/dubeyko/ssdfs-tools.git}. Source code of file system driver is available in \url{https://github.com/dubeyko/ssdfs-driver.git}. Source code of Linux kernel with integrated SSDFS file system driver is available in \url{https://github.com/dubeyko/linux.git}.

\section{ACKNOWLEDGEMENTS}

The author gratefully acknowledge the initial support of the idea by Zvonimir Bandic and Cyril Guyot.






\end{document}